\documentclass[journal,10pt,onecolumn,draftclsnofoot,]{IEEEtran}

\usepackage{enumitem}
\usepackage{threeparttable}
\usepackage{multirow}
\usepackage{multicol}
\usepackage[pdftex]{graphicx}

\usepackage{booktabs}
\usepackage{subfigure}

\usepackage{amssymb}
\usepackage{color}
\usepackage{amsmath}
\usepackage{cite}
\usepackage[utf8]{inputenc}
\usepackage[english]{babel}
 \usepackage{float}
\usepackage{amsthm}
\renewcommand{\vec}[1]{\boldsymbol{#1}}

\newcommand{\tabincell}[2]{\begin{tabular}{@{}#1@{}}#2\end{tabular}}

\newtheoremstyle{mystyle}
  {}
  {}
  {\itshape}
  {}
  {\bfseries}
  {.}
  { }
  {\thmname{#1}\thmnumber{ #2}\thmnote{ (#3)}}

\theoremstyle{mystyle}
\newtheorem{theorem}{Theorem}
\newtheorem*{lemma*}{Lemma}
 \newcommand{\Cov}{\mathrm{Cov}}
\newtheorem{corollary}{Corollary}
\newtheorem{lemma}{Lemma}
\newtheorem{definition}{Definition}

\newtheorem{proposition}{Proposition}
\addto\captionsenglish{}

\ifodd 0
    \newcommand\rev[1]{{\color{blue}#1}}
    \newcommand{\com}[1]{\textbf{\color{red} (COMMENT: #1)}} 
\else
    \newcommand\rev[1]{{#1}}
    
    \newcommand{\com}[1]{}
\fi

\begin{document}
\title{Efficient and Robust Equilibrium Strategies of Utilities in Day-ahead Market with Load Uncertainty}

\author{Tianyu~Zhao, Hanling~Yi, Minghua~Chen, Chenye~Wu, and Yunjian~Xu  
\thanks{Tianyu Zhao and Hanling Yi are with Information Engineering, The Chinese University of Hong Kong. Minghua Chen is with School of Data Science, City University of Hong Kong. Chenye Wu is with School of Science and Engineering, The Chinese University of Hong Kong, Shenzhen. Yunjian Xu is with Mechanical and Automation Engineering, The Chinese University of Hong Kong.}}
\providecommand{\keywords}[1]{\textbf{\textit{Index terms---}} #1}
\markboth{}
{}

\maketitle
\IEEEpeerreviewmaketitle

\begin{abstract}
We consider the scenario where $N$ utilities strategically bid for electricity in the day-ahead market and balance the mismatch between the committed supply and actual demand in the real-time market, with uncertainty in demand and local renewable generation in consideration. 
We model the interactions among utilities as a non-cooperative game, in which each utility aims at minimizing its per-unit electricity cost. \rev{We investigate utilities' optimal bidding strategies and} show that all utilities bidding according to (net load) prediction is a unique pure strategy Nash Equilibrium with two salient properties. First, it incurs no loss of efficiency; hence, competition among utilities does not increase the social cost. Second, it is robust and (0, $N-1$) fault immune. That is, fault behaviors of irrational utilities only help to reduce other rational utilities' costs. \rev{The expected market supply-demand mismatch is minimized simultaneously, which improves the planning and supply-and-demand matching efficiency of the electricity supply chain.} We prove the results hold under the settings of correlated prediction errors and a general class of real-time spot pricing models, which capture the relationship between the spot price, the day-ahead clearing price, and the market-level mismatch. Simulations based on real-world traces corroborate our theoretical findings. \rev{Our study adds new insights to market mechanism design. In particular, we derive a set of fairly general sufficient conditions for the market operator to design real-time pricing schemes so that the interactions among utilities admit the desired equilibrium.}
\end{abstract}

\section{Introduction}

\begin{figure}
\centering
\includegraphics[width = 0.65\linewidth]{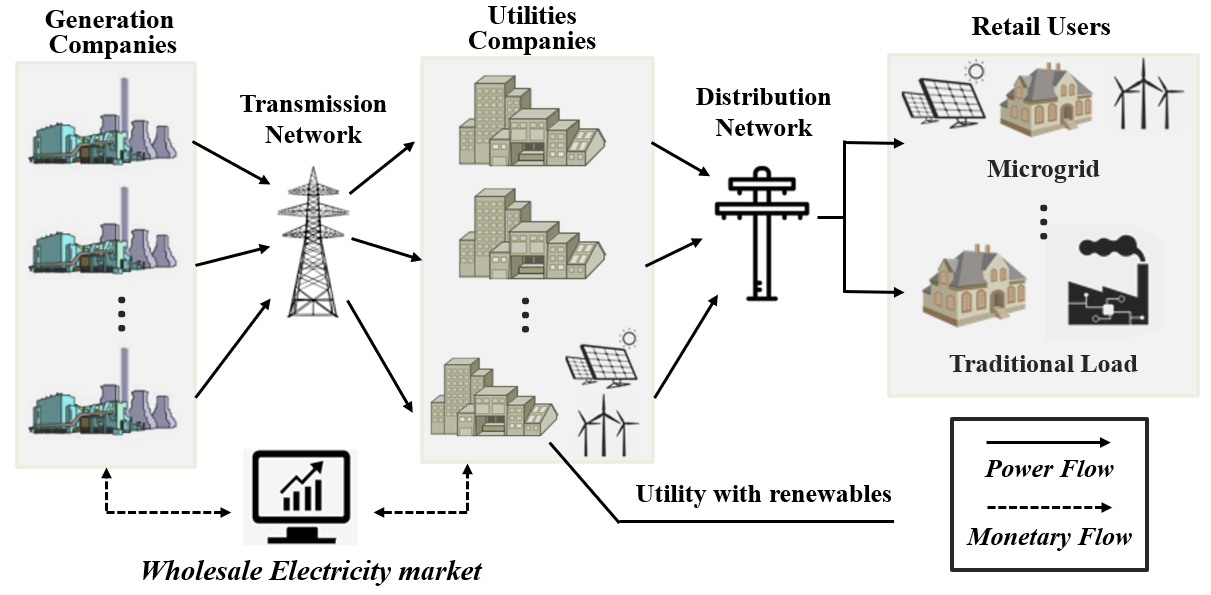}
\caption{The deregulated electricity supply chain.}
\label{fig1}
\end{figure}

Modern power system has been actively practicing deregulated electricity supply chain since the reform in 1990's~\cite{gilbert2007international}. As illustrated in Fig.~\ref{fig1}, the deregulated supply chain usually consists of  generation companies, utility companies, and sectors in charge of transmission and distribution networks~\cite{kirschen2018fundamentals}. In particular, utilities obtain power supply from the regional electricity market and local renewable sources to serve households and microgrids. The market operator (known as independent system operator (ISO), e.g., \rev{ISO New England (ISO-NE)}~\cite{ISONE}) provides a trading place and matches the supply offers and demand bids at two different timescales and prices~\cite{Nordpool,stoft2002power,zhang2010restructured}.
\begin{itemize}
    \item {Day-Ahead Market}: Generation companies (utilities) submit offers (bids) for selling (buying) electricity one day before the actual dispatch, based on generation (net load) forecasting. They are cleared at a market clearing price.
    \item {Real-Time Market}: Real-time market is designed to resolve the imbalance between the actual real-time demand and the committed supply purchased from the day-ahead market, on an hourly basis. We remark that the real-time electricity price depends on both the day-ahead clearing price as well as the real-time market level imbalance.
\end{itemize}
\rev{New York Independent System Operator (NYISO)} reports that while roughly $95\%$ of the electricity load is scheduled 
in the day-ahead market transaction, $5\%$ of that remains to be settled in the real-time market~\cite{ISONEReport}. The cost of the utility is composed of the payment in the day-ahead market and the expense in the real-time market to resolve the imbalance.\footnote{Similar to ~\cite{herranz2012optimal,song2016purchase,fleten2005constructing,hajati2011optimal,nojavan2015optimal,nojavan2015robust}, we do not consider the intra-day market as its trading quantity is negligible as compared to the day-ahead and real-time markets.}

Although the existing electricity market is not a free market and is regulated to a certain degree, utilities can still play strategically to lower their own costs. In particular, in the day-ahead market, utilities can overbuy (respectively underbuy) electricity given the load forecasting results, expecting they can sell the surplus in the real-time market at a higher price (respectively buy the shortage at a lower price). We remark that utilities interact with each other in this process. The cost of a utility depends on both its own bidding strategy and those of all other utilities since the latter affect the real-time market-level imbalance and consequently the real-time spot price.

\subsection{Related Work}\label{ssec:related_work}

\rev{There have been a number of works studying the optimal bidding strategies of generation companies in order to minimize their operating costs on the \emph{supply side} of the electricity supply chain~\cite{xu2015efficient, kazempour2013equilibria, zhang2018competition,rasouli2019efficient,mazzi2017price,steeger2014optimal,shrestha2009generation,yucekaya2013bidding,conejo2002price,vilim2014wind,papavasiliou2013multiarea,pritchard2010single,lee2011firm,lavaei2012competitive,kumbartzky2017optimal,boomsma2014bidding,nazari2019optimal}; see e.g.,~\cite{steeger2014optimal} for a survey. In particular, generation companies derive their optimal decisions by solving the scheduling problems that seek optimal production schedules. For example, study~\cite{kazempour2013equilibria} proposes an equilibrium problem with equilibrium constraints (EPEC) framework to analyze the strategic behaviors of electricity producers owning dispatchable wind power units. Study~\cite{zhang2018competition} formulates the Cournot competition model to analyze the interactions among generation firms with production uncertainty. The impact of coalition on the game equilibrium is explored.
{As the intermediary between consumers and the deregulated electricity market, utilities' optimal bidding strategies and their interactions have also been actively investigated in many studies on the \emph{demand side} of the electricity market, mainly on two aspects.}}

The first is on the optimal bidding strategies of utilities. This includes characterizing the maximum profit for individual utility in the market~\cite{ventosa2005electricity,rebennack2010energy} and designing optimal bidding decisions~\cite{david2000strategic,philpott2006optimizing,atzeni2014noncooperative}. 
\rev{We note that the existing works mainly focus on formulating the demand side optimization problems to obtain utilities' optimal decisions under various scenarios, e.g., time-series methods, data-driven predictions, and genetic algorithms are discussed in~\cite{persson2018simplify, herranz2012optimal}. However, the analysis on the market equilibrium structure and properties may not be included, especially considering demand uncertainty in the two-settlement electricity market. \textit{The first category} is investigating the optimization methods and frameworks considering demand response~\cite{samadi2013tackling,aghaei2013demand, kamyab2016demand,nojavan2015optimal,dowling2017multi,li2015demand,song2016purchase,jin2018microgrid}. For example, a short-term planning model to determine the bidding strategies for utilities with flexible power demand is proposed in~\cite{song2016purchase}.  
\textit{The second category} is applying stochastic programming models to incorporate price and demand uncertainties~\cite{carrion2009bilevel,zavala2017stochastic,khazaei2017single,habibian2020multistage,safdarian2013stochastic}. This includes deriving utilities' optimal strategies and the market clearing mechanism under various probabilistic scenarios. For example, study~\cite{habibian2020multistage} presents a multistage stochastic demand-side management model to solves a large-scale extended time-horizon strategic electricity consumption scheduling problem. The impact of the strategic behavior of a major consumer with flexible load is shown. A stochastic market clearing mechanism that considers deviation costs and demand uncertainty is introduced in~\cite{khazaei2017single}. 
\textit{The third category} focuses on the market equilibrium characterization~\cite{wang2019equilibrium,khazaei2017single,hu2007using,atzeni2014noncooperative,9029514}, which involves utilities' interactions, e.g., the authors in~\cite{hu2007using} employ the EPEC optimization model to represent the interaction between the market participants. The sufficient conditions for the existence of such equilibrium are established. Study~\cite{atzeni2014noncooperative} formulates the day-ahead grid optimization as a generalized Nash equilibrium problem (GNEP). Conditions for the existence of GENP solutions and the converging algorithm are proposed. Besides focusing on attaining the market equilibrium, the authors in~\cite{srinivasan2014bidding} propose a coevolutionary approach to investigate individual and cooperative strategies of utilities in a power market. The impact of cooperative behavior by forming coalitions is shown.

In addition to the three categories described above, there has been a line of research on studying utilities' behaviors~\cite{fang2016strategic,wei2014energy} and market efficiency~\cite{roozbehanit2011analysis} under some specific scenarios. Study~\cite{fang2016strategic} proposes a scheduling model for utilities' energy storage operation with CVaR as a risk metric to cope with price and load uncertainties.
\cite{roozbehanit2011analysis} characterizes a real-time retail pricing model to investigate the competitive electricity market stability and efficiency. The interactions between utilities, consumers, and wholesale electricity market are studied in~\cite{wei2014energy}. 

Due to the uncertain demand and the fast penetration of intermittent renewables, there is a series of studies further study the impact of load uncertainty, i.e., how renewables and demand uncertainty would affect utilities' net load predictions~\cite{yi2018impact} and costs~\cite{cai2013impact}, besides containing such modeled uncertainty~\cite{ventosa2005electricity} in the optimization programs discussed above.  It is observed that renewable penetration is likely to deteriorate load prediction errors~\cite{cai2013impact}, and consequently, net load uncertainty will affect utilities' costs~\cite{yi2018impact,cai2013impact, atia2016sizing, jia2016renewables,lin2011potential,ye2019combined}. {For instance, in our previous work~\cite{yi2018impact}, H. Yi \emph{et al.} show that larger  renewable penetration degrades prediction accuracy,
leading to higher load uncertainty. The corresponding local impact and global impact of such uncertainty are investigated. It is demonstrated that load uncertainty can cause an increase in utilities' costs} }

It should be noted that electricity price forecasting is also
fundamental for energy companies’ decision-making mechanisms~\cite{mohsenian2010optimal,motamedi2012electricity}. Utilities involved in the electricity market are faced with the uncertainty challenge of volatile power market prices in their daily operations, and prior knowledge of such price fluctuations can help them set up corresponding bidding strategies so as to maximize their profit~\cite{wan2013hybrid}. Hourly energy price is a complex signal due to its nonlinearity, nonstationarity, and time-variant behavior~\cite{amjady2008day,anbazhagan2012day}. Multiple probabilistic price forecasting models and frameworks are proposed to approximate the parameters of the probability density function for the hourly price variables based on past prices, power loads, chronological information, etc~\cite{monteiro2018new, jonsson2014predictive,chai2018conditional}.


Our work differs from the existing literature in that, we consider the demand side bidding under the game-theoretical setting with load uncertainty in the two-settlement electricity market and we investigate the uniqueness and efficiency of the market equilibrium according to its structure. Equilibrium robustness and the aggregate impact of all utilities’ bidding strategies on the individual utility’s cost are also studied.
\rev{The explicit form of utilities' optimal bidding strategies and the market equilibrium can be derived. Furthermore, besides designing the optimal decisions of utilities, our study adds new insights to market mechanism design and market operation. In particular, we show that if the market operator can design the real-time pricing scheme to meet a set of fairly general sufficient conditions described in Sec.~\ref{sec:general}, then the utility bidding game will admit a unique pure Nash Equilibrium without loss of efficiency, and it is robust to irrational utilities' fault behaviors. Meanwhile, the expected total market supply-demand mismatch can be minimized simultaneously, which improves the planning and supply-and-demand matching efficiency of the electricity supply chain.}


\subsection{Contributions}
We formulate the interactions among utilities as a non-cooperative game. Utilities aim at minimizing individual costs by optimizing  their own bidding strategies, taking into account uncertainty in load and local renewable generation. We seek answers to three critical questions:
\begin{itemize}
\item What is the outcome of such a utility bidding game? In particular, does there exist a pure strategy Nash  Equilibrium\footnote{A pure strategy corresponds to the common practice that individual utility places one bidding curve (quantity) for each trading hour in day-ahead market.}? If so, is it unique? 
\item Does competition introduce efficiency loss compared with the social optimal under the coordinated setting? That is, what is the loss of efficiency\footnote{The ratio between the social cost at the equilibrium and the social optimal quantifies the loss of efficiency due to competition~\cite{roughgarden2004bounding}.} at the equilibrium?
\item How robust is the equilibrium against irrational fault behaviors\footnote{An equilibrium is ($\epsilon$, $K$) fault immune if
non-fault utilities' expected costs increase by at most $\epsilon$ when up to $K$ other utilities deviate arbitrarily~\cite{gradwohl2008fault}.}? In particular, will rational utilities suffer in a fault-ridden setting with irrational utilities?
\end{itemize}

Answers to these questions provide new understandings of the effectiveness  of  the  electricity  market  design, as well as the impact of load uncertainty. We conduct a comprehensive study and make the following contributions.

To better bring out the insights and intuitions, we first focus on a baseline setting where the load prediction errors across utilities are mutually independent and the spot pricing model, describing the relationship among the spot price, day-ahead clearing price, and real-time market-level mismatch, is a class of piece-wise linear functions similar to the ones in~\cite{skytte1999regulating,Neupane}. 
\begin{itemize}
    \item 
    \rev{We show that all utilities bidding according to (net load) prediction is a unique pure strategy Nash Equilibrium.}
    \item 
    \rev{We show that the  Nash Equilibrium incurs no loss of efficiency, i.e., the social cost of the equilibrium is the same as the optimal one under the coordinated setting. Furthermore, the equilibrium is robust and (0, $N-1$) fault immune. That is, irrational fault behaviors of any subset of the utilities only help reduce the costs of the remaining rational utilities~\cite{gradwohl2008fault}}. 
\end{itemize}
We then generalize the results to the setting with correlated prediction errors and general pricing models.
\begin{itemize}
    \item 
    \rev{We present a set of sufficient conditions on the spot pricing model for observing the unique, efficient, and (0, $N-1$) fault immune robust pure strategy Nash Equilibrium. In particular, we show that our above results hold for correlated prediction errors and  a general class of real-time spot pricing models which can be nonlinear.}
    \end{itemize} 
 In Sec.~\ref{sec:simulation}, we conduct extensive simulations based on price and load data from the ISO-NE electricity market. The  results corroborate our theoretical findings and highlight that it is possible to design effective real-time pricing schemes satisfying the sufficient conditions derived in our paper, such that the interactions among utilities admit a unique, efficient, and robust pure strategy Nash Equilibrium. 
 
 \rev{The structure of this paper is as follows. We first introduce the utility bidding game formulation in Sec.~\ref{sec:formulation}. Based on the proposed game-theoretical model, Sec.~\ref{sec:equilibrium} investigates the desired properties of the equilibrium. Sec.~\ref{sec:general} provides the generalized conditions to still admit the preferred equilibrium. Experimental results are included in Sec.~\ref{sec:simulation}. Sec.~\ref{conclusion} delivers the concluding remarks. Due to the space limitation, all proofs and the summary of theoretical results are presented in the appendix.}

\section{Problem Formulation}\label{sec:formulation}
We consider the scenario where $N$ utilities strategically bid for electricity in the day-ahead market and balance the mismatch between the committed supply and actual demand in the real-time market, with uncertainty in demand and local renewable generation in consideration.  Since the transactions are settled on an hourly basis, we focus on a particular hour without loss of generality. We use $p_{d}$ and $p_{s}$ (unit: \$/MWh) to denote the corresponding day-ahead price and the spot price, respectively. 
Similar to~\cite{hajati2011optimal,nojavan2015robust,herranz2012optimal,fleten2005constructing,carrion2009bilevel,nojavan2015optimal}, we consider the setting where a single utility does not have market power to manipulate the day-ahead clearing prices, i.e., utilities are considered as price-takers in the day-ahead market. This setting is consistent with the situation in many liberalized electricity markets where the trading volume of most utilities is too small to influence the price as compared to the total market turnover~\cite{song2016purchase, fleten2005constructing,bang2012existing}. We also consider that utilities' bidding strategies can affect the real-time market spot price, i.e., utilities are viewed as price-makers in the real-time market. 
\rev{This is reasonable as utilities' strategic bidding behaviors can have significant impact in real-time balancing markets where only a small amount of energy is traded~\cite{herranz2012optimal,arteaga2019price}.} 
In practice, at the time of day-ahead bidding, utilities usually do not know precisely its load and local renewable generation. Instead, they have distributional information of the net load by load forecasting. For ease of analysis, we first focus on the setting that utilities only place one quantity bid for each trading hour in the day-ahead market. 
Our results can be generalized to the case in which utilities submit demand curves as bidding pairs (price, quantity); {see Sec.~\ref{ssec:bidding curve} for a discussion.}


\subsection{Utility bidding and Load Mismatch Modeling}
We define $D_{i}$ (unit: MWh) as utility $i's$ actual {net load} at a particular hour, and it is only revealed to the utility in the real-time manner. When utility $i$ participates in day-ahead market, it has a prediction of $D_{i}$, denoted as $\hat{D_{i}}$, modeled as follows:
\begin{equation}\label{equation1}
\begin{split}
\hat{D_{i}}=D_{i}+\epsilon_{i},
\end{split}
\end{equation}
where $\epsilon_{i}$ is a random variable representing the load prediction error. It is affected by the uncertainties of demand and local renewable generation owned by the utilities and microgrids.

Given the load prediction $\hat{D_{i}}$, utility $i$ can strategically participate in the day-ahead market by bidding a quantity $Q_{i}\triangleq\hat{D}_{i}-\mu_{i},$  where
\begin{itemize}
\item $\mu_{i}=0$: Utility bids precisely according to prediction.

\item $\mu_{i}>0$: Utility strategically underbuys.

\item $\mu_{i}<0$: Utility strategically overbuys.
\end{itemize}

We use $\mu_i$ to represent  the bidding strategy of utility $i$.

The bidding strategy of utility $i$ will affect its mismatch between the real-time actual net load and the day-ahead purchased supply in the real-time market, which is denoted as $\Delta_{i}$ (unit: MWh). By definition, we have
\begin{equation}\label{equation2}
\begin{split}
\Delta_{i}\triangleq D_{i}-Q_{i}=\mu_i-\epsilon_{i}.
\end{split}
\end{equation}

Whenever there is an imbalance, i.e., $\Delta_{i}\neq0$, the utility has to settle this imbalance in the real-time market at the spot price $p_{s}$. It either sells the residual electricity back to the market when $\Delta_{i}<0$, or buys the deficient electricity from the market when $\Delta_{i}>0$. For ease of presentation, we define
\begin{equation}\label{equation3}
\begin{split}
\Delta\triangleq\sum_{i=1}^{N}\Delta_{i}, \ \mbox{and}\  \Delta_{-i}\triangleq\sum_{j:j\neq i}^{N}\Delta_{j}
\end{split}
\end{equation}
as the market-level mismatch and the aggregate mismatch of all other utilities except utility $i$, respectively.

\subsection{Real-time Market Spot Pricing Model}
\begin{figure}  [H]
\centering
\includegraphics[width = 0.35\linewidth]{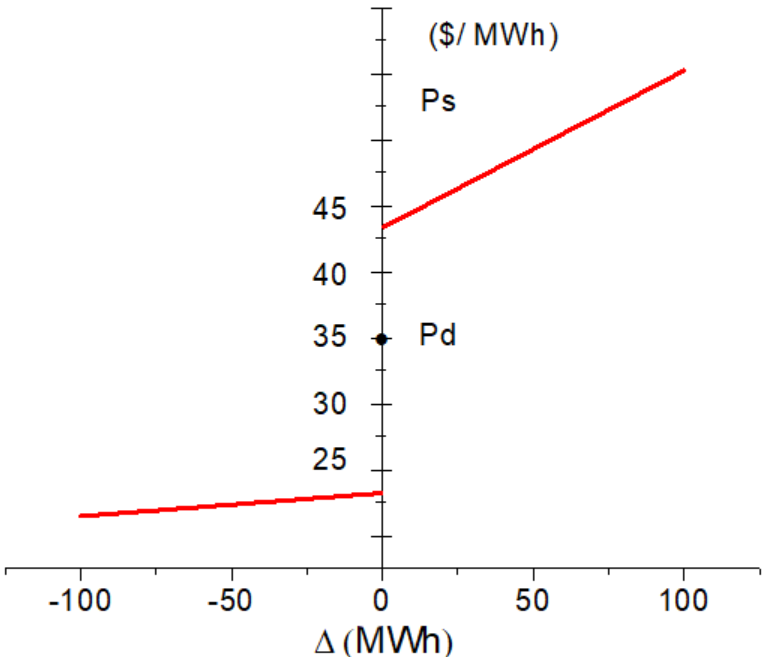}
\caption{Real-time piece-wise linear pricing model of $p_d$ and total imbalance $\Delta$.}
\label{fig2}
\end{figure}

The real-time market price generally depends on the market-level supply and demand imbalance, i.e., the difference between the day-ahead scheduled supply and the real-time actual demand. Deficient supply in the market leads to a higher spot price (up-regulation), whereas excessive supply results in a lower spot price (down-regulation). To capture their relationship, we consider the following piece-wise linear pricing model\footnote{Generation imbalance from generation companies' side also proposes an effect on the real-time market electricity price, e.g., in case of generator failure or the uncertainty from large-scale renewable generation. Such imbalance may increase the market mismatch variability. This paper mainly focuses on the demand side of the electricity market, and the effect of generation uncertainty on the market imbalance is not considered. We leave the incorporation of generation uncertainty into market equilibrium analysis for future study.}~\cite{Neupane}; see Fig.~\ref{fig2} for illustration.
\begin{equation}\label{pricing}
\begin{split}
p_{s}=\begin{cases}
p_{d}, & \Delta=0;\\
(a_{1}\Delta+b_{1})p_{d}, & \Delta>0;\\
(a_{2}\Delta+b_{2})p_{d}, & \Delta<0.
\end{cases}
\end{split}
\end{equation}
Here $a_{1}$, $a_{2}$, $b_{1}$, $b_{2}$ $\in\mathbb{R}_{+}$ are parameters of the pricing model.

\textbf{Remark:} (i) The model was proposed in~\cite{Neupane} by curve-fitting the historical data. In general, $a_{1}\neq a_{2}, b_{1}\neq b_{2}$. Specifically, \cite{Neupane} suggests $a_1=0.0034, a_2=0.0005, b_1=1.2378$, and $b_2=0.6638$. (ii) The spot price function is discontinuous at $\Delta=0$, i.e., $b_1>1>b_2$. This discontinuity models the \textit{premium of readiness} that utilities need to pay for the generation companies, since they have to generate urgent regulating power~\cite{skytte1999regulating}.
(iii) In this paper, we first focus on the piece-wise linear symmetric pricing model, i.e., $a_{1}=a_{2}>0$, $(b_{1}p_d+b_{2}p_d)/2=p_d$, and $b_1> b_2$. Then in Sec.~\ref{sec:general}, we generalize our results to a larger class of pricing models, which can be nonlinear or continuous at the origin.

\subsection{Cost Function of Utilities and Strategic Bidding Game}
For utility $i$, its electricity procurement is settled in two timescales: (i) an amount of $D_{i}-\Delta_{i}$ is settled in the day-ahead market at price $p_{d}$, and (ii) the remaining amount $\Delta_{i}$ is settled in the real-time market at the spot price $p_{s}$. Hence the total electricity cost of utility $i$  is given as follows (unit: \$):
\begin{equation}\label{equation5}
\begin{split}
 \tilde{C}_i=p_{d}\left(\hat{D}_{i}-\mu_{i}\right)+p_{s}\Delta_{i}=p_{d}\left(D_{i}-\Delta_{i}\right)+p_{s}\Delta_{i}.
\end{split}
\end{equation}
The Average Buying Cost  per unit electricity ($\textsf{ABC}$) for utility $i$ is simply
\begin{equation}\label{equation6}
\begin{split}
\textsf{ABC}_{i}	&\triangleq\frac{\tilde{C}_{i}}{D_{i}}=\frac{1}{D_{i}}\left[p_{d}\left(D_{i}-\Delta_{i}\right)+p_{s}\Delta_{i}\right].
\end{split}
\end{equation}
Consider the net load uncertainty, the cost function for utility $i$ is defined as the expected $\textsf{ABC}_i$, i.e.,
\begin{equation}\label{equation7}
\begin{split}
{{C}_{i}}(\mu_{i},\mu_{-i})\triangleq    \mathbb{E}[\textsf{ABC}_{i}].
\end{split}
\end{equation}

Note that the cost of utility $i$ not only depends on its own strategy $\mu_{i}$ but also depends on $\mu_{-i}\triangleq\sum_{j:j\neq i}^{N}\mu_{j}$, the aggregate strategies of all other utilities. The underlying reason is that the real-time spot price is determined by the market-level mismatch, and other utilities' strategic behavior can affect the cost of utility $i$ through the spot price $p_{s}$.

Given the model of strategic behaviors and utilities' cost functions, we model their interactions as {a non-cooperative }game with $N$ utilities where each utility aims to minimize {its own cost ${C}_{i}(\mu_{i},\mu_{-i})$} by choosing a strategy represented by $\mu_i$. Formally, a strategy profile $\boldsymbol{\mu^{*}}=(\mu_{1}^{*},\mu_{2}^{*},...,\mu_{N}^{*}) $ constitutes a \textbf{Nash Equilibrium} if for each $i=1,2,...,N$,
\begin{equation}\label{equation8}
\begin{split}
{{C}_{i}}(\mu_{i}^{*},\mu_{-i}^{*})\le {{C}_{i}}(\mu_{i},\mu_{-i}^{*}),\forall\mu_{i}\in\mathbb{R},
\end{split}
\end{equation}
where $\mathbb{R}$ is the set of real numbers.

\section{Main Results: Market Equilibrium Analysis }\label{sec:equilibrium}
Under the  two-settlement market mechanism, we study the \emph{Existence}, \emph{Uniqueness}, \emph{Efficiency}, and \emph{Robustness} of the equilibrium of the game among utilities. {In this section, we consider the scenario that the net load prediction errors $ \epsilon_1,..., \epsilon_N$ are mutually independent and the real-time spot pricing model is piece-wise linear symmetric as defined in (\ref{pricing}) with $a_1 = a_2$, $(b_1p_d + b_2p_d)/2=p_d$, and $b_1 > b_2$.}



These two {settings} allow us to better highlight the impacts of utilities' strategies and the market equilibrium characteristics. 
We later extend the results to the case with correlated prediction errors and general pricing models in Sec.~\ref{sec:general}.

\subsection{Load Imbalance Distribution and Analysis of Utilities' Strategic Behaviors}

We model the net load {prediction} error $\epsilon_{i}$ as a general \emph{symmetric unimodal} random variable with zero mean and variance $\sigma_{i}^{2}$. 

\begin{definition}\label{proper}
A  probability density function $f_{X}(\cdot)$ is symmetric unimodal if it is

(i) Symmetric w.r.t. its mean $\xi$: \begin{equation}\label{equation9}
\begin{split}
f_{X}(\xi+x)=f_{X}(\xi-x), \ \forall x \in \mathbb{R};
\end{split}
\end{equation}

(ii) Central dominant: 
\begin{equation}\label{equation10}
\begin{split}
f_{X}(x)\leq f_{X}(y), \ \mbox{if} \ |x-\xi|\geq|y-\xi|, \ \forall x, y \in \mathbb{R}.
\end{split}
\end{equation}

Here $\xi\triangleq \int^{+\infty}_{-\infty}x\cdot f_{X}(x)dx$ is the expected value of $X$.
\end{definition}

Many prediction error distributions are symmetric unimodal, including  Gaussian distribution and Laplace distribution. We say that a random variable is \emph{symmetric unimodal} if it follows a \emph{symmetric unimodal} distribution. The specific meaning of \emph{central dominant} comes from that utilities tend to make larger prediction errors with smaller probability compared with the case of smaller errors with larger probability. \emph{Symmetric distribution} implies that utilities have equal chances to encounter positive or negative prediction errors; 
see Sec.~\ref{ssec:distribution} for details.
\begin{figure}  [H]
\centering
\includegraphics[width = 0.5\linewidth]{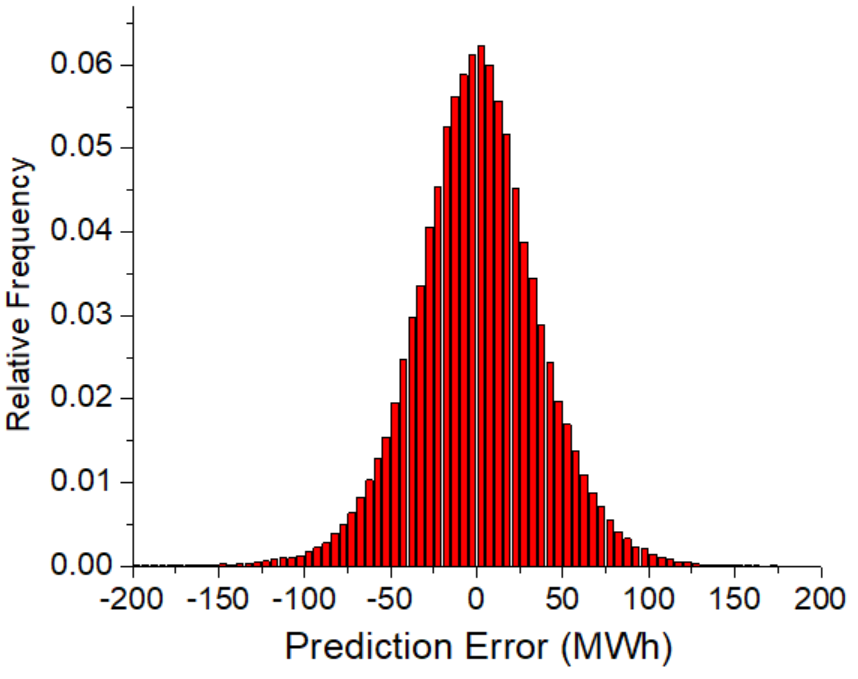}
\caption{Load prediction error histogram from case study based on real-world data.}
\label{fig3}
\end{figure}

Given a utility's bidding strategy $\mu_i$, the real-time imbalance $\Delta_i$ follows a symmetric unimodal distribution with mean $\mu_i$ and variance $\sigma_{i}^{2}$,
\begin{equation}\label{equation11}
\begin{split}
\Delta_{i}\sim\mathcal{P}_i(\mu_{i},\sigma_{i}^{2}),\quad i=1,2,...,N.
\end{split}
\end{equation}



\begin{lemma}\label{lemma1}
The sum of independent symmetric unimodal random variables is still symmetric unimodal. That is, suppose that there are $N$ independent random variables $\Delta_1,..., \Delta_N$, and for each $ i=1,2,...,N$, $\Delta_i$ follows a symmetric unimodal distribution with mean $\mu_i$, then their sum  $\Delta\triangleq \sum^{N}_{i=1}\Delta_i$ follows a symmetric unimodal distribution with mean $\mu= \sum^{N}_{i=1}\mu_i$. In addition, if there exists one $\Delta_i$ whose probability density function is strictly central dominant, i.e.,
\begin{equation}\label{equation12}
\begin{split}f_{\Delta_i}(x)< f_{\Delta_i}(y), \ \mbox{if} \ |x-\xi|>|y-\xi|,\ \forall x, y \in \mathbb{R},
\end{split}
\end{equation}
then the probability density function of $\Delta$ is also strictly central dominant.
\end{lemma}


Lemma~\ref{lemma1} states the property of the convolutions of symmetric unimodal random variables~\cite{purkayastha1998simple}. We present here a stronger observation on such distribution convolutions. Applying Lemma~\ref{lemma1}, we have
\begin{equation}\label{equation13}
\begin{split}
\Delta \sim\mathcal{P}(\mu,\sigma^{2}), \ \mbox{and}\   \Delta_{-i}\sim\mathcal{P}_{-i}(\mu_{-i},\sigma_{-i}^{2}),
\end{split}
\end{equation}
where $\mu\triangleq\sum_{i=1}^{N}\mu_{i}$, $\mu_{-i}\triangleq\sum_{j:j\neq i}^{N}\mu_{j}$,
$\sigma^{2}\triangleq\sum_{i=1}^{N}\sigma_{i}^{2}$, and $\sigma_{-i}^{2}\triangleq\sum_{j:j\neq i}^{N}\sigma_{j}^{2}$.

Based on the above observations, Theorem~\ref{theorem1} characterizes the cost function {${C}_{i}(\mu_{i},\mu_{-i})=\mathbb{E}[\textsf{ABC}_{i}]$} of utility $i$.

\begin{theorem}\label{theorem1}
{Under independent prediction errors and piece-wise linear symmetric spot
pricing model}, the expectation of $\mathbf{\textsf{ABC}}_i$ is given as:
\begin{equation}\label{equation14}
\begin{split}
\mathbb{E}[\mathbf{\textsf{ABC}}_{i}]	=&p_{d}+\frac{p_{d}}{D_{i}}\left[\frac{a_{1}+a_{2}}{2}(\mu_{i}\mu_{-i}+\sigma_{i}^{2}+\mu_{i}^{2})\right. \\
	& \left.+(b_{1}-b_{2})\mathbb{E}\left[\Delta_{i}\cdot\tilde{F}(\Delta_i)\right]\right],
\end{split}
\end{equation}
where $\tilde{F}(\Delta_i)\triangleq \int_{-\Delta_{i}}^{\mu_{-i}}f_{\Delta_{-i}}(\delta_{-i})d\delta_{-i}$, $f_{\Delta_{-i}}(\cdot)$ is the PDF of $\Delta_{-i}$ with mean $\mu_{-i}$, and coefficients $a_{1}$,$a_{2}$,$b_{1}$,$b_{2}$ are parameters of the spot pricing model defined in (\ref{pricing}).
\end{theorem}

\textsf{Remarks:} \rev{Utility estimates its expected cost via (\ref{equation14}) considering load uncertainty.}
The expectation of $\mathbf{\textsf{ABC}}_i$ in (\ref{equation14}) depends on three terms. The first term is simply the day-ahead market clearing price. The second and the third terms represent the real-time market operation cost to balance the mismatch. It is clear that the strategic behavior of utility $i$ and the aggregation of all other utilities' strategies affect both the second and the third terms. In addition, the second term reveals the influence of the pricing model slope, and the third term presents the discontinuous part of the pricing model. Meanwhile, given $\mu_{-i}=0$, both term two and term three are positive and $\mathbb{E}[\mathbf{\textsf{ABC}}_{i}]$ will increase under the following conditions.
\begin{itemize}
\item The day-ahead market clearing price $p_d$ increases.
\item The slope of the real-time market pricing model increases, i.e., $a_1, a_2$ increases, 
\item The discontinuity gap of the pricing model $b_1-b_2$ increases.
\end{itemize}

In Sec.~\ref{ssec:sensitivity}, our simulation results verify these observations. The results show that utilities suffer higher costs under higher day-ahead clearing price and larger real-time market sensitivity.
\rev{Note that in day-ahead market operations, the actual net load $D_i$ is not realized precisely to utilities, and they only estimate their expected cost via (\ref{equation14}). Our following results show that utilities' optimal bidding strategy is independent of the specific value of $D_i$. Therefore, utilities can still derive their optimal operations even in absence of the full knowledge of $D_i$; accordingly, the market equilibrium can be described.}
\subsection{Existence and Uniqueness of the Nash Equilibrium}\label{ssec:existence}

A Nash Equilibrium in the utility bidding game is a strategic profile in which all utilities choose the optimal strategy that minimizes its own cost given others' behaviors. We start by understanding the characteristics of the third term in (\ref{equation14}).

\begin{lemma}\label{lemma2}
Given $\mu_{-i}=0$, the optimal $\mu^*_i$ that minimizes $\mathbb{E}\left[\Delta_{i}\cdot\tilde{F}(\Delta_i)\right]$ is 0, and it is strictly increasing w.r.t. $|\mu_i|$, the absolute value of $\mu_i$.
\end{lemma}
Lemma~\ref{lemma2} shows that the third term in (\ref{equation14}) that related to the discontinuous part of the spot pricing model will increase if the utility deviates from bidding according to prediction, given that the aggregation of all other utilities' strategies is zero. 

Recall that utilities only place one quantity for one bid in the market; it is sufficient to focus on the pure strategy Nash Equilibrium. With Theorem~\ref{theorem1} and Lemma~\ref{lemma2}, we present the following necessary condition for all pure strategy Nash Equilibria.

\begin{theorem}\label{theorem2}
{Under independent prediction errors and piece-wise linear symmetric spot
pricing model}, a strategy profile $\boldsymbol{\mu^{*}}=(\mu_{1}^{*},\mu_{2}^{*},...,\mu_{N}^{*})$ constitutes a pure strategy Nash  Equilibrium only if for all $ i=1,2,...,N$,
\begin{equation}\label{equation15}
\begin{split}
&\bullet  \ \mu_{i}^{*}=0,\ \mbox{if} \ \mu_{-i}^{*}=0;\\
&\bullet \ \mu^*_{i}\in\left(-\mu_{-i}^{*}, 0\right),\ \mbox{if}\ \mu_{-i}^{*}>0;\\
&\bullet \ \mu^*_{i}\in\left(0, -\mu_{-i}^{*}\right),\ \mbox{if}\ \mu_{-i}^{*}<0.
\end{split}
\end{equation}
\end{theorem}

\textsf{Remarks:} \rev{Theorem~\ref{theorem2} shows the structure of utilities' optimal bidding strategies.} It says that if $\mu_{-i}=0$, then the best response of utility $i$ is to choose $\mu^*_i=0$. If $\mu_{-i}\neq0$, then utility $i's$ optimal strategy will always be opposite to this value.

{The insights behind Theorem~\ref{theorem2} \rev{and utilities' optimal bidding strategies at equilibrium} can be revealed from the perspective of cost minimization and best response. We note that when the utility's real-time mismatch has the same sign as the market-level mismatch, the utility will suffer a loss; otherwise, it will gain. For example, when the market-level mismatch is positive, the real-time price is higher than the day-ahead price according to the pricing model. If the utility's mismatch is negative, it means that the utility buys excessive energy in the day-ahead market and it can sell it back to the market at a higher price. Thus the utility will gain.

With this in mind, let us look at the case when utility $i$ chooses the bidding strategy $\mu_i > 0$, given $\mu_{-i}=0$. Recall that the utility's prediction error follows a symmetric unimodal distribution with mean zero, which indicates it has the same possibility to encounter particular positive or negative errors. Consequently, if the utility strategically underbuys when participating in the day-ahead market, i.e., $\mu_i>0$, its real-time imbalance will tend to be negative. Since the market-level mismatch follows a symmetric unimodal distribution $\mathcal{P}(\mu_{i},\sigma^{2})$, when $\mu_{i}>0$, the market-level mismatch and the utility $i's$ mismatch tend to be positive simultaneously, thus the utility $i$ tends to suffer a loss. Similarly, when $\mu_{i}<0$, given $\mu_{-i}=0$, the utility $i$ will also suffer a loss.}

Then we consider the cases when $\mu_{-i}\neq0$. If $\mu_{-i}>0$, the utility $i$ will not choose $\mu_i>0$ since this will make the market imbalance have more tendency to be positive.  A similar result holds for the case when $\mu_i<-\mu_{-i}$. These two situations expose the utility to the risk that its imbalance has more possibility to have the same sign with the market-level imbalance. Furthermore, choosing $\mu_i$ to be less than $0$ and greater than $-\mu_{-i}$ will always be better than $\mu_i=0$ and $\mu_i=-\mu_{-i}$. The optimal $\mu^*_{i}$ comes from the trade-off between the price and the amount. Similarly, when $\mu_{-i}<0$, utility $i$ will choose $\mu_i\in (0, -\mu_{-i})$.

Suppose $(\mu_{1}^{*},\mu_{2}^{*},...,\mu_{N}^{*})$ is a pure strategy Nash Equilibrium profile with $m$ non-zero elements, where $1\leq m\leq N$. According to Theorem~\ref{theorem2}, for these $m$ elements, denoted as $ \left(\mu_{1}^{*},\mu_{2}^{*},...,\mu_{m}^{*}\right)$ without loss of generality, we define 
\begin{equation}
\begin{split}\alpha_i\triangleq\frac{\mu_{i}^{*}}{\mu_{-i}^{*}}\in(-1, 0),\ \forall i\in\{1, 2,..., m\}.
\end{split}
\end{equation}

It is straightforward to derive the following condition:
\begin{equation}
\begin{split}
\left[\begin{array}{cccc}
     -1 & \alpha_1&\dots&\alpha_1 \\
    \alpha_2 & -1&\dots&\alpha_2\\
    \vdots& \vdots& \ddots& \vdots \\
\alpha_{m} & \alpha_{m}&\dots&-1
\end{array}\right]
\begin{bmatrix}
  \mu_{1}^{*} \\
  \vdots   \\
\vdots   \\
\mu_{m}^{*}
 \end{bmatrix}=
\begin{bmatrix}
  0 \\
  \vdots   \\
\vdots   \\
0
 \end{bmatrix},
\end{split}
\end{equation}
which describes the second and the third conditions in (\ref{equation15}) for all non-zero $\mu_i$. Let $M$ be the left-hand side $m\times m$ matrix.
We have the following results.

\begin{theorem}\label{theorem3}
{Under independent prediction errors and piece-wise linear symmetric spot
pricing model}, the matrix $M$ is a full rank matrix, and consequently $\boldsymbol{\mu^{*}}=\vec{0}$ is the unique pure strategy Nash Equilibrium.
\end{theorem} 
Theorem~\ref{theorem2} and Theorem~\ref{theorem3} imply that for any day-ahead clearing price $p_d$,\footnote{The day-ahead clearing price is not known precisely to utilities beforehand. See Sec.~\ref{ssec:related_work} for discussions on electricity price forecasting and uncertainty.} if all utilities except utility $i$ bid according to prediction, then bidding according to prediction is the best response of utility $i$ (see Fig.~\ref{fig7} based on real-world data for illustration). 
Consequently, all utilities bid exactly according to (net load) prediction is the unique pure strategy Nash Equilibrium. 
Conventionally, Nash Equilibrium indicates that a utility does not benefit from deviating from the equilibrium, assuming other utilities keep their strategies unchanged. In Corollary~\ref{corollary1}, we show a stronger characteristic of the equilibrium. That is, the cost of the utility is strictly increasing w.r.t. the deviation distance between the strategy chosen and the equilibrium. 

\begin{corollary}\label{corollary1}
{Under independent prediction errors and piece-wise linear symmetric spot
pricing model}, given $\mu_{-i}=0$, the optimal $\mu_{i}^{*} $ that minimizes $\mathbb{E}[\mathbf{\textsf{ABC}}_{i}]$ is 0, and $\mathbb{E}[\mathbf{\textsf{ABC}}_{i}]$ is strictly increasing w.r.t. $|\mu_i|$.
\end{corollary}

 \rev{In Sec.~\ref{sec:general}, we generalize the results to the case of correlated prediction errors (across utilities), general pricing models, and the setting in which utilities submit bidding curves considering the uncertainty of the day-ahead clearing prices.}

\subsection{Efficiency and Robustness of the Nash Equilibrium}\label{ssec:efficiency}

We have shown that $\boldsymbol{\mu^{*}}=(u^*_{1}, u^*_{2},..., u^*_{N})=\vec{0}$ is the unique pure strategy Nash Equilibrium. The next natural question is: what is the corresponding loss of efficiency? Recall that loss of efficiency is characterized as the gap between the social costs under the game-theoretical strategic setting, i.e., the social cost at the equilibrium, and the coordinated setting.

The optimal social cost under the coordinated setting is obtained by solving the following cost minimization problem:
\begin{equation}\label{equation18}
\begin{split}
\min \limits_{\mu_i\in\mathbb{R}, \forall i\in\{1,2,..., N\}} \quad \mathbb{E}[\mathbf{\textsf{ABC}}_{total}],
\end{split}
\end{equation}
where $\textsf{ABC}_{total}$ is defined as:
\begin{equation}
\begin{split}
\textsf{ABC}_{total}	&\triangleq\frac{C_{total}}{D_{total}}=\frac{\sum_{i}D_i\cdot\textsf{ABC}_{i}}{\sum_{i}D_i}\\ 
	&=\frac{1}{D_{total}}\left[p_{d}\left(D_{total}-\Delta\right)+p_{s}\Delta\right].
\end{split}
\end{equation}
Here $\Delta$ follows a symmetric unimodal distribution with mean $\mu=\sum_{i=1}^{N}\mu_{i}$, and $\textsf{ABC}_{total}$ (unit: \$/MWh) can be interpreted as the unit cost of the market to settle $D_{total}\triangleq\sum_{i=1}^{N}D_{i}$ amount of electricity.


\begin{theorem}\label{theorem4}
{Under independent prediction errors and piece-wise linear symmetric spot
pricing model}, $\mathbb{E}[\mathbf{\textsf{ABC}}_{total}]$ is minimized at $\mu^*=0$, and $\mathbb{E}[\mathbf{\textsf{ABC}}_{total}]$ is strictly increasing w.r.t. $|\mu|$. Consequently, the unique pure strategy Nash Equilibrium $\boldsymbol{\mu^{*}}=\vec{0}$ incurs no loss of efficiency.
\end{theorem}

\textsf{Remarks:} \rev{Theorem~\ref{theorem4} shows the market operator that the social cost at the equilibrium is the same as the optimal one under the coordinated setting in (\ref{equation18}).} 
The intuition behind Theorem~\ref{theorem4} lies in that we can treat the whole market as an entity. From the market-level perspective, an amount of $(D_{total}-\Delta)$ power is committed at the price $p_d$ and the imbalance $\Delta$ is settled at the spot market price $p_s$. When the market has a particular positive (respectively negative) real-time mismatch, this mismatch is settled at a price $p_s>p_d$ (respectively $p_s<p_d$); hence the market will suffer a loss.

Since the market mismatch $\Delta$ follows a symmetric unimodal distribution $\mathcal{P}(\mu,\sigma^{2})$, when $\mu>0$, the market-level mismatch tends to be positive, thus the market tends to suffer a loss facing a higher spot price. Similar analysis can be applied to the case when $\mu<0$. We conclude that the unique pure strategy Nash Equilibrium is efficient; see Fig.~\ref{fig7} for illustration.

The guarantee provided by a Nash equilibrium is that each player’s strategy is optimal, assuming all others play their designated strategies. However, there could exist fault behaviors, wherein some utilities are irrational, and their actions can be arbitrary or even adversarial. In the following theorem, we illustrate a strong observation on the robustness of the equilibrium. That is, the sub-optimal response faults behaviors of irrational utilities not only increase their own costs but also benefit other rational utilities.

\begin{theorem}\label{theorem5}
{Under independent prediction errors and piece-wise linear symmetric spot
pricing model}, the equilibrium $\boldsymbol{\mu^{*}}=\vec{0}$ is (0, $N-1$) fault immune. That is, consider all utilities except a group of utilities $S$, $S\subset\{1, 2,..., N\}$ and $1\leq|S|\leq N-1$. Given $\mu_{j}=0$, $\forall j\in\{1, 2,..., N\}\backslash \{S\}$, $\mathbb{E}[\mathbf{\textsf{ABC}}_{j}]$ is non-increasing w.r.t. $|\mu_{S}|$, where $\mu_{S}=\sum_{i\in S}\mu_i$. In addition, if there exists one $\epsilon_k$, $k\in\{1, 2,..., N\}$, whose probability density function is strictly central dominant,\footnote{see Lemma~\ref{lemma1} for the definition of strictly central dominant.} then $\mathbb{E}[\mathbf{\textsf{ABC}}_{j}]$ is strictly decreasing w.r.t. $|\mu_{S}|$.
\end{theorem}

\begin{figure}[H]
\centering
\includegraphics[width = 0.5\linewidth]{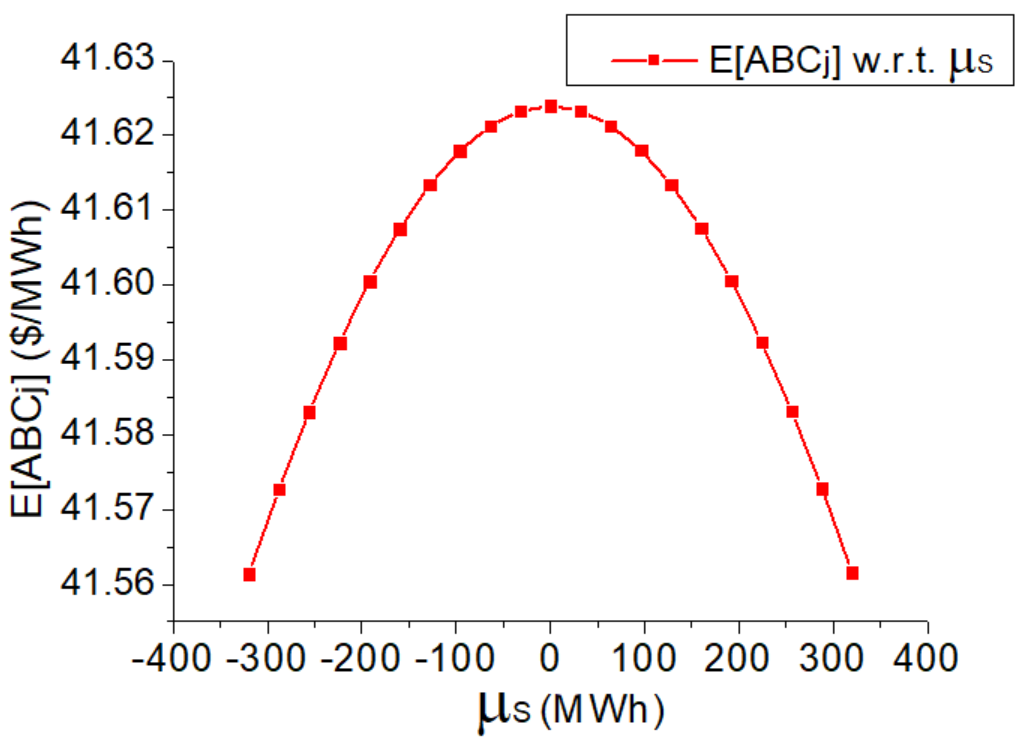}
\caption{Illustration of the (0, $N-1$) fault immune robustness of the pure strategy Nash equilibrium under piece-wise linear symmetric pricing model and Gaussian distributed prediction errors.}
\label{fig4}
\end{figure}

Theorem~\ref{theorem5} shows the robustness of the unique pure strategy Nash Equilibrium $\boldsymbol{\mu^{*}}=\vec{0}$. This is a desirable property as discussed in~\cite{gradwohl2008fault}. \rev{It provides the utility an understanding of its cost change characteristic w.r.t. irrational market participants' actions. The result indicates that fault behaviors of irrational utilities will not increase the costs of other rational utilities; see Fig.~\ref{fig4} for illustration.} 


\section{Main results: Equilibrium Generalization}\label{sec:general}

Previous analysis focuses on the scenario where the real-time market pricing model is {piece-wise linear symmetric} and prediction errors of utilities are {mutually independent} and utilities only place one {quantity bid} in the day ahead market. \rev{In this section, we extend our results to the setting with correlated prediction errors, general pricing models, and utilities submitting bidding curves as bidding pairs ((price, quantity) considering the uncertain day-ahead prices.}
\subsection{Beyond Piece-Wise Linear Symmetric Spot Pricing Model}

The uniqueness, efficiency, and robustness of the pure strategy Nash Equilibrium hold for a large class of pricing models which can be nonlinear or continuous at the origin.
\begin{theorem}\label{theorem6}
{Suppose the net load prediction errors are mutually independent.} Denote the pricing model as:
\begin{equation}\label{equation20}
\begin{split}
p_{s}=\begin{cases}
p_{d}, & \Delta=0;\\
p(\Delta)+b_1p_d, & \Delta>0;\\
p(\Delta)+b_2p_d, & \Delta<0.
\end{cases}
\end{split}
\end{equation}
Here $b_1+b_2=2$, $b_1\geq b_2$, $p(\Delta)$ is a non-decreasing odd function, i.e., $p(\Delta)=-p(-\Delta)$ and $p(x)\geq p(y), \forall x\geq y$, and $p(\Delta)$ is continuous at $\Delta = 0$. The following statements are true:

(1)  Given $\mu_{-i}=0$, the optimal $\mu_{i}^{*} $ that minimizes $\mathbb{E}[\mathbf{\textsf{ABC}}_{i}]$ is 0, and $\mathbb{E}[\mathbf{\textsf{ABC}}_{i}]$ is non-decreasing w.r.t. $|\mu_i|$. Consequently, $\boldsymbol{\mu^{*}}=(\mu_{1}^{*},\mu_{2}^{*},...,\mu_{N}^{*})=\vec{0}$ is a pure strategy Nash Equilibrium. In addition, if $p(\Delta)$ is strictly increasing w.r.t. $\Delta$ or $b_1> b_2$, then $\mathbb{E}[\mathbf{\textsf{ABC}}_{i}]$ is strictly increasing w.r.t. $|\mu_i|$.

(2) If $p(\cdot)$ is differentiable for all $x\in\mathbb{R}$, and
\begin{equation}\label{equation21}
\begin{split}
&p'(x_{1})\geq p'(x_{2})>0,\quad \forall x_{1}>x_{2}\geq0,\\
&p'(x_{1})\geq p'(x_{2})>0,\quad \forall x_{1}<x_{2}\leq0,
\end{split}
\end{equation}
then $\boldsymbol{\mu^{*}}=\vec{0}$ is the unique pure strategy Nash Equilibrium.

(3) $\mathbb{E}[\mathbf{\textsf{ABC}}_{total}]$ is minimized at $\mu^{*}=0$, and $\mathbb{E}[\mathbf{\textsf{ABC}}_{total}]$ is non-increasing w.r.t. $|\mu|$. Consequently, the pure strategy Nash Equilibrium
$\boldsymbol{\mu^{*}}=\vec{0}$ incurs no loss of efficiency. In addition, if $p(\Delta)$ is strictly increasing w.r.t. $\Delta$ or $b_1> b_2$, then $\mathbb{E}[\mathbf{\textsf{ABC}}_{total}]$ is strictly increasing w.r.t. $|\mu|$.

(4) If $p(\cdot)$ is differentiable for all $x\in\mathbb{R}$, and either
\begin{equation}\label{equation22}
\begin{split}
& \quad \quad \quad \quad \quad  b_1> b_2,\\
&p'(x_{2})\geq p'(x_{1})\geq0,\quad \forall x_{1}>x_{2}\geq0,\\
&p'(x_{2})\geq p'(x_{1})\geq0,\quad \forall x_{1}<x_{2}\leq0,
\end{split}
\end{equation}
or 
\begin{equation}\label{equation23}
\begin{split}
&p'(x_{2})> p'(x_{1})\geq0,\quad \forall x_{1}>x_{2}\geq0,\\
&p'(x_{2})> p'(x_{1})\geq0,\quad \forall x_{1}<x_{2}\leq0,
\end{split}
\end{equation}
then the pure strategy Nash Equilibrium  $\boldsymbol{\mu^{*}}=\vec{0}$ 
is (0, $N-1$) fault immune.
\end{theorem}

\begin{figure} [ht]
  \centering
  \subfigure[Discontinuous symmetric pricing model: $b_1=1.2378$, $b_2=0.7622$.]{
    \includegraphics[width = 0.4\linewidth]{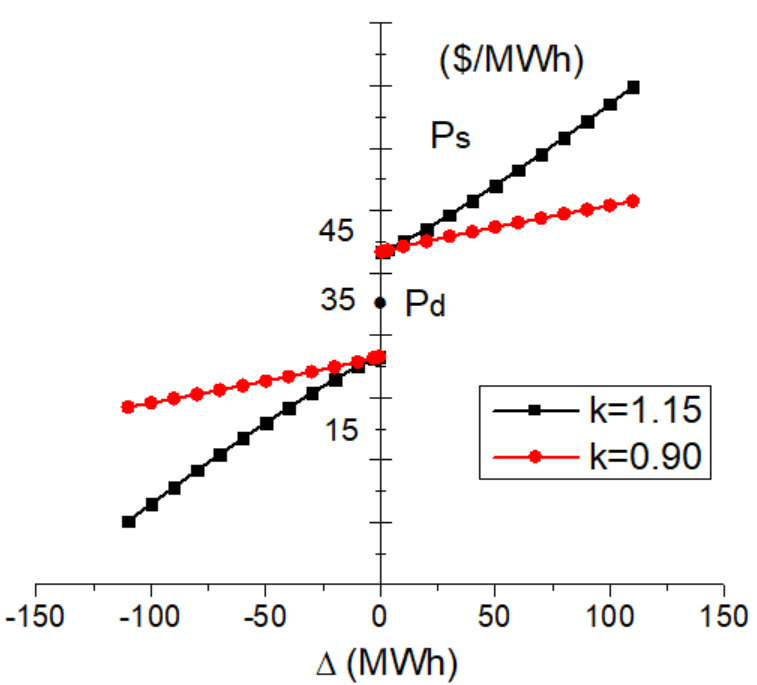}
  }
  \subfigure[Continuous symmetric pricing model: $b_1=b_2=1$.]{
    \includegraphics[width = 0.4\linewidth]{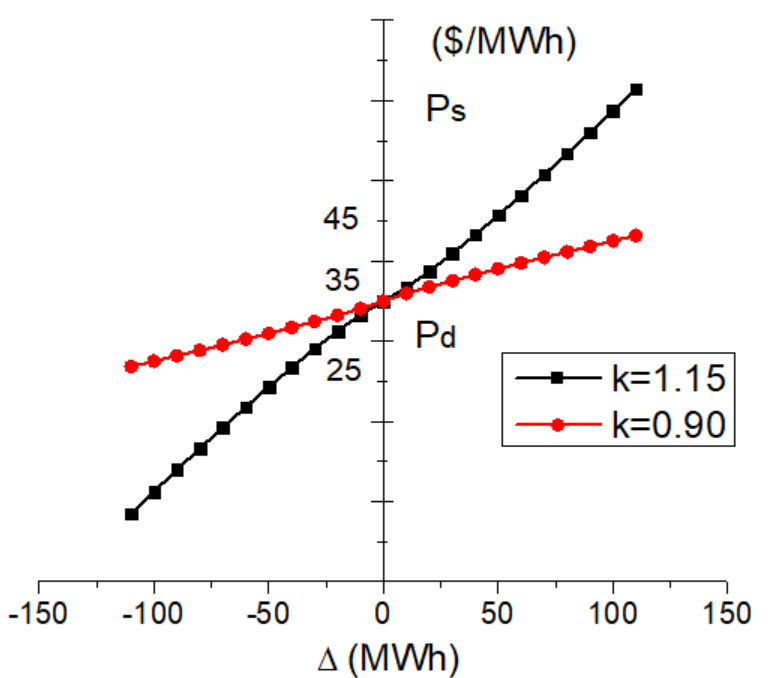}
  }
  \caption{Nonlinear pricing function $p(x)=p_{d}\times (1_{\{x>0\}}(a_{1}x^{k}+b_{1})+1_{\{x<0\}}(-a_{2}(-x)^{k}+b_{2})+1_{\{x=0\}})$ with $a_1=a_2=0.0034$, $p_d=35$, $k=1.15$ satisfying (\ref{equation21}) and $k=0.9$ satisfying (\ref{equation23}), where $1_{\{\cdot\}}$ is the indicator function. To show the nonlinear relationship more clearly and let the price within the reasonable range, we vary $\Delta$ from -0.1GWh to 0.1GW.}\label{fig5}
\end{figure}

\textsf{Remarks:}
(i) The general pricing models defined in (\ref{equation20}) are described by the  discontinuous gap $(b_1-b_2)\cdot p_d$ and the imbalance related term $p(\Delta)$. The results in Sec.~\ref{sec:equilibrium} are for the special case of Theorem~\ref{theorem6} of the piece-wise linear symmetric pricing model.  (ii) Theorem~\ref{theorem6} indicates that the results of the uniqueness, efficiency, and (0, $N-1$) fault immune robustness of the equilibrium obtained in Sec.~\ref{sec:equilibrium} can be extended to the case with general pricing models, which can be nonlinear or continuous at the origin. (iii)  
Pricing models satisfying (\ref{equation21}) are convex when $\Delta>0$ and concave when $\Delta<0$. The pricing models satisfying (\ref{equation22}) or (\ref{equation23}) are concave when $\Delta>0$ and convex when $\Delta<0$; see Fig. \ref{fig5} for illustration. Therefore, the piece-wise linear symmetric pricing model with discontinuous gap at the origin is the only one that satisfies both (\ref{equation21}) and (\ref{equation22}) or (\ref{equation23}) 
and thus admits the unique, efficient, and (0, $N-1$) fault immune robust pure strategy Nash Equilibrium. 

\subsection{Beyond Independent Prediction Errors}

In practice, load prediction errors among utilities may be correlated; see e.g., \cite{umbrellareport}. In this subsection, {we generalize our results} and consider the case that utilities' prediction errors $\epsilon_i$ follow correlated Gaussian distributions with non-negative correlation coefficients $\rho_{i,j}\triangleq \Cov\left(\epsilon_i, \epsilon_j\right)/\sigma_{i}\sigma_{j}
\geq 0$. We obtain the following results.

\begin{theorem}\label{theorem7}
{Suppose the real-time spot pricing model is piece-wise linear symmetric, and $\epsilon_i$. $ i=1,2,...,N$, are zero mean Gaussian random variables with non-negative correlations,} then the strategy profile $\boldsymbol{\mu^{*}}=\vec{0}$ is the unique, efficient, and (0, $N-1$) fault immune robust pure strategy Nash Equilibrium.
\end{theorem}

{Practically, the load prediction errors are usually observed to be positively correlated~\cite{yi2018impact, umbrellareport}. Theorem~\ref{theorem7} highlights that with non-negatively correlated Gaussian prediction errors, the market still admits the equilibrium with the desired properties.}

\subsection{Beyond Submitting Quantity Bid}\label{ssec:bidding curve}
In modern deregulated electricity market, market participants are always allowed to react to the uncertain market prices since they may encounter short-term price fluctuation~\cite{fleten2005constructing, song2016purchase,nojavan2015optimal,wei2014energy}. Specifically, generation companies and utilities submit their selling and buying bidding curves, then the ISO compiles these submitted bidding curves and calculates the clearing price for each trading period. A certain volume of electricity is agreed for a utility according to the clearing price and its bidding curve. The bidding curve of each individual utility represents its willingness for procuring different amounts of electricity from the market under different prices. In this subsection, we extend our results to the setting in which utilities submit bidding pairs (price, quantity) which form their bidding curves for each hour considering the uncertain market clearing prices. . 

Similar to the previous formulation, let $D_i$ and $\hat{D_i}$ be the actual net load and the net load prediction. Furthermore, we use $Q_i(p_d)$ to denote the bidding curve of utility $i$, which represents the desired buying quantity under different day-ahead uncertain clearing price ${p_d}$. Therefore, utility $i's$ bidding strategy and the aggregate strategies of all other utilities can be represented by $\mu_i(p_d)\triangleq\hat{D_i}-Q_i({p_d})$ and $\mu_{-i}(p_d)\triangleq\sum^{N}_{j\neq i}\mu_j(p_d)$ respectively. We model $\mu_i(p_d)$ as piece-wise continuous functions defined on $\mathbb{R^+}$ for each $i=1,2,...,N$. 

Since the day-ahead clearing price $p_d$ is not known to the utilities beforehand,\footnote{The day-ahead clearing price is not known precisely to the utilities beforehand. Utilities can have a probabilistic density forecast of electricity price by multiple probabilistic price forecasting models. We refer to Sec.~\ref{ssec:related_work} for a discussion on electricity price forecasting.} when considering the randomization of $p_d$, utilities report demand bidding curves and aim at minimizing their expected costs, the probability-weighted sum of $\mathbb{E}[\mathbf{\textsf{ABC}}_{i}|\, p_d=\hat{p_d}]$, i.e.,
\begin{equation}
    C_i(\mu_i(p_d), \mu_{-i}(p_d))\triangleq\int^{+\infty}_{0}\mathbb{E}[\mathbf{\textsf{ABC}}_{i}|\, p_d=\hat{p_d}]f_i(\hat{p_d})d\hat{p_d},
\end{equation}
where $f_i(\cdot)$ is a piece-wise continuous probability density function defined on the set of positive real number $R^+$, which represents utility $i's$ estimation on the day-ahead price $p_d$. Here we use $(\mathbf{\textsf{ABC}}_{i}|p_d=\hat{p_d})$ to denote utility's cost at day-ahead price $\hat{p_d}$. Note that we focus on the setting where utilities are price-takers in the day-ahead market and therefore their individual strategies will not affect $f_i(\cdot)$, their estimations of $p_d$.  

Consequently, under the setting in which utilities submit bidding curves, a strategy profile $\boldsymbol{\mu^{*}(p_d)}=(\mu_{1}^{*}(p_d),\mu_{2}^{*}(p_d),\\...,\mu_{N}^{*}(p_d)) $ constitutes a Nash Equilibrium if for each $ i=1,2,...,N$,
\begin{equation}
    C_{i}(\mu_{i}^{*}(p_d),\mu_{-i}^{*}(p_d))\le C_{i}(\mu_{i}(p_d),\mu_{-i}^{*}(p_d)),\forall\mu_{i}(p_d)\in\mathbb{F^+},
\end{equation}
where $\mathbb{F^+}$ is the set of all piece-wise continuous functions defined on $\mathbb{R^+}$.

We further define the corresponding optimal social cost under coordinated setting, which is obtained by solving the following social cost minimization problem: 
\begin{equation}\label{equation26}
\begin{split}
\min \limits_{\mu_i(p_d),  i= 1,2,..., N}  C_{total}(\mu(p_d))=\int^{+\infty}_{0}\mathbb{E}[\mathbf{\textsf{ABC}}_{total}|\, p_d=\hat{p_d}]f(\hat{p_d})d\hat{p_d},
\end{split}
\end{equation}
where $f(\cdot)$ is a piece-wise continuous probability density function defined on $R^+$, which represents the ISO's estimation on $p_d$. Similarly, $\textsf{ABC}_{total}$ at price $\hat{p_d}$ is defined as:
\begin{equation}
\begin{split}
(\textsf{ABC}_{total}|p_d=\hat{p_d})	&\triangleq\frac{C_{total}}{D_{total}}=\frac{\sum_{i}D_i\cdot(\mathbf{\textsf{ABC}}_{i}|p_d=\hat{p_d})}{\sum_{i}D_i}\\ 
	&=\frac{1}{D_{total}}\left[p_{d}\left(D_{total}-\Delta(\hat{p_d})\right)+p_{s}\Delta(\hat{p_d})\right].
\end{split}
\end{equation}
Here $\Delta(\hat{p_d})$ follows a symmetric unimodal distribution with mean $\mu(\hat{p_d})=\sum_{i=1}^{N}\mu_{i}(\hat{p_d})$, and $(\textsf{ABC}_{total}|p_d=\hat{p_d})$ (unit: \$/MWh) can be interpreted as the unit cost of the market to settle $D_{total}\triangleq\sum_{i=1}^{N}D_{i}$ amount of electricity at price $\hat{p_d}$.

Based on the above game-theoretical formulation, we have the following results.
\begin{theorem}\label{theorem8}
{Suppose the net load prediction errors are mutually independent.} Denote the non-decreasing symmetric pricing model as (\ref{equation20}) and $f_i({p_d})>0, \forall {p_d}>0$. The following statements are true:

(1)  Given $\mu_{-i}(p_d)=0$, the optimal $\mu_{i}^{*}(p_d) $ that minimizes $C_{i}(\mu_{i}(p_d),0)$ is 0, and $C_i(\bar{\mu}_i(p_d), 0)\geq C_i(\hat{\mu}_i(p_d), 0)$ if $|\bar{\mu}_i(p_d)|\geq |\hat{\mu}_i(p_d)|, \forall p_d \in R^+$. Consequently, $\boldsymbol{\mu^{*}(p_d)}=(\mu_{1}^{*}(p_d),\mu_{2}^{*}(p_d),...,\mu_{N}^{*}(p_d))=\vec{0}$ is a pure strategy Nash Equilibrium. In addition, if $p(\Delta)$ is strictly increasing w.r.t. $\Delta$ or $b_1> b_2$, then $C_i(\bar{\mu}_i(p_d), 0)>C_i(\hat{\mu}_i(p_d), 0)$ if $|\bar{\mu}_i(p_d)|\geq |\hat{\mu}_i(p_d)|, \forall p_d \in R^+$ and there exits an interval $(s, t)$ such that $|\bar{\mu}_i(p_d)|> |\hat{\mu}_i(p_d)|, \forall p_d\in(s,t)$.

(2) If (\ref{equation21}) holds, $\mu_i(x)$ is piece-wise continuous on $\mathbb{R^+}$, and there does not exist a point $x_0$ on $\mu_i(x)$ such that $\mu_i(x_0)\neq \lim _{{x\to x_{0}^{{-}}}}\mu_i(x)$ and $\mu_i(x_0)\neq \lim _{{x\to x_{0}^{{+}}}}\mu_i(x)$, $\forall i \in \{1,2,...,N\}$, then $\boldsymbol{\mu^{*}(p_d)}=\vec{0}$ is the unique pure strategy Nash Equilibrium.

(3) $C_{total}(\mu(p_d))$ is minimized at $\mu^*(p_d)=0$, and $C_{total}(\bar{\mu}(p_d))\geq C_{total}(\hat{\mu}(p_d))$ if $|\bar{\mu}(p_d)|\geq |\hat{\mu}(p_d)|, \forall p_d \in R^+$. Consequently, the pure strategy Nash Equilibrium $\boldsymbol{\mu^{*}(p_d)}=\vec{0}$ incurs no loss of efficiency.  In addition, if $p(\Delta)$ is strictly increasing w.r.t. $\Delta$ or $b_1> b_2$, then $C_{total}(\bar{\mu}(p_d))>C_{total}(\hat{\mu}(p_d))$ if $|\bar{\mu}(p_d)|\geq |\hat{\mu}(p_d)|, \forall p_d \in R^+$ and there exits an interval $(s, t)$ such that $|\bar{\mu}(p_d)|> |\hat{\mu}(p_d)|, \forall p_d\in(s,t)$.

(4) If (\ref{equation22}) or (\ref{equation23}) holds, then the pure strategy Nash Equilibrium  $\boldsymbol{\mu^{*}(p_d)}=\vec{0}$ is (0, $N-1$) fault immune.\footnote{Under the setting of utilities submitting bidding curves, the corresponding (0, $N-1$) fault immune robustness can be expressed as: consider all utilities except a group of utilities $S$, $S\subset\{1, 2, ..., N\}$ and $1\leq|S|\leq N-1$. Given $\mu_{j}(p_d)=0$, $\forall j\in\{1, 2, ..., N\}\backslash \{S\}$, $C_j(\mu_S(p_d), \mu_{-S}(p_d))$ is no greater than $C_j(\mu^*_S(p_d), \mu_{-S}(p_d))$, where $\mu_{S}(p_d)=\sum_{i\in S}\mu_i({p_d})$, $\mu_{-S}(p_d)=\sum_{j\in \{1, 2, ..., N\}\backslash \{S\}}\mu_j({p_d})=0$, and $\mu^*_{S}(p_d)=0$. In addition, if there exists one $\epsilon_k$, $k\in\{1, 2, ..., N\}$, whose probability density function is strictly central dominant (see Lemma~\ref{lemma1} for definition), then $C_j(\bar{\mu}_S(p_d), 0)<C_j(\hat{\mu}_S(p_d), 0)$ if $|\bar{\mu}_S(p_d)|\geq |\hat{\mu}_i(p_d)|, \forall p_d \in R^+$ and there exits an interval $(s, t)$ such that $|\bar{\mu}_S(p_d)|> |\hat{\mu}_S(p_d)|, \forall p_d\in(s,t).$} 

\end{theorem}

\begin{figure} [ht]
  \centering
  \subfigure[Bidding curve with inelastic load.]{
    \label{fig6a} 
    \includegraphics[width = 0.4\linewidth]{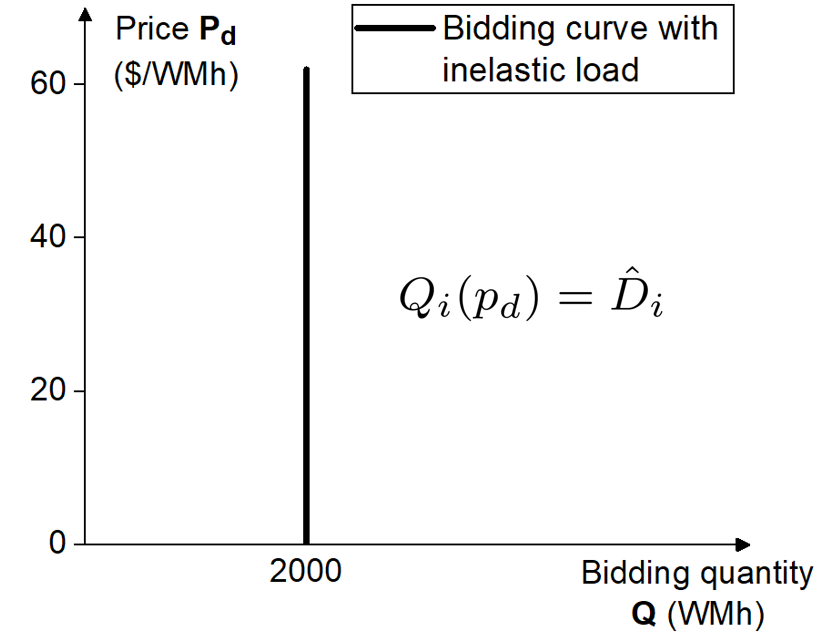}
  }
  \subfigure[Bidding curve with flexible load.]{
    \label{fig6b} 
    \includegraphics[width = 0.4\linewidth]{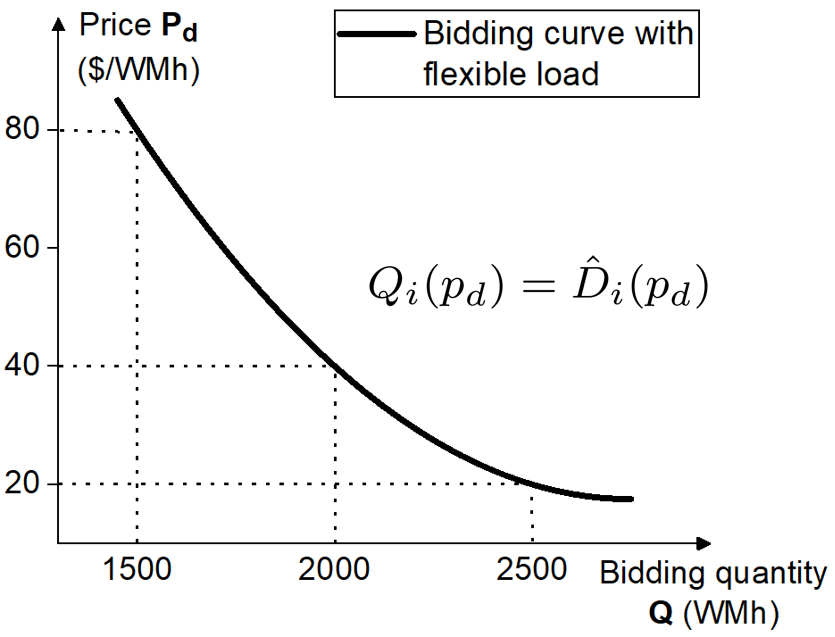}
  }
  \caption{Illustration of the pure strategy Nash Equilibrium strategies under the setting that utilities submit bidding curves at the predicted inelastic/flexible net load.}\label{fig6}
\end{figure}

Theorem~\ref{theorem8} states that the strategy profile that all utilities submit vertical bidding curves\footnote{We focus on the setting that the net load $D_i$ is inelastic and does not change with the day-ahead clearing price $p_d$. In reality, utilities may have price-related flexible demands due to dynamic pricing~\cite{song2016purchase}. Therefore, the net loads $D_i(p_d)$ may change with price $p_d$. We model $D_i(p_d) $ to be piece-wise continuous for $i=1,2,...,N$, and utilities' net load prediction errors $\epsilon_i$ as \emph{symmetric unimodal} random variables with zero mean as defined in Definition~\ref{proper}. Our results still hold under such setting in which utilities submit bidding curves exactly at $Q_i(p_d)=\hat{D}_i(p_d)$ that react to the market prices with flexible demands and can be generalize to the scenario with correlated prediction errors and the general pricing models; see Fig.~\ref{fig6b} for illustration.} exactly at the predicted demand is a (0, $N-1$) fault immune robust pure strategy Nash Equilibrium which incurs no loss of efficiency. Furthermore, under mild continuity condition\footnote{The continuity condition actually requires that the strategy curves $\mu_i(x)$ do not have discontinuity points isolated from the function, which includes the continuous and piece-wise continuous bidding curves as a special case. In our model, it means that utilities will not \emph{deviate} from the strategy curves at some particular points.} on $\mu_i(x)$, the above efficient and robust equilibrium is unique; see Fig.~\ref{fig6a} for illustration.

We further extend the results to the setting in which utilities submit bidding curves and their prediction errors $\epsilon_i$ follow correlated Gaussian distributions with non-negative correlation coefficients $\rho_{i,j}\triangleq \Cov\left(\epsilon_i, \epsilon_j\right)/\sigma_{i}\sigma_{j}
\geq 0$.

\begin{theorem}\label{theorem9}
{Suppose the real-time spot pricing model is piece-wise linear symmetric, and $\epsilon_i$. $ i=1,2,...,N$, are zero mean Gaussian random variables with non-negative correlations,} then the strategy profile $\boldsymbol{\mu^{*}(p_d)}=\vec{0}$ is an efficient and (0, $N-1$) fault immune robust pure strategy Nash Equilibrium. In addition, if $\mu_i(x)$ is piece-wise continuous on $\mathbb{R^+}$, and there does not exist a point $x_0$ on $\mu_i(x)$ such that $\mu_i(x_0)\neq \lim _{{x\to x_{0}^{{-}}}}\mu_i(x)$ and $\mu_i(x_0)\neq \lim _{{x\to x_{0}^{{+}}}}\mu_i(x)$, $\forall i  \in \{1,2,...,N\}$, then the equilibrium is unique.
\end{theorem}

Theorem~\ref{theorem9} highlights that the equilibrium with desired properties is still admitted when utilities submit bidding curves with non-negatively correlated Gaussian prediction errors.

\rev{
\subsection{Impact of Load Uncertainty on Equilibrium Strategy and Equilibrium Social Cost}

 Note that load uncertainty is captured by the variance of the load prediction error. In this part, we discuss the impact of load uncertainty on market equilibrium strategy and social cost.

 \begin{corollary}\label{corollary2}
  The utilities' market equilibrium strategies $\boldsymbol{\mu^{*}}$ are independent of load uncertainty variances $\sigma^2_i$, and equilibrium social cost $\mathbb{E}[\mathbf{\textsf{ABC}}_{total}]$ is increasing w.r.t. each $\sigma_i$.
\end{corollary}

Though load uncertainty affects utilities' costs~\cite{yi2018impact}, their optimal bidding strategies at equilibrium only depend on the mean of the load prediction error. In particular, for zero-mean prediction error, bidding according prediction is exactly the equilibrium strategy and do not rely on the variance. We remark that it is straightforward to generalize our results to the scenario that utilities' net load prediction errors have non-zero means. Under such a condition, the equilibrium with desired properties is still admitted when all utilities bid at their specific prediction error mean, i.e., $\mu^*_i=\mathbb{E}[\epsilon_i], i=1,2,...,N$; see Fig.~\ref{fig19} for illustration. This implies that each utility aims to make its individual real-time mismatch $\Delta_i$ have zero mean bias, which is invariant with load uncertainty variances $\sigma^2_i$. Thus, the expected cost is minimized, and the market equilibrium can be attained.

We further investigate the market social cost $\mathbb{E}[\mathbf{\textsf{ABC}}_{total}]$. Note that for independent or non-negatively correlated Gaussian load prediction errors, the increase of $\sigma_i$ will lead to higher $\sigma^2$, the variance of the market level total mismatch $\Delta$. By applying the approach in~\cite{yi2018impact} to (\ref{equation15}), it can be observed that the increase load uncertainty will increase $\mathbb{E}[\mathbf{\textsf{ABC}}_{total}]$. 
}

\section{Simulation Results}\label{sec:simulation}

\subsection{Simulation setting}
We conduct simulations of the ISO-NE electricity market with 8 virtual utilities, each in charge of a state in New England region. 
The piece-wise linear pricing model is obtained from~\cite{Neupane}. 
We study the market equilibrium under symmetric pricing model and the impact of the asymmetricity of the pricing model whose parameters are presented in Table~\ref{table1}. The hourly electricity net loads of the 8 utilities from January 1, 2011 to December 31, 2018 are obtained from ISO-NE market. The day-ahead price is obtained from the mean of the hourly day-ahead prices from ISO-NE market, which is $35$\$/MWh. 
Utilities' costs during each time slot are calculated according to (\ref{equation5}). \rev{Here we investigate the market equilibrium at a specific day-ahead clearing price. Generalizing the results to the setting of utilities reporting bidding curves considering the uncertain market clearing prices is straightforward as discussed in Sec.~\ref{ssec:bidding curve}.} 
We focus on the utility in Maine state to present the results. 
We observe similar results for other utilities. 
\begin{table}[ht]  
\centering  
\caption{Parameters of pricing models} 
\label{table1}
 \begin{tabular}{c|cccc}  
     \hline
     \hline
       Pricing model and Parameters& $a_1$ & $ a_2$ & $b_1$ & $ b_2$  \\ [1ex]  
       \hline 
Symmetric pricing model& 0.0034& 0.0034 & 1.2378& 0.7622 \\
Asymmetric pricing model& 0.0034& 0.0005 & 1.2378& 0.6638 \\
       \hline
       \hline
   \end{tabular}
\end{table}

\subsection{Prediction Error Distribution}\label{ssec:distribution}
Our theoretical analysis focuses on the scenario that the prediction errors of utilities follow symmetric unimodal distributions with zero mean, i.e., symmetric w.r.t. zero and central dominant. In this part, we verify these two conditions.

In practice, utilities predict their demands based on historical loads, weather information, holiday/weekend information, etc.~\cite{umbrellareport}. Multiple types of  load prediction methods are applied in short-term demand forecasting, e.g., similar day method, Artificial Neural Network (ANN), and time series regression~\cite{ISONE}. In our study, we use ANN to forecast demands. Prediction errors are computed as the difference between the predicted values and the actual demands correspondingly. We plot the load prediction error histogram of the utility in Maine state in Fig.~\ref{fig3}.  Similar error histograms can be observed for other utilities.

In our simulation, we observe that the sample mean of the load prediction error is -0.068 (MWh), and the sample standard deviation is 38.7 (MWh). Note that we model the load prediction error $\epsilon_i$ as a symmetric unimodal random variable with zero mean. For testing the symmetry around a specific center, we apply \emph{Two-sample Kolmogorov-Smirnov test} on the prediction error samples~\cite{zheng2010bootstrap,csorgHo1987testing}. It is used to test whether two underlying one-dimensional probability distributions differ.  Let $\{X_1,..., X_n\}$ be the observed values of $\epsilon_i$. The result shows that the null hypothesis (i.e., ``The sample data $\{X_1,..., X_n\}$ and $\{-X_1,..., -X_n\}$ are from the same continuous distribution'') is not rejected at the $5\%$ significance level. This observation verifies the  symmetry setting on the load prediction error, i.e., $\epsilon_i$ follows a symmetric distribution with zero mean. The \emph{central dominant} condition can also be justified. As seen from the error histogram in Fig.~\ref{fig3}, when
the amplitude of the
prediction error becomes larger, it has a smaller frequency of occurrence accordingly. We have similar observations for other utilities. Note that here we do not assume the independence of utilities' load prediction errors $\epsilon_i$. It is observed that there are positive correlation coefficients in $\epsilon_i$ among utilities, as large as 0.66. In the following subsection, we find that our previous results still hold under this scenario, which shows the robustness of our analysis.

\subsection{Market Equilibrium and Efficiency Performance}
We now investigate the impact of the strategic behaviors of utilities on the $\textsf{ABC}$ and the market equilibrium. We obtained the empirical $\textsf{ABC}_i$ with respect to the bidding strategy $\mu_i$ of the utility in Maine state by calculating the two-settlement average hourly cost across the consecutive 8 years, for different values of $\mu_i$.
\begin{figure}[H]
\centering
\includegraphics[width = 0.5\linewidth]{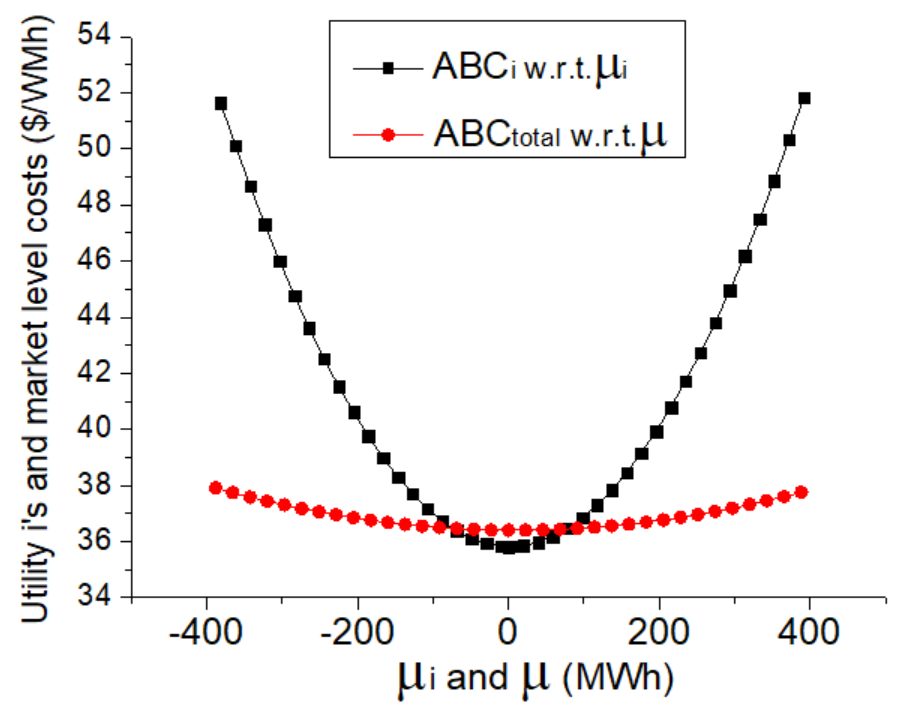}
\caption{Utility's and market-level costs w.r.t. $\mu_i$ and $\mu$.}
\label{fig7}
\end{figure}

 As seen in Fig.~\ref{fig7}, the utility's cost takes the minimum when the utility bids according to prediction, given all others are bidding at prediction. Thus the strategy profile that all utilities bid according to prediction is a pure strategy Nash equilibrium. 
 Furthermore, we investigate the market-level  $\textsf{ABC}_{total}$, which is computed as the aggregate cost over aggregate demand of all utilities. It is easy to justify that the market-level $\textsf{ABC}_{total}$ only relates to $\mu=\sum^{N}_{i=1}\mu_i$. For a particular $\hat{\mu}$, we randomly decompose it such that $\hat{\mu}=\sum^{N}_{i=1}\hat{\mu_i}$. The corresponding market-level $\textsf{ABC}_{total}$ is computed as the average of $\textsf{ABCs}$ of different strategy profiles.  {{We observe that the social cost under the game-theoretical strategic setting is the same as the optimal one under the coordinated setting, i.e., the equilibrium strategy profile $\boldsymbol{\mu^{*}}=\vec{0}$ is the optimal solution that minimizes the social cost.} Fig.~\ref{fig7} shows that this equilibrium incurs no loss of efficiency with respect to the market-level $\textsf{ABC}_{total}$. {We remark that when a utility deviates from the equilibrium, its cost will increase while all other utilities' costs will decrease; see Fig.~\ref{fig7} and Fig.~\ref{fig4} for illustration. Furthermore, the market-level $\textsf{ABC}_{total}$, average of all utilities' costs, will also increase. This corresponds to our theoretical results in Theorem~\ref{theorem4} and implies that deviating from equilibrium leads to efficiency loss.}}

\subsection{Market Size Analysis and Sensitivity of Real-time Market}\label{ssec:sensitivity}

\begin{figure*}[!t]
    \minipage{.225\textwidth}
    \includegraphics[width=1.15\linewidth]{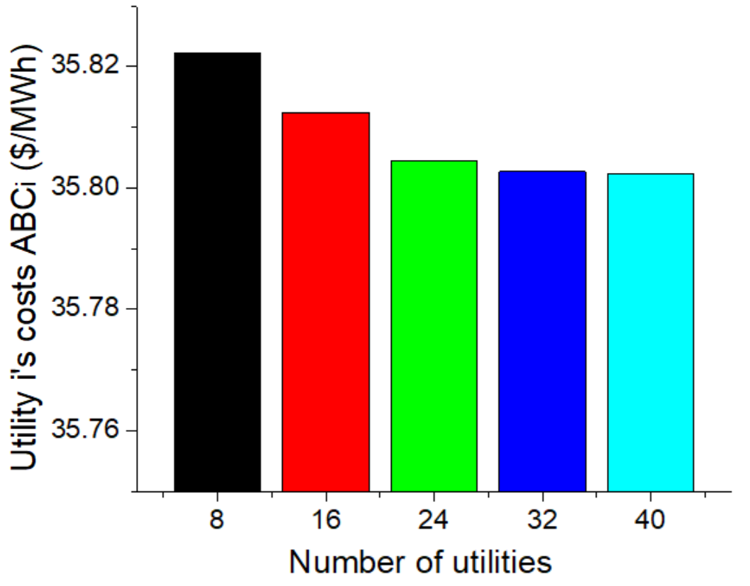}
	\caption{Impact of the market size on utility's cost under equilibrium.}
	\label{fig8}
	\endminipage\hfill
	\minipage{.225\textwidth}
    \includegraphics[width=1.15\linewidth]{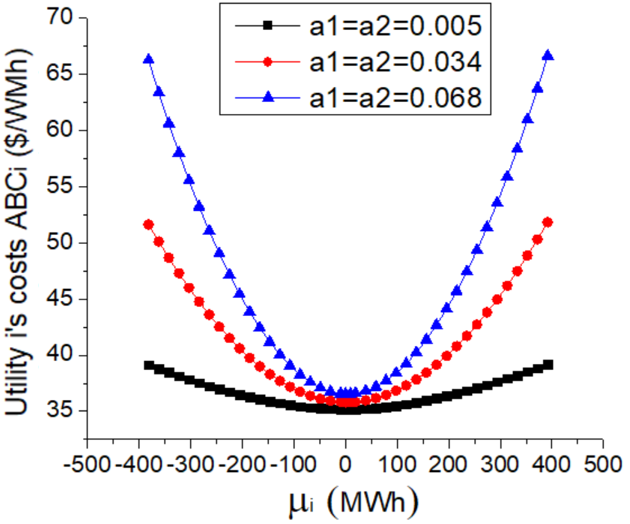}
	\caption{Impact of $a_1$ and $a_2$ on utility's cost.}
	\label{fig9}
	\endminipage\hfill
	\minipage{.225\textwidth}
	\includegraphics[width=1.15\linewidth]{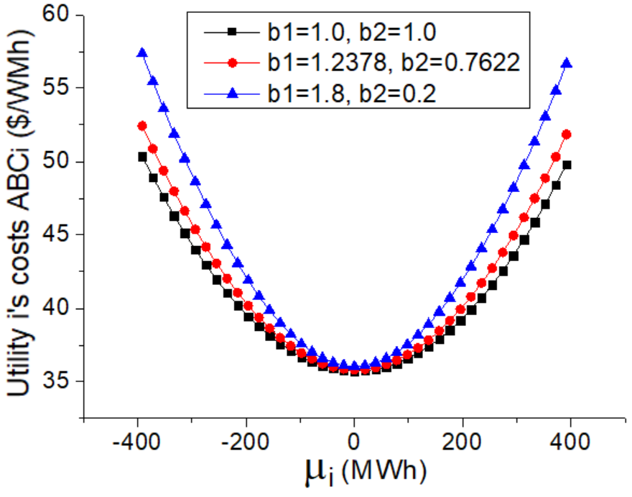}
	\caption{Impact of  $b_1-b_2$ on utility's cost.}
	\label{fig10}
	\endminipage\hfill
	\minipage{.225\textwidth}
	\includegraphics[width=1.15\linewidth]{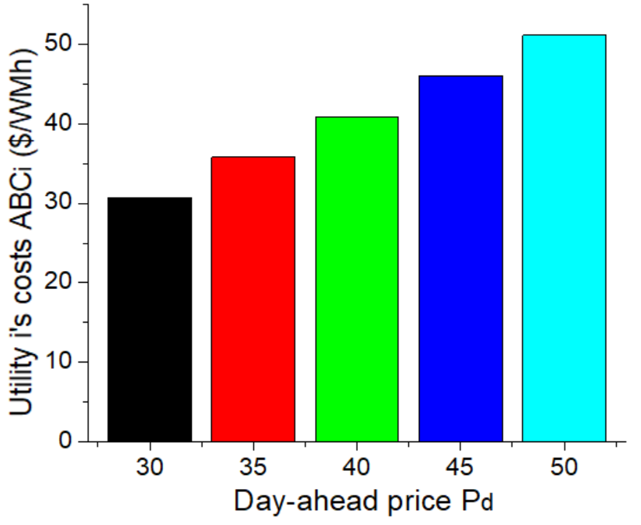}
	\caption{Impact of $p_d$ on utility's cost under equilibrium.}
	\label{fig11}
	\endminipage\hfill
\end{figure*}

The strategic potential of utilities arises from the dynamically changed real-time market electricity price. The sensitivity of the real-time market and the market size may propose impacts on utilities' costs. These two aspects of the real-time market can be regarded as the price-changing characteristics with respect to the total imbalance, i.e., the slope and the discontinuous gap of the pricing model, and the number of participants $N$ respectively. We study the impact of these three parameters. Toward this end, we equally separate each one of the 8 utilities into 2 to 5 sub-utilities as expanding the market size and calculate each new utility's \textsf{ABC}.  Meanwhile, we vary the slope of the symmetric pricing model to be 0.005, 0.034 and 0.068, and we change $(b_1, b_2)$ to be $(1, 1)$, $(1.2378, 0.7622)$ and $(1.8, 0.2)$, which are sufficient to illustrate the impacts of the real-time market sensitivity to imbalance. {The corresponding utility's cost is studied.} The trend of cost change can be observed when the market size expands and the market sensitivity increases. We use a utility split from the original utility in Maine state as an example to show the cost change trend. {Previously, we have shown that when a utility deviates from the equilibrium, both its cost and the market-level cost will increase. In order to study the impacts of market size and market sensitivity on the aggregate cost of all utilities, we further investigate the market-level cost $\textsf{ABC}_{total}$ change under different market settings both at the equilibrium and the strategy deviation conditions.}

\begin{figure*}[t]
    \minipage{.225\textwidth}
    \includegraphics[width=1.15\linewidth]{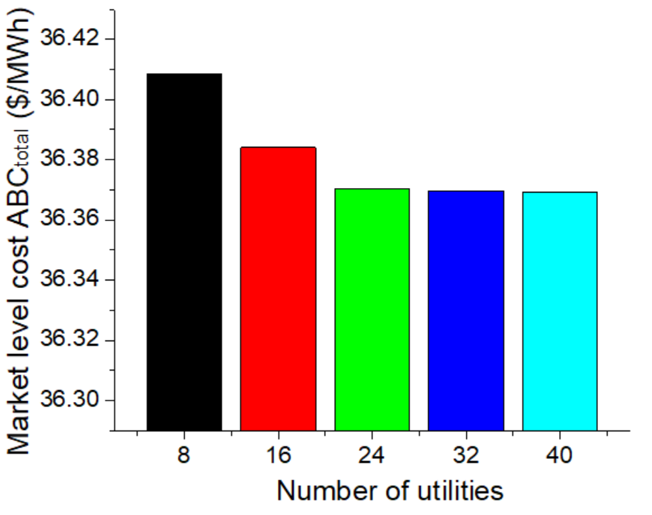}
	\caption{Impact of the market size on market-level cost under equilibrium.}
	\label{fig12}
	\endminipage\hfill
	\minipage{.225\textwidth}
	\includegraphics[width=1.15\linewidth]{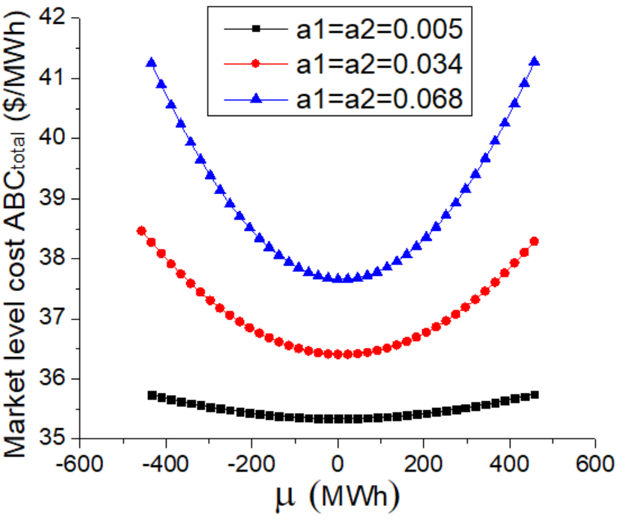}
	\caption{Impact of $a_1$ and $a_2$ on market-level cost.}
	\label{fig13}
	\endminipage\hfill
	\minipage{.225\textwidth}
    \includegraphics[width=1.15\linewidth]{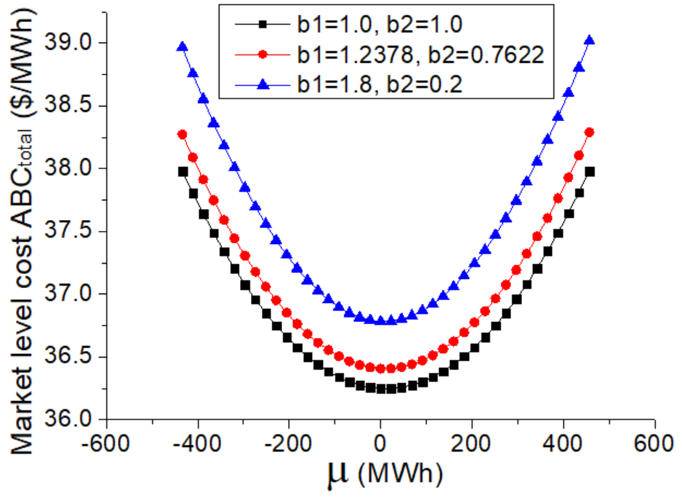}
	\caption{Impact of $b_1-b_2$ on market-level cost.}
	\label{fig14}
	\endminipage\hfill
	\minipage{.225\textwidth}
	\includegraphics[width=1.15\linewidth]{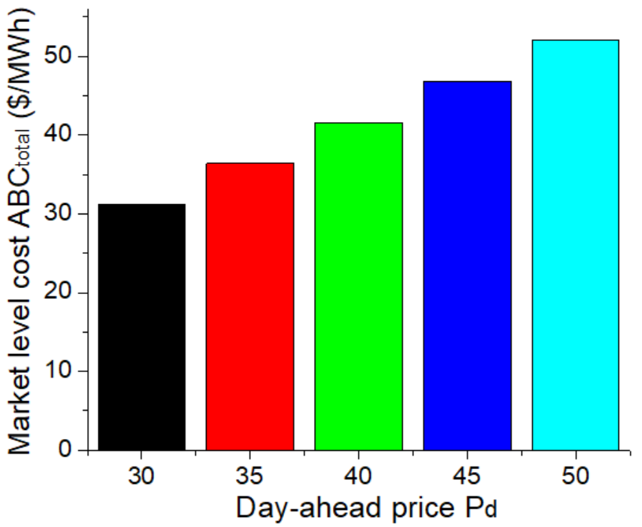}
	\caption{Impact of $p_d$ on market-level cost under equilibrium.}
	\label{fig15}
	\endminipage\hfill
\end{figure*}

As seen, Fig.~\ref{fig8} and Fig.~\ref{fig12} demonstrate that expanding the market size, i.e., increasing the number of utilities $N$, contributes to decreasing both the utility's cost and the market-level cost under the equilibrium. This characteristic implies that competition improves efficiency. Based on these observations, market designers have an economic incentive to allow competition and expand market access in order to lower both utilities' costs and the market-level cost.

Meanwhile, Fig.~\ref{fig9} depicts that when the slope of the piece-wise linear symmetric spot pricing model becomes larger, i.e., larger $a_1$ and $a_2$, given $\mu_{-i}=0$, the utility suffers a larger cost given the same deviation quantity $\mu_i$. The market-level $\textsf{ABC}_{total}$ presents a similar cost-strategy relationship with respect to $\mu$. In addition, Fig.~\ref{fig10} and Fig.~\ref{fig14} show that when the premium of readiness increases, i.e., larger $b_1-b_2$, both the utility and the market observe an increase in their costs under the same strategy deviation quantity $\mu_i$ and $\mu$ respectively.

{Furthermore, it is worth noticing that compared with shrinking $b_1-b_2$, decreasing $a_1$ and $a_2$ presents a more significant reduction in the deterioration rate (defined as the ratio between the cost increase and the cost under the equilibrium) for  both the utility and the market under the same deviation quantity $\mu_i$ and $\mu$. This observation implies that when the real-time market becomes more robust to the total imbalance (corresponding to smaller $a_1$ and $a_2$), the market equilibrium tends to be less sensitive to the fault behaviors of utilities.}

We also study the impacts of day-ahead clearing price $p_d$. Fig.~\ref{fig11} and Fig.~\ref{fig15} show that when the day-ahead clearing price $p_d$ increases, both the utility and the market have lager $\textsf{ABCs}$. Since the costs of utilities are proportional to $p_d$, we observe that there exist linear relationships between $\textsf{ABC}_i$ and $\mu_i$, and between $\textsf{ABC}_{total}$ and $\mu$.

These observations correspond to our theoretical results in Theorem~\ref{theorem1} and Theorem~\ref{theorem4}. The above study suggests that improving the level of competition and the resilience of the spot pricing against the market-level mismatch can not only benefit utilities but also reduce the social cost.

\rev{

\subsection{Performance under Asymmetric Load Uncertainty}\label{aysmmetric_load}
 Recall that the probability density functions of net load uncertainties are modeled as symmetric and unimodal. We further study the impact of asymmetric load prediction errors on utilities' optimal bidding strategies and the market equilibrium. Multi-peak asymmetric probability densities are observed when conducting short-term wind power predictions~\cite{ye2019combined,ge2018parameter}, which can be fitted by piece-wise exponential distribution, Beta distribution, or Gaussian Mixture Model (GMM). In this study, we adopt the GMM to describe such prediction errors, which is the combination of several Gaussian distribution components. We form net load prediction errors according to the GMMs in~\cite{ye2019combined} and scale them to zero mean; see Fig.~\ref{fig16} for illustration. 
 Here the GMMs are weighted combinations of three different Gaussian distributions, i.e.,
 \begin{equation}
     \begin{split}
      f_{\text{GMM}}(x)=\sum^{3}_{k=1}\omega_k\cdot f_{(\mu_k,\sigma_k)}(x),
     \end{split}
 \end{equation}
where $\omega_k$, $\mu_k$, and $\sigma_k$ denote the weight, mean value, and standard deviation of the $k$-th Gaussian density component $f_{(\mu_k,\sigma_k)}(x)$ respectively. 
 The GMM parameters are listed in Table~\ref{table2}. Utility's cost to strategy curve and the market-level cost curve are presented in Fig.~\ref{fig17}.
 
 It is observed that with such multi-peak asymmetric load errors, the utility's cost still takes the minimum when it bids at prediction, given all others are bidding according to prediction. Therefore, all utility bidding according to prediction is a pure strategy Nash Equilibrium. We further investigate the market-level $\textsf{ABC}_{total}$. The market-level cost curve indicates that the social cost is minimized at the equilibrium, and hence the equilibrium incurs no loss of efficiency. 
 The above results show that the market could still admit an efficient equilibrium even with the multi-peak and asymmetric load uncertainty. 
 
 \begin{figure}[!t]
\centering
\begin{minipage}[t]{0.225\textwidth}
\centering
\includegraphics[width=1.05\linewidth]{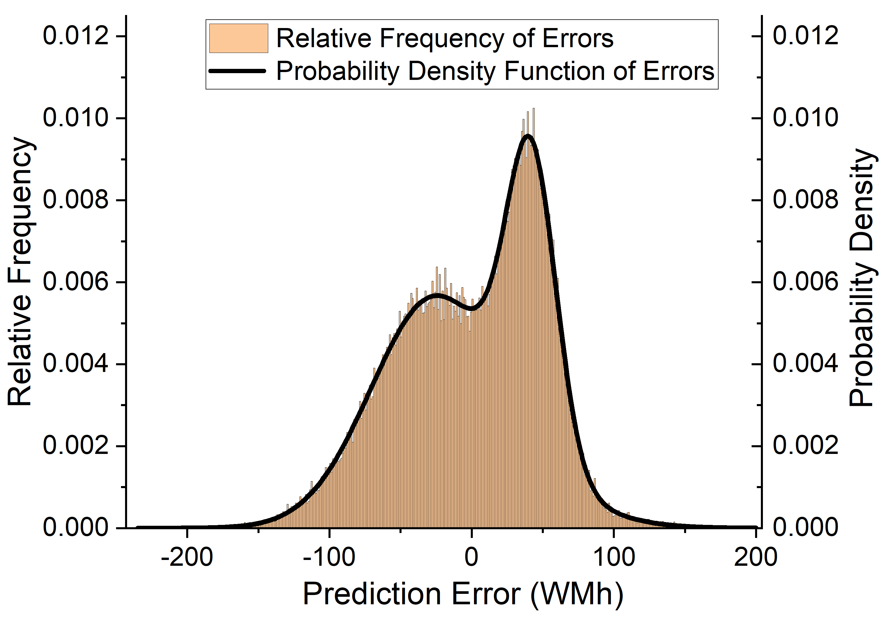}
\caption{Multi-peak and asymmetric load uncertainties.}\label{fig16}
\end{minipage}
\begin{minipage}[t]{0.225\textwidth}
\centering
\includegraphics[width=0.92\linewidth]{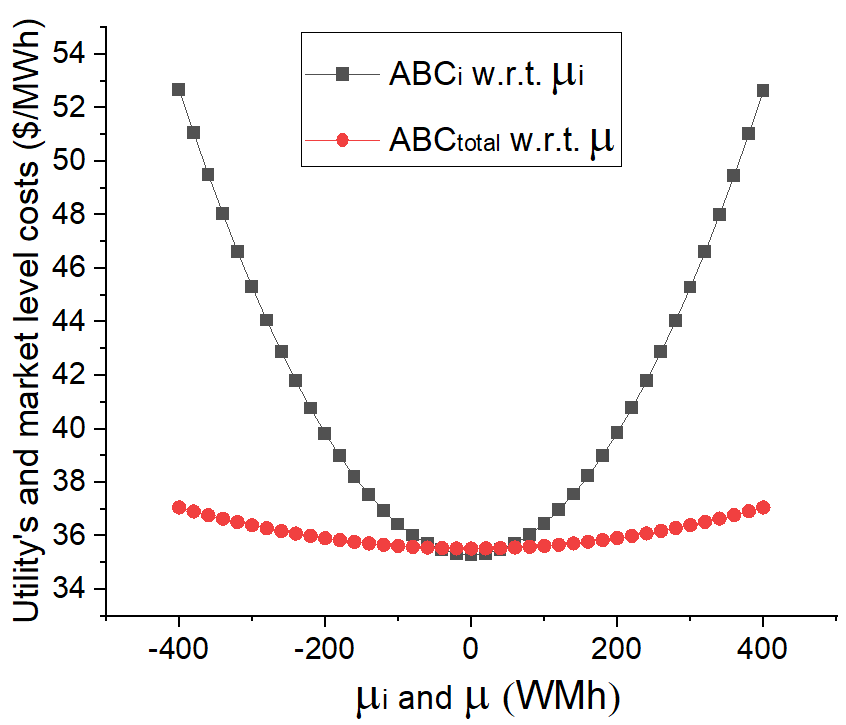}
\caption{Utility's and market-level costs w.r.t. $\mu_i$ and $\mu$.}\label{fig17}
\end{minipage}
\begin{minipage}[t]{0.225\textwidth}
\centering
\includegraphics[width=0.92\linewidth]{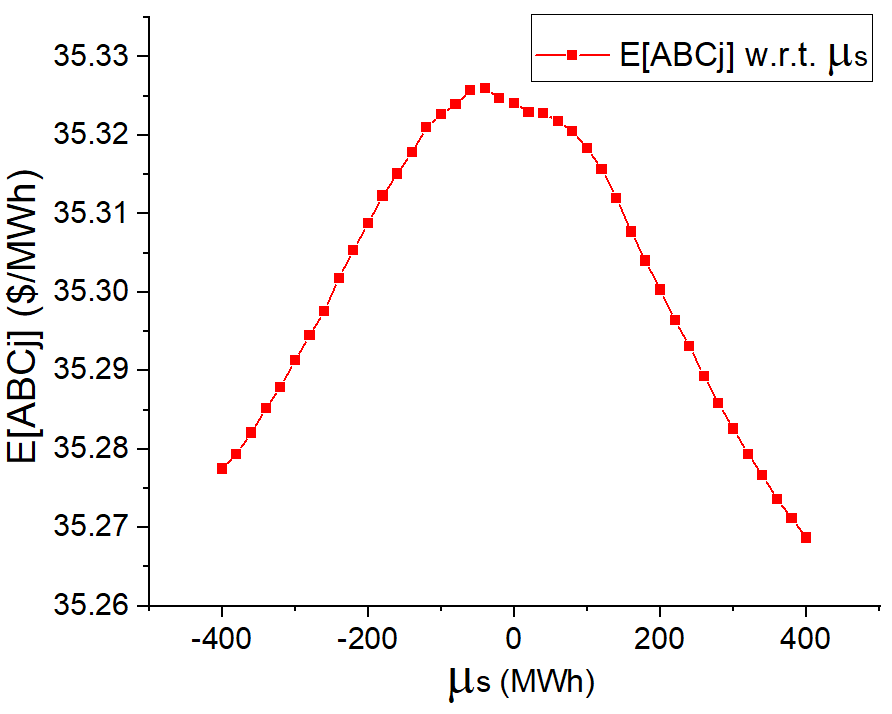}
\caption{Equilibrium robustness.}\label{fig18}
\end{minipage}
\begin{minipage}[t]{0.225\textwidth}
\centering
\includegraphics[width=0.92\linewidth]{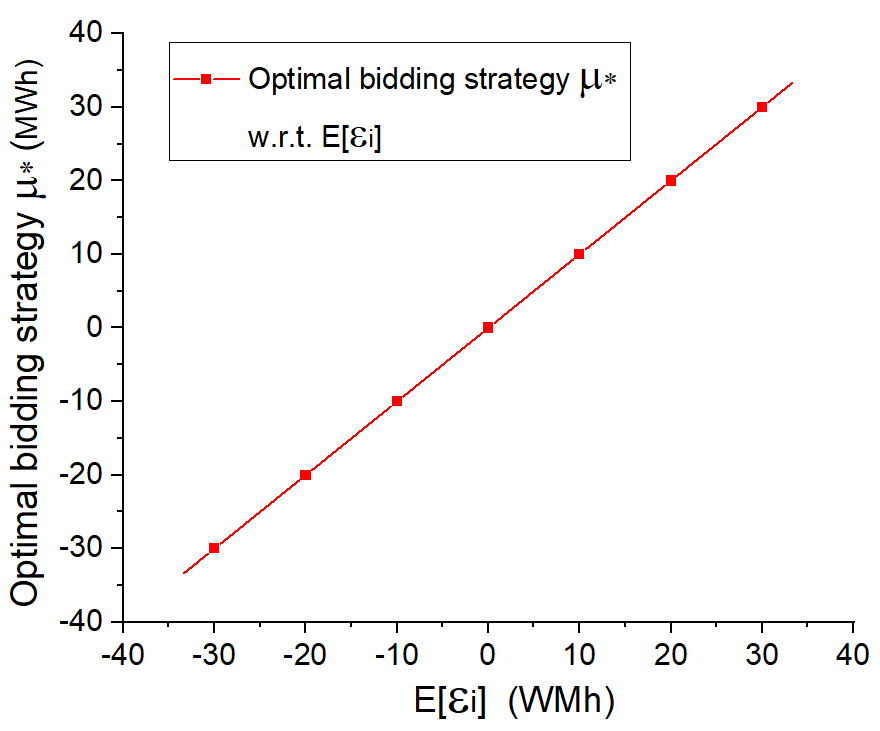}
\caption{Bidding w.r.t. $\mathbb{E}[\epsilon_i]$}\label{fig19}
\end{minipage}
\end{figure}

\begin{table}[!t]  
\centering  
\caption{Parameters of GMMs for each utility} 
\label{table2}
 \begin{tabular}{c|ccccccccc}  
     \hline
     \hline
       Utility Location& $\omega_1$ & $ \omega_2$ & $\omega_3$ & $ \mu_1$ &$\mu_2$ & $ \mu_3$ & $\sigma_1$ & $ \sigma_2$ & $ \sigma_3$  \\ [1ex]  
       \hline 
ME	& 0.3156	& 0.1414	& 0.5430&	42.0089&	19.9489&	-29.6111&	17.494&	51.2893&	43.7818\\
NH	& 0.3156	& 0.1414	& 0.5430&	42.6092&	20.234&	-30.0342&	17.744&	52.0221&	44.4074\\
VT	& 0.3156	& 0.1414	& 0.5430&	20.6125&	9.78833&	-14.5292&	8.58377&	25.1661&	21.4824\\
CT	& 0.3156	& 0.1414	& 0.5430&	111.516&	52.9562&	-78.6051&	46.4394&	136.152&	116.223\\
RI	& 0.3156	& 0.1414	& 0.5430&	29.8041&	14.1532&	-21.0082&	12.4115&	36.3882&	31.0619\\
SEMA	& 0.3156	& 0.1414	& 0.5430&	54.1433&	25.7112&	-38.1642&	22.5472&	66.1042&	56.4282\\
WCMA	& 0.3156	& 0.1414	& 0.5430&	62.827&	29.8349&	-44.2852&	26.1634&	76.7063&	65.4784\\
NEMA	& 0.3156	& 0.1414	& 0.5430&	92.4694&	43.9113&	-65.1794&	38.5075&	112.897&	96.3718\\

       \hline
       \hline
   \end{tabular}
\end{table}

  Towards the robustness of the equilibrium, we demonstrate the result in Fig.~\ref{fig18}. We observe that different from the case of symmetric unimodal prediction errors in Fig.~\ref{fig4}, the utility's cost curve is distorted may not exactly decrease w.r.t. other utilities' irrational behaviors, though the trends performs similarly. This observation provides utilities an incentive to enhance the prediction models towards symmetric unimodal forecasting errors so that the equilibrium can be robust to any fault actions. Meanwhile, we find that fault behaviors cause at most $0.005\%$ cost increase compared with the cost at equilibrium among 8 utilities, indicating the adaptability of equilibrium robustness. We leave the theoretical analysis for multi-peak and asymmetric load uncertainty for further work.
 
 }
 
\subsection{Performance under Asymmetric Pricing Model}

We now study the impact of the asymmetric pricing model on the market equilibrium and the utility's cost. The pricing model parameters are listed in Table~\ref{table1}. We present the cost to strategy curve of the utility in Maine State. We observe similar cost to strategy relationships for other utilities.

\begin{figure}[H]
\centering
\includegraphics[width = 0.5\linewidth]{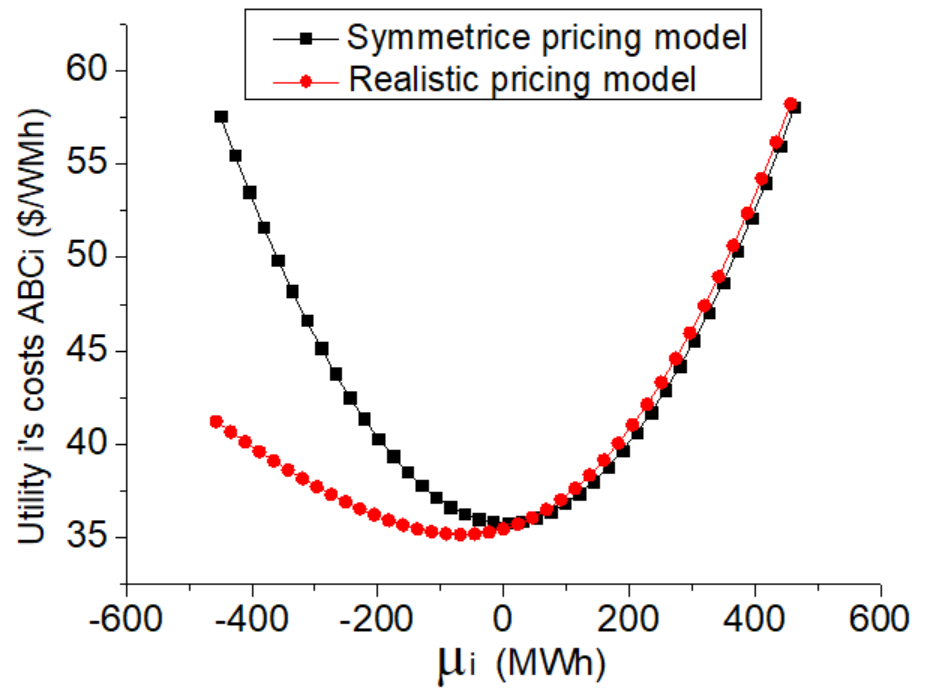}
\caption{Impact of the asymmetricity of the pricing model, where parameters are shown in Table~\ref{table1}.}
\label{fig20}
\end{figure}

As seen, Fig.~\ref{fig20} shows that when all other utilities bid according to prediction, the utility has an incentive to deviate from bidding according to prediction, which implies that the strategy profile that all utilities bid according to prediction is no longer a Nash Equilibrium. Under the asymmetric pricing model, a utility can reduce its cost by bidding higher than the predicted net load. This can be explained intuitively as follows: when the real-time market performs less sensitive to the negative imbalance, i.e., $a_1>a_2$, utilities can overbuy in the day-ahead market to sell the surplus at a higher price compared with the symmetric pricing model case that we choose.  

From the cost to strategy curve in Fig.~\ref{fig20}, the optimal non-zero bidding strategy for the utility is to choose $\mu^*_{i}=-68.5 \ (\text{MWh})$. Under this case, the utility only witnesses a $0.84\%$ cost reduction compared with choosing $\mu_i=0$, which indicates that under the realistic asymmetric pricing model suggested in~\cite{Neupane}, the utility does not have much incentive to deviate from bidding according to prediction given all other utilities bid according to prediction.

\section{Conclusion}\label{conclusion}

We study the strategic bidding behaviors of utilities under a game-theoretical setting in the deregulated electricity market, with  the uncertainty in demand and local renewable generation in consideration. 
We show that all utilities bidding according to (net load) prediction is the unique pure strategy Nash Equilibrium. Furthermore, it incurs no loss of efficiency and is (0, $N-1$) fault immune to irrational fault behaviors. We extend the results to the cases with correlated prediction errors and a general class of real-time spot pricing models. Our simulation results suggest that market designers may improve the level of competition and the resilience of the spot pricing against the market-level mismatch to reduce utilities' costs and the social cost. Our study highlights that the market operator can design real-time pricing schemes according to certain conditions, such that the interactions among utilities admit a unique, efficient, and robust pure strategy Nash Equilibrium.

{This study is based on the setting that utilities are price-takers in the day-ahead market but price-makers in the real-time market. Under such a scenario, we show that the market admits a unique equilibrium with salient properties. Different from this setting, as it has often been pointed out, in some existing electricity markets, the day-ahead market clearing price is dominated by several large corporations~\cite{maekawa2019speculative}. In this situation, our results still hold if we focus on the interactions between small-scale utilities. The interactions between price-maker utilities, the market clearing price, and the corresponding market equilibrium can be studied in the future. In addition, this work focuses on the demand side game-theoretical economic analysis, we note that bringing the power network constraints and the time-coupling energy storage operations and the generation imbalance uncertainty into market equilibrium analysis requires further investigation.}

\bibliographystyle{IEEEtran}
\bibliography{IEEEabrv,reference}

\begin{thebibliography}{10}
\providecommand{\url}[1]{#1}
\csname url@samestyle\endcsname
\providecommand{\newblock}{\relax}
\providecommand{\bibinfo}[2]{#2}
\providecommand{\BIBentrySTDinterwordspacing}{\spaceskip=0pt\relax}
\providecommand{\BIBentryALTinterwordstretchfactor}{4}
\providecommand{\BIBentryALTinterwordspacing}{\spaceskip=\fontdimen2\font plus
\BIBentryALTinterwordstretchfactor\fontdimen3\font minus
  \fontdimen4\font\relax}
\providecommand{\BIBforeignlanguage}[2]{{%
\expandafter\ifx\csname l@#1\endcsname\relax
\typeout{** WARNING: IEEEtran.bst: No hyphenation pattern has been}%
\typeout{** loaded for the language `#1'. Using the pattern for}%
\typeout{** the default language instead.}%
\else
\language=\csname l@#1\endcsname
\fi
#2}}
\providecommand{\BIBdecl}{\relax}
\BIBdecl

\bibitem{gilbert2007international}
R.~J. Gilbert and E.~P. Kahn, \emph{International Comparisons of Electricity
  Regulation}.\hskip 1em plus 0.5em minus 0.4em\relax Cambridge University
  Press, 2007.

\bibitem{kirschen2018fundamentals}
D.~S. Kirschen and G.~Strbac, \emph{Fundamentals of Power System
  Economics}.\hskip 1em plus 0.5em minus 0.4em\relax John Wiley \& Sons, 2018.

\bibitem{ISONE}
{New England ISO}, ``{ISO New England electricity market},'' 2020,
  \url{http://www.iso-ne.com/} Accessed 20 April 2020.

\bibitem{Nordpool}
{Nord Pool}, ``{Nord Pool power market},'' 2020,
  \url{https://www.nordpoolgroup.com/the-power-market/} Accessed 20 April 2020.

\bibitem{stoft2002power}
S.~Stoft, ``Power system economics,'' \emph{Journal of Energy Literature},
  vol.~8, pp. 94--99, 2002.

\bibitem{zhang2010restructured}
X.-P. Zhang, \emph{Restructured Electric Power Systems: Analysis of Electricity
  Markets with Equilibrium Models}.\hskip 1em plus 0.5em minus 0.4em\relax John
  Wiley \& Sons, 2010, vol.~71.

\bibitem{ISONEReport}
{New York ISO}, ``{2018 state of the market report for the New York ISO
  markets},'' 2018,
  \url{https://www.nyiso.com/documents/20142/2223763/2018-State-of-the-Market-Report.pdf}
  Accessed 20 April 2020.

\bibitem{herranz2012optimal}
R.~Herranz, A.~M. San~Roque, J.~Villar, and F.~A. Campos, ``Optimal demand-side
  bidding strategies in electricity spot markets,'' \emph{IEEE Transactions on
  Power Systems}, vol.~27, no.~3, pp. 1204--1213, 2012.

\bibitem{song2016purchase}
M.~Song and M.~Amelin, ``Purchase bidding strategy for a retailer with flexible
  demands in day-ahead electricity market,'' \emph{IEEE Transactions on Power
  Systems}, vol.~32, no.~3, pp. 1839--1850, 2016.

\bibitem{fleten2005constructing}
S.-E. Fleten and E.~Pettersen, ``Constructing bidding curves for a price-taking
  retailer in the norwegian electricity market,'' \emph{IEEE Transactions on
  Power Systems}, vol.~20, no.~2, pp. 701--708, 2005.

\bibitem{hajati2011optimal}
M.~Hajati, H.~Seifi, and M.~K. Sheikh-El-Eslami, ``Optimal retailer bidding in
  a da market--a new method considering risk and demand elasticity,''
  \emph{Energy}, vol.~36, no.~2, pp. 1332--1339, 2011.

\bibitem{nojavan2015optimal}
S.~Nojavan, B.~Mohammadi-Ivatloo, and K.~Zare, ``Optimal bidding strategy of
  electricity retailers using robust optimisation approach considering
  time-of-use rate demand response programs under market price uncertainties,''
  \emph{IET Generation, Transmission \& Distribution}, vol.~9, no.~4, pp.
  328--338, 2015.

\bibitem{nojavan2015robust}
------, ``Robust optimization based price-taker retailer bidding strategy under
  pool market price uncertainty,'' \emph{International Journal of Electrical
  Power \& Energy Systems}, vol.~73, pp. 955--963, 2015.

\bibitem{xu2015efficient}
Y.~Xu and S.~H. Low, ``An efficient and incentive compatible mechanism for
  wholesale electricity markets,'' \emph{IEEE Transactions on Smart Grid},
  vol.~8, no.~1, pp. 128--138, 2015.

\bibitem{kazempour2013equilibria}
S.~J. Kazempour and H.~Zareipour, ``Equilibria in an oligopolistic market with
  wind power production,'' \emph{IEEE Transactions on Power Systems}, vol.~29,
  no.~2, pp. 686--697, 2013.

\bibitem{zhang2018competition}
B.~Zhang, R.~Johari, and R.~Rajagopal, ``Competition and efficiency of
  coalitions in cournot games with uncertainty,'' \emph{IEEE Transactions on
  Control of Network Systems}, vol.~6, no.~2, pp. 884--896, 2018.

\bibitem{rasouli2019efficient}
M.~Rasouli and D.~Teneketzis, ``An efficient market design for electricity
  networks with strategic users possessing local information,'' \emph{IEEE
  Transactions on Control of Network Systems}, 2019.

\bibitem{mazzi2017price}
N.~Mazzi, J.~Kazempour, and P.~Pinson, ``Price-taker offering strategy in
  electricity pay-as-bid markets,'' \emph{IEEE Transactions on Power Systems},
  vol.~33, no.~2, pp. 2175--2183, 2017.

\bibitem{steeger2014optimal}
G.~Steeger, L.~A. Barroso, and S.~Rebennack, ``Optimal bidding strategies for
  hydro-electric producers: A literature survey,'' \emph{IEEE Transactions on
  Power Systems}, vol.~29, no.~4, pp. 1758--1766, 2014.

\bibitem{shrestha2009generation}
G.~Shrestha and Q.~Songbo, ``Generation scheduling for a price taker genco in
  competitive power markets,'' in \emph{2009 IEEE/PES Power Systems Conference
  and Exposition}.\hskip 1em plus 0.5em minus 0.4em\relax IEEE, 2009, pp. 1--6.

\bibitem{yucekaya2013bidding}
A.~Yucekaya, ``Bidding of price taker power generators in the deregulated
  turkish power market,'' \emph{Renewable and Sustainable Energy Reviews},
  vol.~22, pp. 506--514, 2013.

\bibitem{conejo2002price}
A.~J. Conejo, F.~J. Nogales, and J.~M. Arroyo, ``Price-taker bidding strategy
  under price uncertainty,'' \emph{IEEE Transactions on Power Systems},
  vol.~17, no.~4, pp. 1081--1088, 2002.

\bibitem{vilim2014wind}
M.~Vilim and A.~Botterud, ``Wind power bidding in electricity markets with high
  wind penetration,'' \emph{Applied Energy}, vol. 118, pp. 141--155, 2014.

\bibitem{papavasiliou2013multiarea}
A.~Papavasiliou and S.~S. Oren, ``Multiarea stochastic unit commitment for high
  wind penetration in a transmission constrained network,'' \emph{Operations
  Research}, vol.~61, no.~3, pp. 578--592, 2013.

\bibitem{pritchard2010single}
G.~Pritchard, G.~Zakeri, and A.~Philpott, ``A single-settlement, energy-only
  electric power market for unpredictable and intermittent participants,''
  \emph{Operations research}, vol.~58, no. 4-part-2, pp. 1210--1219, 2010.

\bibitem{lee2011firm}
Y.-Y. Lee, R.~Baldick, and J.~Hur, ``Firm-based measurements of market power in
  transmission-constrained electricity markets,'' \emph{IEEE Transactions on
  Power Systems}, vol.~26, no.~4, pp. 1962--1970, 2011.

\bibitem{lavaei2012competitive}
J.~Lavaei and S.~Sojoudi, ``Competitive equilibria in electricity markets with
  nonlinearities,'' in \emph{2012 American Control Conference (ACC)}.\hskip 1em
  plus 0.5em minus 0.4em\relax IEEE, 2012, pp. 3081--3088.

\bibitem{kumbartzky2017optimal}
N.~Kumbartzky, M.~Schacht, K.~Schulz, and B.~Werners, ``Optimal operation of a
  {CHP} plant participating in the german electricity balancing and day-ahead
  spot market,'' \emph{European Journal of Operational Research}, vol. 261,
  no.~1, pp. 390--404, 2017.

\bibitem{boomsma2014bidding}
T.~K. Boomsma, N.~Juul, and S.-E. Fleten, ``Bidding in sequential electricity
  markets: The nordic case,'' \emph{European Journal of Operational Research},
  vol. 238, no.~3, pp. 797--809, 2014.

\bibitem{nazari2019optimal}
M.~Nazari and M.~Ardehali, ``Optimal bidding strategy for a {GENCO} in
  day-ahead energy and spinning reserve markets with considerations for
  coordinated wind-pumped storage-thermal system and {CO2} emission,''
  \emph{Energy Strategy Reviews}, vol.~26, p. 100405, 2019.

\bibitem{ventosa2005electricity}
M.~Ventosa, A.~Ba{\i}llo, A.~Ramos, and M.~Rivier, ``Electricity market
  modeling trends,'' \emph{Energy Policy}, vol.~33, no.~7, pp. 897--913, 2005.

\bibitem{rebennack2010energy}
S.~Rebennack, J.~Kallrath, and P.~M. Pardalos, ``Energy portfolio optimization
  for electric utilities: case study for germany,'' in \emph{Energy, Natural
  Resources and Environmental Economics}.\hskip 1em plus 0.5em minus
  0.4em\relax Springer, 2010, pp. 221--246.

\bibitem{david2000strategic}
A.~K. David and F.~Wen, ``Strategic bidding in competitive electricity markets:
  a literature survey,'' in \emph{IEEE Power Engineering Society Summer
  Meeting}.\hskip 1em plus 0.5em minus 0.4em\relax Seattle, USA, July, 2000,
  pp. 2168--2173.

\bibitem{philpott2006optimizing}
A.~B. Philpott and E.~Pettersen, ``Optimizing demand-side bids in day-ahead
  electricity markets,'' \emph{IEEE Transactions on Power Systems}, vol.~21,
  no.~2, pp. 488--498, 2006.

\bibitem{atzeni2014noncooperative}
I.~Atzeni, L.~G. Ord{\'o}{\~n}ez, G.~Scutari, D.~P. Palomar, and J.~R.
  Fonollosa, ``Noncooperative day-ahead bidding strategies for demand-side
  expected cost minimization with real-time adjustments: A {GNEP} approach,''
  \emph{IEEE Transactions on Signal Processing}, vol.~62, no.~9, pp.
  2397--2412, 2014.

\bibitem{persson2018simplify}
S.~Persson, ``Simplify bidding on the day-ahead electricity market nordpool
  through structured time-series,'' \emph{Dissertation}, 2018.

\bibitem{samadi2013tackling}
P.~Samadi, H.~Mohsenian-Rad, V.~W. Wong, and R.~Schober, ``Tackling the load
  uncertainty challenges for energy consumption scheduling in smart grid,''
  \emph{IEEE Transactions on Smart Grid}, vol.~4, no.~2, pp. 1007--1016, 2013.

\bibitem{aghaei2013demand}
J.~Aghaei and M.-I. Alizadeh, ``Demand response in smart electricity grids
  equipped with renewable energy sources: A review,'' \emph{Renewable and
  Sustainable Energy Reviews}, vol.~18, pp. 64--72, 2013.

\bibitem{kamyab2016demand}
F.~Kamyab, M.~Amini, S.~Sheykhha, M.~Hasanpour, and M.~M. Jalali, ``Demand
  response program in smart grid using supply function bidding mechanism.''
  \emph{IEEE Transactions on Smart Grid}, vol.~7, no.~3, pp. 1277--1284, 2016.

\bibitem{dowling2017multi}
A.~W. Dowling, R.~Kumar, and V.~M. Zavala, ``A multi-scale optimization
  framework for electricity market participation,'' \emph{Applied Energy}, vol.
  190, pp. 147--164, 2017.

\bibitem{li2015demand}
N.~Li, L.~Chen, and M.~A. Dahleh, ``Demand response using linear supply
  function bidding,'' \emph{IEEE Transactions on Smart Grid}, vol.~6, no.~4,
  pp. 1827--1838, 2015.

\bibitem{jin2018microgrid}
M.~Jin, W.~Feng, C.~Marnay, and C.~Spanos, ``Microgrid to enable optimal
  distributed energy retail and end-user demand response,'' \emph{Applied
  Energy}, vol. 210, pp. 1321--1335, 2018.

\bibitem{carrion2009bilevel}
M.~Carri{\'o}n, J.~M. Arroyo, and A.~J. Conejo, ``A bilevel stochastic
  programming approach for retailer futures market trading,'' \emph{IEEE
  Transactions on Power Systems}, vol.~24, no.~3, pp. 1446--1456, 2009.

\bibitem{zavala2017stochastic}
V.~M. Zavala, K.~Kim, M.~Anitescu, and J.~Birge, ``A stochastic electricity
  market clearing formulation with consistent pricing properties,''
  \emph{Operations Research}, vol.~65, no.~3, pp. 557--576, 2017.

\bibitem{khazaei2017single}
J.~Khazaei, G.~Zakeri, and S.~S. Oren, ``Single and multisettlement approaches
  to market clearing under demand uncertainty,'' \emph{Operations Research},
  vol.~65, no.~5, pp. 1147--1164, 2017.

\bibitem{habibian2020multistage}
M.~Habibian, A.~Downward, and G.~Zakeri, ``Multistage stochastic demand-side
  management for price-making major consumers of electricity in a co-optimized
  energy and reserve market,'' \emph{European Journal of Operational Research},
  vol. 280, no.~2, pp. 671--688, 2020.

\bibitem{safdarian2013stochastic}
A.~Safdarian, M.~Fotuhi-Firuzabad, and M.~Lehtonen, ``A stochastic framework
  for short-term operation of a distribution company,'' \emph{IEEE Transactions
  on Power Systems}, vol.~28, no.~4, pp. 4712--4721, 2013.

\bibitem{wang2019equilibrium}
X.~Wang, K.~Zhang, S.~Zhang, and L.~Wu, ``Equilibrium analysis of electricity
  market with demand response exchange to counterbalance bid deviations of
  renewable generators,'' \emph{IEEE Systems Journal}, vol.~14, no.~2, pp.
  2713--2724, 2019.

\bibitem{hu2007using}
X.~Hu and D.~Ralph, ``Using {EPECs} to model bilevel games in restructured
  electricity markets with locational prices,'' \emph{Operations research},
  vol.~55, no.~5, pp. 809--827, 2007.

\bibitem{9029514}
P.~{You}, D.~F. {Gayme}, and E.~{Mallada}, ``The role of strategic load
  participants in two-stage settlement electricity markets,'' in \emph{2019
  IEEE 58th Conference on Decision and Control (CDC)}, 2019, pp. 8416--8422.

\bibitem{srinivasan2014bidding}
D.~Srinivasan, L.~T. Trung, and C.~Singh, ``Bidding and cooperation strategies
  for electricity buyers in power markets,'' \emph{IEEE Systems Journal},
  vol.~10, no.~2, pp. 422--433, 2014.

\bibitem{fang2016strategic}
X.~Fang, F.~Li, Y.~Wei, and H.~Cui, ``Strategic scheduling of energy storage
  for load serving entities in locational marginal pricing market,'' \emph{IET
  Generation, Transmission \& Distribution}, vol.~10, no.~5, pp. 1258--1267,
  2016.

\bibitem{wei2014energy}
W.~Wei, F.~Liu, and S.~Mei, ``Energy pricing and dispatch for smart grid
  retailers under demand response and market price uncertainty,'' \emph{IEEE
  Transactions on Smart Grid}, vol.~6, no.~3, pp. 1364--1374, 2014.

\bibitem{roozbehanit2011analysis}
M.~Roozbehanit, M.~Rinehart, M.~Dahleh, S.~Mitter, D.~Obradovic, and
  H.~Mangesius, ``Analysis of competitive electricity markets under a new model
  of real-time retail pricing,'' in \emph{2011 8th International Conference on
  the European Energy Market (EEM)}.\hskip 1em plus 0.5em minus 0.4em\relax
  IEEE, 2011, pp. 250--255.

\bibitem{yi2018impact}
H.~Yi, M.~H. Hajiesmaili, Y.~Zhang, M.~Chen, and X.~Lin, ``Impact of the
  uncertainty of distributed renewable generation on deregulated electricity
  supply chain,'' \emph{IEEE Transactions on Smart Grid}, vol.~9, no.~6, pp.
  6183--6193, 2018.

\bibitem{cai2013impact}
D.~W. Cai, S.~Adlakha, S.~H. Low, P.~De~Martini, and K.~M. Chandy, ``Impact of
  residential {PV} adoption on retail electricity rates,'' \emph{Energy
  Policy}, vol.~62, pp. 830--843, 2013.

\bibitem{atia2016sizing}
R.~Atia and N.~Yamada, ``Sizing and analysis of renewable energy and battery
  systems in residential microgrids,'' \emph{IEEE Transactions on Smart Grid},
  vol.~7, no.~3, pp. 1204--1213, 2016.

\bibitem{jia2016renewables}
L.~Jia and L.~Tong, ``Renewables and storage in distribution systems:
  Centralized vs. decentralized integration,'' \emph{IEEE JSAC}, vol.~34,
  no.~3, pp. 665--674, 2016.

\bibitem{lin2011potential}
J.~Lin, ``Potential impact of solar energy penetration on pjm electricity
  market,'' \emph{IEEE Systems Journal}, vol.~6, no.~2, pp. 205--212, 2011.

\bibitem{ye2019combined}
L.~Ye, Y.~Zhang, C.~Zhang, P.~Lu, Y.~Zhao, and B.~He, ``Combined gaussian
  mixture model and cumulants for probabilistic power flow calculation of
  integrated wind power network,'' \emph{Computers \& Electrical Engineering},
  vol.~74, pp. 117--129, 2019.

\bibitem{mohsenian2010optimal}
A.-H. Mohsenian-Rad and A.~Leon-Garcia, ``Optimal residential load control with
  price prediction in real-time electricity pricing environments,'' \emph{IEEE
  Transactions on Smart Grid}, vol.~1, no.~2, pp. 120--133, 2010.

\bibitem{motamedi2012electricity}
A.~Motamedi, H.~Zareipour, and W.~D. Rosehart, ``Electricity price and demand
  forecasting in smart grids,'' \emph{IEEE Transactions on Smart Grid}, vol.~3,
  no.~2, pp. 664--674, 2012.

\bibitem{wan2013hybrid}
C.~Wan, Z.~Xu, Y.~Wang, Z.~Y. Dong, and K.~P. Wong, ``A hybrid approach for
  probabilistic forecasting of electricity price,'' \emph{IEEE Transactions on
  Smart Grid}, vol.~5, no.~1, pp. 463--470, 2013.

\bibitem{amjady2008day}
N.~Amjady and F.~Keynia, ``Day-ahead price forecasting of electricity markets
  by mutual information technique and cascaded neuro-evolutionary algorithm,''
  \emph{IEEE Transactions on Power Systems}, vol.~24, no.~1, pp. 306--318,
  2008.

\bibitem{anbazhagan2012day}
S.~Anbazhagan and N.~Kumarappan, ``Day-ahead deregulated electricity market
  price forecasting using recurrent neural network,'' \emph{IEEE Systems
  Journal}, vol.~7, no.~4, pp. 866--872, 2012.

\bibitem{monteiro2018new}
C.~Monteiro, I.~J. Ramirez-Rosado, L.~A. Fernandez-Jimenez, and M.~Ribeiro,
  ``New probabilistic price forecasting models: Application to the iberian
  electricity market,'' \emph{International Journal of Electrical Power \&
  Energy Systems}, vol. 103, pp. 483--496, 2018.

\bibitem{jonsson2014predictive}
T.~J{\'o}nsson, P.~Pinson, H.~Madsen, and H.~A. Nielsen, ``Predictive densities
  for day-ahead electricity prices using time-adaptive quantile regression,''
  \emph{Energies}, vol.~7, no.~9, pp. 5523--5547, 2014.

\bibitem{chai2018conditional}
S.~Chai, Z.~Xu, and Y.~Jia, ``Conditional density forecast of electricity price
  based on ensemble elm and logistic emos,'' \emph{IEEE Transactions on Smart
  Grid}, vol.~10, no.~3, pp. 3031--3043, 2018.

\bibitem{roughgarden2004bounding}
T.~Roughgarden and {\'E}.~Tardos, ``Bounding the inefficiency of equilibria in
  nonatomic congestion games,'' \emph{Games and Economic Behavior}, vol.~47,
  no.~2, pp. 389--403, 2004.

\bibitem{gradwohl2008fault}
R.~Gradwohl and O.~Reingold, ``Fault tolerance in large games,'' in
  \emph{Proceedings of the 9th ACM Conference on Electronic Commerce}.\hskip
  1em plus 0.5em minus 0.4em\relax ACM, 2008, pp. 274--283.

\bibitem{skytte1999regulating}
K.~Skytte, ``The regulating power market on the nordic power exchange nord
  pool: an econometric analysis,'' \emph{Energy Economics}, vol.~21, no.~4, pp.
  295--308, 1999.

\bibitem{Neupane}
B.~Neupane, T.~B. Pedersen, and B.~Thiesson, ``Evaluating the value of
  flexibility in energy regulation markets,'' in \emph{Proceedings of the ACM
  e-Energy}, 2015, pp. 131--140.

\bibitem{bang2012existing}
C.~Bang, F.~Fock, and M.~Togeby, ``The existing nordic regulating power market:
  {FlexPower WP1—Report 1},'' \emph{EA-Energy analyses}, 2012.

\bibitem{arteaga2019price}
J.~Arteaga and H.~Zareipour, ``A price-maker/price-taker model for the
  operation of battery storage systems in electricity markets,'' \emph{IEEE
  Transactions on Smart Grid}, vol.~10, no.~6, pp. 6912--6920, 2019.

\bibitem{purkayastha1998simple}
S.~Purkayastha, ``Simple proofs of two results on convolutions of unimodal
  distributions,'' \emph{Statistics \& probability letters}, vol.~39, no.~2,
  pp. 97--100, 1998.

\bibitem{umbrellareport}
{Umbrella Project}, ``{Report on uncertainty modelling},'' 2013,
  \url{http://www.e-umbrella.eu/download/82/} Accessed 20 April 2020.

\bibitem{zheng2010bootstrap}
T.~Zheng and J.~L. Gastwirth, ``On bootstrap tests of symmetry about an unknown
  median,'' \emph{Journal of Data Science}, vol.~8, no.~3, p. 413, 2010.

\bibitem{csorgHo1987testing}
S.~CS{\"O}RG{\H{O}} and C.~Heathcote, ``Testing for symmetry,''
  \emph{Biometrika}, vol.~74, no.~1, pp. 177--184, 1987.

\bibitem{ge2018parameter}
F.~Ge, Y.~Ju, Z.~Qi, and Y.~Lin, ``Parameter estimation of a gaussian mixture
  model for wind power forecast error by {Riemann L-BFGS} optimization,''
  \emph{IEEE Access}, vol.~6, pp. 38\,892--38\,899, 2018.

\bibitem{maekawa2019speculative}
J.~Maekawa and K.~Shimada, ``A speculative trading model for the electricity
  market: Based on {Japan} electric power exchange,'' \emph{Energies}, vol.~12,
  no.~15, p. 2946, 2019.

\bibitem{sachs2012applied}
L.~Sachs, \emph{Applied statistics: a handbook of techniques}.\hskip 1em plus
  0.5em minus 0.4em\relax Springer Science \& Business Media, 2012.

\bibitem{rao2018measure}
M.~M. Rao, \emph{Measure theory and integration}.\hskip 1em plus 0.5em minus
  0.4em\relax CRC Press, 2018.

\end{thebibliography}

\newpage
\section*{\large \textbf{Supplementary Materials}}

\begin{appendices}

\section{Summary of theoretical results}
We would like to briefly summarize the theoretical results here for easy understanding. 

The theoretical results can be understood from the following aspects:

\noindent
\textbf{Different roles in the electricity market}

For utilities, we discuss
\begin{itemize}
    \item how to design the optimal bidding strategies (Lemma~\ref{lemma2}, Theorem~\ref{theorem1}, Theorem~\ref{theorem2}, Corollary~\ref{corollary1},Theorem~\ref{theorem6} (1), Theorem~\ref{theorem7}, Theorem~\ref{theorem8} (1), and Theorem~\ref{theorem9}.);
    \item how would the utility’s cost change in the face of other irrational market participants (Theorem~\ref{theorem5}, Theorem~\ref{theorem6} (4), Theorem~\ref{theorem7}, Theorem~\ref{theorem8} (4), and Theorem~\ref{theorem9}).
\end{itemize}

 For the market operator, we discuss
 \begin{itemize}
 \item[$\circ$] the properties of equilibrium, e.g., the uniqueness and the efficiency of the equilibrium (Theorem~\ref{theorem3}, Theorem~\ref{theorem4}, Theorem~\ref{theorem6} (2)-(3), Theorem~\ref{theorem7}, Theorem~\ref{theorem8} (2)-(3), and Theorem~\ref{theorem9});
 \item[$\circ$] how to design the real-time pricing scheme such that the desired equilibrium can be obtained (Theorem~\ref{theorem6}, Theorem~\ref{theorem8}).
 \end{itemize}
 
 We further provide the layout of the theoretical results/theorems under \textit{different settings} as follows:

\vspace{0.2in}
\noindent
\textbf{Under the setting of independent prediction errors and piece-wise linear symmetric pricing model:}
\begin{itemize}
    \item Theorem~\ref{theorem1} characterizes the cost function of utilities.
    \item Lemma~\ref{lemma2} and Theorem~\ref{theorem2} present the structure of the optimal bidding strategies of utilities.
    \item Theorem~\ref{theorem3}, Theorem~\ref{theorem4}, and Theorem~\ref{theorem5} discuss the existence and uniqueness, efficiency, and robustness of the equilibrium. Utilities' optimal bidding strategies are investigated, and the structure of the utility's cost function w.r.t its bidding strategy is shown in Corollary~\ref{corollary1}.
\end{itemize}
\textbf{We extend our results to general settings:}
\begin{itemize}
    \item Theorem~\ref{theorem6} shows that the desired equilibrium is admitted under a set of general pricing models, which provides the system operator an insight on how to design the real-time pricing scheme;
    \item Theorem~\ref{theorem7} shows the extension to correlated Gaussian prediction errors;
    \item  Theorem~\ref{theorem8} and Theorem~\ref{theorem9} generalize the results to the case of utilities submitting bidding curves in the face of price uncertainty with inelastic/flexible loads.
    \item Corollary~\ref{corollary2} further discusses the impact of load uncertainty on utilities' optimal bidding strategies and the market social cost.
\end{itemize}

\begin{table*}
	\centering
	\fontsize{9}{12}\selectfont
	\caption{Summary of theoretical results under different settings from different market roles perspective.}
	\begin{threeparttable}
	\label{tablesummary}
	\begin{tabular}{c|c|c|c|c|c}
	\toprule
		\hline
		\multirow{2}{*}{\tabincell{c}{Bidding \\ protocol}}& \multirow{2}{*}{\tabincell{c}{Pricing scheme / \\ Load uncertainty model}}&
		\multicolumn{2}{c|}{Utility}&\multicolumn{2}{c}{Market operator}  \cr\cline{3-6}
		&  & Optimal bidding strategy & Robustness & Uniqueness & Efficiency  \cr
		\hline
		\multirow{5}{*}{\tabincell{c}{Submit \\ quantity bid}}
		& \tabincell{c}{Piece-wise linear symmetric / \\ Independent symmetric unimodal }&\tabincell{c}{Lemma~\ref{lemma2}, Theorem~\ref{theorem1} \\Theorem~\ref{theorem2}, Corollary~\ref{corollary1} }&Theorem~\ref{theorem5}&Theorem~\ref{theorem3}&Theorem~\ref{theorem4}\cr\cline{2-6}
		& \tabincell{c}{General pricing model / \\ Independent symmetric unimodal }&\tabincell{c}{Theorem~\ref{theorem6} (1)}&\tabincell{c}{Theorem~\ref{theorem6} (4)}&\tabincell{c}{Theorem~\ref{theorem6} (2)}&\tabincell{c}{Theorem~\ref{theorem6} (3)}\cr\cline{2-6}
		& \tabincell{c}{Piece-wise linear symmetric / \\ Correlated Gaussian error } &Theorem~\ref{theorem7}&Theorem~\ref{theorem7}&Theorem~\ref{theorem7}&Theorem~\ref{theorem7}\cr\hline
		\multirow{3}{*}{\tabincell{c}{Submit \\ bidding curve}}
		&\tabincell{c}{General pricing model / \\ Symmetric Unimodal}    &\tabincell{c}{Theorem~\ref{theorem8} (1)}&\tabincell{c}{Theorem~\ref{theorem8} (4)}&\tabincell{c}{Theorem~\ref{theorem8} (2)}&\tabincell{c}{Theorem~\ref{theorem8} (3)}\cr\cline{2-6}
		& \tabincell{c}{Piece-wise linear symmetric / \\ Correlated Gaussian error } &\tabincell{c}{Theorem~\ref{theorem9}}&\tabincell{c}{Theorem~\ref{theorem9}}&\tabincell{c}{Theorem~\ref{theorem9}}&\tabincell{c}{Theorem~\ref{theorem9}}\cr\hline
		\bottomrule
	\end{tabular}
	\begin{tablenotes}
			\footnotesize
            \item[*] {Piece-wise linear symmetric and General pricing model stand for the piece-wise linear symmetric pricing model define in (\ref{pricing}) and the general pricing model defined in (\ref{equation20}) respectively. Note that the piece-wise linear symmetric pricing model can be a special case of the general pricing model.}
            \item[*] {Independent symmetric unimodal and Correlated Gaussian error stand for the independent symmetric unimodal load prediction error defined in Definition~\ref{proper} and non-negatively correlated Gaussian uncertainty defined in Theorem~\ref{theorem7} respectively. }
		\end{tablenotes}
	\end{threeparttable}
\end{table*}

We summarize the relationships of the theoretical results under different settings from different market roles in Table~\ref{tablesummary}. We remark that Corollary~\ref{corollary2} holds for all the five possible combinations of pricing schemes and load uncertainty models list in Table~\ref{tablesummary}.

\section{Proof of theoretical results}\label{s-appendic}

\subsection{Proof of Lemma~\ref{lemma1}}\label{prooflemma1}
\proof{}
Without loss of generality, we may first consider two random variables, $\Delta_1$ and $\Delta_2$, both have zero mean. For $\Delta\triangleq \Delta_1+\Delta_2$, we have:
$$f_{\Delta}(z)=\int_{-\infty}^{+\infty} f_{\Delta_1}(x) f_{\Delta_2}(z-x)dx.$$
Let us first study the symmetry of $f_{\Delta}(\cdot)$, we have
\begin{equation}
\begin{split}
f_{\Delta}(-z)&=\int_{-\infty}^{+\infty} f_{\Delta_1}(x) f_{\Delta_2}(-z-x)dx\\
	&=	\int_{-\infty}^{+\infty} f_{\Delta_1}(-x) f_{\Delta_2}(-z+x)dx\\
	&=	\int_{-\infty}^{+\infty}  f_{\Delta_1}(x)f_{\Delta_2}(z-x)\\
   &=f_{\Delta}(z).
\end{split}
\end{equation}
From the above equality, we know that $f_{\Delta}(z)$ is an even function and hence is symmetric around zero. Therefore, we only need to prove the following central dominant condition:
$$f_{\Delta}(z_1)\leq f_{\Delta}(z_2) \quad \quad \mbox{if} \quad \quad z_1> z_2 \geq0,$$
where
$$f_{\Delta}(z_1)=\int_{-\infty}^{+\infty} f_{\Delta_1}(x) f_{\Delta_2}(z_1-x)dx,$$
and
$$f_{\Delta}(z_2)=\int_{-\infty}^{+\infty} f_{\Delta_1}(x) f_{\Delta_2}(z_2-x)dx.$$
Let us compute
$$f_{\Delta}(z_1)-f_{\Delta}(z_2)=\int_{-\infty}^{+\infty} f_{\Delta_1}(x) [f_{\Delta_2}(z_1-x)-f_{\Delta_2}(z_2-x)]dx.$$
By the introducing the following $h(\cdot)$ function $$h(x)=f_{\Delta_2}(z_1-x)-f_{\Delta_2}(z_2-x),$$
we have
\begin{equation}
\begin{split}
h(z_1+z_2-x)&=	f_{\Delta_2}(-z_2+x)-f_{\Delta_2}(-z_1+x)\\
&=	f_{\Delta_2}(z_2-x)-f_{\Delta_2}(z_1-x)\\
 &=-h(x).
\end{split}
\end{equation}
Let us consider two points $\frac{z_1+z_2}{2}+x$ and  $\frac{z_1+z_2}{2}-x$ with $x\geq0$. From the symmetry of $h(\cdot)$, we have:
$$h(\frac{z_1+z_2}{2}+x)=-h(\frac{z_1+z_2}{2}-x)\geq 0,$$
and
$$f_{\Delta_1}(\frac{z_1+z_2}{2}+x)\leq f_{\Delta_1}(\frac{z_1+z_2}{2}-x).$$
Hence we conclude that $f_{\Delta}(z_1)-f_{\Delta}(z_2)\leq 0$. Next we will prove the strictly central dominant part. 
Let us compute
$$f_{\Delta}(z_1)-f_{\Delta}(0)=\int_{-\infty}^{+\infty} f_{\Delta_1}(x) [f_{\Delta_2}(z_1-x)-f_{\Delta_2}(-x)]dx.$$
Without loss of generality, we assume $\Delta_2$ satisfies the strictly central dominant condition.
Then $f_{\Delta}(z_1)-f_{\Delta}(0)<0$ since there exist some $x>0$ such that $f_{\Delta_1}(\frac{z_1}{2}+x)<f_{\Delta_1}(\frac{z_1}{2}-x)$ together with the continuity property. Similarly, let us compute
$$f_{\Delta}(z_1)-f_{\Delta}(z_2)=\int_{-\infty}^{+\infty} f_{\Delta_1}(x) [f_{\Delta_2}(z_1-x)-f_{\Delta_2}(z_2-x)]dx.$$
Then $f_{\Delta}(z_1)-f_{\Delta}(z_2)<0$ since there exist some $x>0$ such that $f_{\Delta_1}(\frac{z_1+z_2}{2}+x)<f_{\Delta_1}(\frac{z_1+z_2}{2}-x)$.
By induction, we show that $\Delta\triangleq \sum^{N}_{i=1}\Delta_i$ follows a symmetric unimodal distribution with mean zero if all $\Delta_i$ are symmetric unimodal random variables with mean zero. Following the similar approach, we can prove that $\Delta\triangleq \sum^{N}_{i=1}\Delta_i$ follows a symmetric unimodal distribution with mean with mean $\mu= \sum^{N}_{i=1}\mu_i$.

This completes the proof of Lemma~\ref{lemma1}. 
\endproof

\subsection{Proof of Theorem~\ref{theorem1}}
\proof{}
According to pricing model, given the day-ahead market price $p_{d}$, the spot price $p_{s}$ is a linear step function of $\Delta$. Namely, 
\begin{equation}
\begin{split}
p_{s} & =1_{\{\Delta>0\}}\cdot p_{d}(a_{1}\Delta+b_{1})+1_{\{\Delta<0\}}\cdot p_{d}(a_{2}\Delta+b_{2})+p_d\cdot1_{\{\Delta=0\}}\\
 & =\xi_{1}\Delta+\xi_{2},
\end{split}
\end{equation}
where $1_{\{\cdot\}}$ is indicator function, and $$\xi_{1}\triangleq a_{1}p_{d}1_{\{\Delta>0\}}+a_{2}p_{d}1_{\{\Delta<0\}}+0\cdot1_{\{\Delta=0\}},$$and $$\xi_{2}\triangleq b_{1}p_{d}1_{\{\Delta>0\}}+b_{2}p_{d}1_{\{\Delta<0\}}+p_{d}1_{\{\Delta=0\}}.$$Then we can compute the expectation of $\mathsf{ABC}_{i}$ in the following way: 
\begin{equation}
\begin{split}
\mathbb{E}[\mathsf{ABC}_{i}] & =\mathbb{E}\left[p_{d}+\frac{\Delta_{i}(p_{s}-p_{d})}{D_{i}}\right]=p_{d}+\frac{\mathbb{E}[p_{s}\cdot\Delta_{i}]}{D_{i}}-\frac{p_{d}}{D_{i}}\mu_{i}.
\end{split}
\end{equation}
It remains to compute $\mathbb{E}[p_{s}\cdot\Delta_{i}]$:
\begin{equation}
\begin{split}
\mathbb{E}[p_{s}\cdot\Delta_{i}]	&=	\mathbb{E}[(\xi_{1}\Delta+\xi_{2})\Delta_{i}]\\
	&=	\mathbb{E}[(\xi_{1}(\Delta_{i}+\Delta_{-i})+\xi_{2})\Delta_{i}]\\
	&=	\mathbb{E}[\xi_{1}\Delta_{i}^{2}+\xi_{1}\Delta_{i}\Delta_{-i}+\xi_{2}\Delta_{i}].
\end{split}
\end{equation}
Therefore, $\mathbb{E}[p_{s}\cdot\Delta_{i}]$ can be divided into three terms. In the following, we will compute these three terms one by one.

 It is easy to verify that when the price model is piece-wise linear symmetric, we have:
$$\mathbb{E}[\xi_{1}\Delta_{i}^{2}]=\frac{(a_{1}+a_{2})p_{d}}{2}(\sigma_{i}^{2}+\mu_{i}^{2}),$$
and
$$\mathbb{E}[\xi_{1}\Delta_{i}\Delta_{-i}]=\frac{a_{1}+a_{2}}{2}p_{d}\mu_{i}\mu_{-i}.$$

It remains to compute $\mathbb{E}[\xi_{2}\Delta_{i}]$:
\begin{equation}
\begin{split}
\mathbb{E}[\xi_{2}\Delta_{i}]&=\int_{-\infty}^{+\infty}\delta_{i}\left(\int_{-\delta_{i}}^{+\infty}b_{1}p_{d}f_{\Delta_{-i}}(\delta_{-i})d\delta_{-i}+\int_{-\infty}^{-\delta_{i}}b_{2}p_{d}f_{\Delta_{-i}}(\delta_{-i})d\delta_{-i}\right)f_{\Delta_{i}}(\delta_{i})d\delta_{i}\\
&=\mathbb{E}\left[\Delta_{i}\left(\int_{-\Delta_{i}}^{+\infty}b_{1}p_{d}f_{\Delta_{-i}}(\delta_{-i})d\delta_{-i}+\int_{-\infty}^{-\Delta_{i}}b_{2}p_{d}f_{\Delta_{-i}}(\delta_{-i})d\delta_{-i}\right)\right],
\end{split}
\end{equation}
where in the last equality, the expectation is taken with respect to $\Delta_{i}$. Further, we have \begin{equation}
\begin{split}
\mathbb{E}[\xi_{2}\Delta_{i}]  &=\mathbb{E}\left[b_{1}p_{d}\Delta_{i}\left(1-\int_{-\infty}^{-\Delta_{i}}f_{\Delta_{-i}}(\delta_{-i})d\delta_{-i}\right)+b_{2}p_{d}\Delta_{i}\int_{-\infty}^{-\Delta_{i}}f_{\Delta_{-i}}(\delta_{-i})d\delta_{-i}\right]\\
 & =b_{1}p_{d}\mu_{i}+\mathbb{E}\left[\Delta_{i}(b_{2}-b_{1})p_{d}\int_{-\infty}^{-\Delta_{i}}f_{\Delta_{-i}}(\delta_{-i})d\delta_{-i})\right]\\
 &=b_{1}p_{d}\mu_{i}+\mathbb{E}\left[\Delta_{i}(b_{2}-b_{1})p_{d}(\int_{-\infty}^{\mu_{-i}}f_{\Delta_{-i}}(\delta_{-i})d\delta_{-i}+\int_{\mu_{-i}}^{-\Delta_{i}}f_{\Delta_{-i}}(\delta_{-i})d\delta_{-i})\right]\\
 &=p_{d}\mu_{i}\frac{b_{1}+b_{2}}{2}+(b_{1}-b_{2})p_{d}\mathbb{E}\left[\Delta_{i}\tilde{F}(\Delta_i)\right],
\end{split}
\end{equation}
where 
$$\tilde{F}(\Delta_i)\triangleq -\int_{\mu_{-i}}^{-\Delta_{i}}f_{\Delta_{-i}}(\delta_{-i})d\delta_{-i}=\int_{-\Delta_i}^{\mu_{-i}}f_{\Delta_{-i}}(\delta_{-i})d\delta_{-i}.$$
To sum it up, when the pricing model is symmetric, i.e., $a_{1}=a_{2}$, $b_{1}+b_{2}=2$, and the prediction errors are independent, the expectation of $\mathbf{\textsf{ABC}}_{i}$ is given as:
\begin{equation}
\begin{split}
\mathbb{E}[\mathbf{\textsf{ABC}}_{i}]=p_{d}+\frac{p_{d}}{D_{i}}\left[\frac{a_{1}+a_{2}}{2}(\sigma_{i}^{2}+\mu_{i}^{2}+\mu_i\mu_{-i})+{(b_{1}-b_{2})}\mathbb{E}\left[\Delta_{i}\tilde{F}(\Delta_i)\right]\right],
\end{split}
\end{equation}
where $\tilde{F}$ is defined as above and $\mathbb{E}[\mathbf{\textsf{ABC}}_{i}]$ takes the minimum value when $\mu_i=0$.

This completes the proof of Theorem~\ref{theorem1}. 
\endproof

\subsection{Proof of Lemma~\ref{lemma2}}
\proof{}
 For ease of presentation, we define
$$U(\mu_{i})=\mathbb{E}\left[\Delta_{i}\tilde{F}(\Delta_i)\right].$$
Notice that the above $\tilde{F}(\cdot)$ is an odd function when $\mu_{-i}=0$. Therefore
\begin{equation}
\begin{split}
U(-\mu_{i})&=\int_{-\infty}^{+\infty}\delta_i f^{-\mu_i}_{\Delta_i}(\delta_i)\cdot\tilde{F}(\delta_i)\\
&=\int_{-\infty}^{+\infty}(-\delta_i) f^{-\mu_i}_{\Delta_i}(-\delta_i)\cdot\tilde{F}(-\delta_i)\\
&=\int_{-\infty}^{+\infty}\delta_i f^{\mu_i}_{\Delta_i}(\delta_i)\cdot\tilde{F}(-\delta_i)\\
&=U(\mu_{i}).
\end{split}
\end{equation}
From the above equality, we know that given $\mu_{-i}=0$, $U(\cdot)$ is an even function. It is sufficient to consider the case when $\mu_{i}>0$. Let us compute
\begin{equation}
\begin{split}
U(\mu_{i})-U(0)=\int_{-\infty}^{+\infty}\delta_{i}\tilde{F}(\delta_i)\cdot(f^{\mu_i}_{\Delta_i}(\delta_i)-f^{0}_{\Delta_i}(\delta_i)).
\end{split}
\end{equation}
Let
$$g_1(x)=x\tilde{F}(x)+(x-\mu_i)\tilde{F}(x-\mu_i),$$
$$g_2(x)=-x\tilde{F}(x)+(x-\mu_i)\tilde{F}(x-\mu_i),$$
and
$$g_3(x)=f^{\mu_i}_{\Delta_{i}}(x)-f^{0}_{\Delta_{i}}(x).$$
Then we have
$$g_1(\mu_i-x)=(\mu_i-x)\tilde{F}(\mu_i-x)+(-x)\tilde{F}(-x)=g_1(x),$$
$$g_2(\mu_i-x)=-(\mu_i-x)\tilde{F}(\mu_i-x)+(-x)\tilde{F}(-x)=-g_2(x),$$
and
$$g_3(x)=f^{\mu_i}_{\Delta_{i}}(\mu_i-x)-f^{0}_{\Delta_{i}}(\mu_i-x)=f^{0}_{\Delta_{i}}(x)-f^{\mu_i}_{\Delta_{i}}(x)=-g_3(x).$$
Therefore, we have
\begin{equation}
\begin{split}
U(\mu_i&)-U(0)=\frac{1}{2}\int_{-\infty}^{+\infty}[g_1(\delta_i)-g_2(\delta)]\cdot g_3(\delta_i)d\delta_i\\
&=\frac{1}{2}\int_{-\infty}^{+\infty}g_1(\delta_i)\cdot g_3(\delta_i)d\delta_i-\frac{1}{2}\int_{-\infty}^{+\infty}g_2(\delta_i)\cdot g_3(\delta_i)d\delta_i\\
&=-\int_{\frac{\mu_i}{2}}^{+\infty}g_2(\delta_i)\cdot g_3(\delta_i)d\delta_i.
\end{split}
\end{equation}
Since $f_{\Delta_i}(\cdot)$ is symmetric and central dominant, it is easy to see that $g_3(\delta_i)\geq 0$ when $\delta_i \geq \frac{\mu_i}{2}$. Next we will show that when $\delta_i\geq\frac{\mu_i}{2}$, $g_{2}(\delta_i)\leq0$. To see this, consider the following two cases: (i) when $\delta_i\geq \mu_i$, we have $0\leq \delta_i-\mu_i< \delta_i$, then $g_{2}(\delta_i)<0$; (ii) when $\frac{\mu_i}{2}< \delta_i \leq \mu_i$, we have $0\leq\mu_i-\delta_i< \delta_i$, then $g_{2}(\delta_i)<0$.
Therefore, when $\delta_i>\frac{\mu_i}{2}$, we always have $g_{3}(\delta_i)\geq0$ and $g_{2}(\delta_i)<0$.
Finally, we get 
\begin{equation}
\begin{split}
U(\mu_i)-U(0)&=\frac{1}{2}\int_{-\infty}^{+\infty}[g_1(\delta_i)-g_2(\delta_i)]\cdot g_3(\delta_i)d\delta_i\\
&=-\int_{\frac{\mu_i}{2}}^{+\infty}g_2(\delta_i)\cdot g_3(\delta_i)d\delta_i\\
&> 0.
\end{split}
\end{equation}
The last strict inequality holds since $g_3(\delta_i)>0$ for some $\delta_i$. 

Next we prove the strictly increasing property of $U(\cdot)$. Consider $\mu_1>\mu_2\geq0$, we have
\begin{equation}
\begin{split}
U(\mu_1)-U(\mu_2)&=\int_{-\infty}^{+\infty}\delta_i (f^{\mu_1}_{\Delta_i}(\delta_i)-f^{\mu_2}_{\Delta_i}(\delta_i))\cdot\tilde{F}(\delta_i).
\end{split}
\end{equation}
Consider $\frac{\mu_1+\mu_2}{2}+\delta_i$ and $\frac{\mu_1+\mu_2}{2}-\delta_i$, where $\delta_i>0$. Denote $f^{\mu_1}_{\Delta_{i}}(\frac{\mu_1+\mu_2}{2}+\delta_i)-f^{\mu_2}_{\Delta_{i}}(\frac{\mu_1+\mu_2}{2}+\delta_i)$ as $\Delta f_{\Delta_i}(\delta_i)$. We have
\begin{equation}
\begin{split}
&(\frac{\mu_1+\mu_2}{2}+\delta_i)\Delta f_{\Delta_i}(\delta_i)\cdot\tilde{F}(\frac{\mu_1+\mu_2}{2}+\delta_i)-(\frac{\mu_1+\mu_2}{2}-\delta_i)\Delta f_{\Delta_i}(\delta_i)\cdot\tilde{F}(\frac{\mu_1+\mu_2}{2}-\delta_i)\\
=&\Delta f_{\Delta_i}(\delta_i)\cdot\left((\frac{\mu_1+\mu_2}{2}+\delta_i)\tilde{F}(\frac{\mu_1+\mu_2}{2}+\delta_i)-(\frac{\mu_1+\mu_2}{2}-\delta_i)\tilde{F}(\frac{\mu_1+\mu_2}{2}-\delta_i)\right)\\\geq&0.
\end{split}
\end{equation}
It is easy to verify that the second term is strictly positive and the first term is non-negative. Since the first term is strictly positive for some $\delta_i$, after taking integral, we will always have $U(\mu_1)>U(\mu_2)$.

This completes the proof of Lemma~\ref{lemma2}. 

\endproof

\subsection{Proof of Theorem~\ref{theorem2}}
\proof{}
Recall that under piece-wise linear symmetric pricing model, we have
\begin{equation}
\begin{split}
\mathbb{E}&[\mathbf{\textsf{ABC}}_{i}]	=p_{d}+\frac{p_{d}}{D_{i}}\left[\frac{a_{1}+a_{2}}{2}(\sigma_{i}^{2}+\mu_{i}^{2}+\mu_i\mu_{-i})+{(b_{1}-b_{2})}\mathbb{E}\left[\Delta_{i}\tilde{F}(\Delta_i)\right]\right]\\&=p_{d}+\frac{p_{d}}{D_{i}}[\frac{a_{1}+a_{2}}{2}(\mu_{i}\mu_{-i}+\sigma_{i}^{2}+\mu_{i}^{2})+(b_1-b_2)\int_{-\infty}^{+\infty}\delta_{i}f^{\mu_i}_{\Delta_i}(\delta_i)\int_{-\delta_i}^{\mu_{-i}}f^{0}_{\Delta_{-i}}(\delta_{-i})d\delta_{-i}d\delta_{i}]\\&=
p_{d}+\frac{p_{d}}{D_{i}}[\frac{a_{1}+a_{2}}{2}(\mu_{i}\mu_{-i}+\sigma_{i}^{2}+\mu_{i}^{2})+(b_1-b_2)\int_{-\infty}^{+\infty}\delta_{i}f^{\mu_i}_{\Delta_i}(\delta_i)\int_{0}^{\delta_{i}+\mu_{-i}}f^{0}_{\Delta_{-i}}(\delta_{-i})d\delta_{-i}d\delta_{i}],
\end{split}
\end{equation}
where $f^{k}(\cdot)$ represents the probability density function centered at $k$.

Notice that the first condition of Theorem~\ref{theorem2} is a direct result of Lemma~\ref{lemma2} and Theorem~\ref{theorem1}. We only need to prove the second and the third conditions. Let us define
$$U(\mu_i)=\int_{-\infty}^{+\infty}\delta_{i}f^{\mu_i}_{\Delta_i}(\delta_i)\int_{0}^{\delta_{i}+\mu_{-i}}f^{0}_{\Delta_{-i}}(\delta_{-i})d\delta_{-i}d\delta_{i}.$$
Let us first prove the second condition when $\mu_{-i}>0$. In order the get the best response of utility, we prove that under the following condition, the strategy $\mu_i$ of utility $i$ can not be optimal.

i) When $\mu_{i}>0$, consider
\begin{equation}
\begin{split}
U(\mu_i)-U(0)=&\int_{-\infty}^{+\infty}\delta_{i}\left[f^{\mu_i}_{\Delta_i}(\delta_i)-f^{0}_{\Delta_i}(\delta_i)\right]\cdot\int_{0}^{\delta_{i}+\mu_{-i}}f^{0}_{\Delta_{-i}}(\delta_{-i})d\delta_{-i}d\delta_{i}.
\end{split}
\end{equation}
By an abuse of notation, let us use $\Delta f_{\Delta_i}(\delta_i)$ to denote $f^{\mu_i}_{\Delta_i}(\delta_i)-f^{0}_{\Delta_i}(\delta_i)$. Consider two symmetric points $\frac{\mu_i}{2}+\delta_i$ and $\frac{\mu_i}{2}-\delta_i$, where $\delta_i\geq0$, then we have
\begin{equation}
\begin{split}
&(\frac{\mu_i}{2\hspace{-0.1em}}+\delta_i)\Delta f_{\Delta_i}(\frac{\mu_i}{2}\hspace{-0.1em}+\delta_i)\hspace{-0.3em}\int_{0}^{\frac{\mu_i}{2}+\delta_i+\mu_{-i}}\hspace{-0.5em}f^{0}_{\Delta_{-i}}(\delta_{-i})d\delta_{-i}\hspace{-0.1em}+\hspace{-0.1em}(\frac{\mu_i}{2}\hspace{-0.1em}-\delta_i)\Delta f_{\Delta_i}(\frac{\mu_i}{2}\hspace{-0.1em}-\delta_i)\hspace{-0.3em}\int_{0}^{\frac{\mu_i}{2}-\delta_i+\mu_{-i}}\hspace{-0.5em}f^{0}_{\Delta_{-i}}(\delta_{-i})d\delta_{-i}\\
&=\frac{\mu_i}{2}\Delta f_{\Delta_i}(\frac{\mu_i}{2}+\delta_i)\int_{\frac{\mu_i}{2}-\delta_i+\mu_{-i}}^{\frac{\mu_i}{2}+\delta_i+\mu_{-i}}f^{0}_{\Delta_{-i}}(\delta_{-i})d\delta_{-i}+\delta_i\Delta f_{\Delta_i}(\frac{\mu_i}{2}+\delta_i)\int_{-\frac{\mu_i}{2}+\delta_i-\mu_{-i}}^{\frac{\mu_i}{2}+\delta_i+\mu_{-i}}f^{0}_{\Delta_{-i}}(\delta_{-i})d\delta_{-i}.
\end{split}
\end{equation}
The above equality holds since $\Delta f_{\Delta_i}(\frac{\mu_i}{2}+\delta_i)=-\Delta f_{\Delta_i}(\frac{\mu_i}{2}-\delta_i)$ and it is easy to verify that the upper formula is no less than zero. Hence we have:
$U(\mu_i)-U(0)\geq0$ when $\mu_{i}>0$. This implies that when $\mu_{-i}>0$, the optimal strategy $\mu_i$ for utility $i$ can not be greater than zero.

ii) When $\mu_i<-\mu_{-i}$, consider
\begin{equation}
\begin{split}
U(\mu_i)-U(-\mu_{-i})=&\int_{-\infty}^{+\infty}\delta_{i}\left[f^{\mu_i}_{\Delta_i}(\delta_i)-f^{-\mu_{-i}}_{\Delta_i}(\delta_i)\right]\cdot\int_{0}^{\delta_{i}+\mu_{-i}}f^{0}_{\Delta_{-i}}(\delta_{-i})d\delta_{-i}d\delta_{i}.
\end{split}
\end{equation}
By an abuse the notation, we use $\Delta f_{\Delta_i}(\delta_i)$ to denote $f^{\mu_i}_{\Delta_i}(\delta_i)-f^{-\mu_{-i}}_{\Delta_i}(\delta_i)$.

Consider two symmetric points $\frac{\mu_i-\mu_{-i}}{2}+\delta_i$ and $\frac{\mu_i-\mu_{-i}}{2}-\delta_i$, where $\delta_i\geq0$, we have
\begin{equation}
\begin{split}
&(\frac{\mu_i-\mu_{-i}}{2}+\delta_i)\Delta f_{\Delta_i}(\frac{\mu_i-\mu_{-i}}{2}+\delta_i)\cdot\int_{0}^{\frac{\mu_i+\mu_{-i}}{2}+\delta_i}f^{0}_{\Delta_{-i}}(\delta_{-i})d\delta_{-i}
\\&+(\frac{\mu_i-\mu_{-i}}{2}-\delta_i)\cdot\Delta f_{\Delta_i}(\frac{\mu_i-\mu_{-i}}{2}-\delta_i)\int_{0}^{\frac{\mu_i+\mu_{-i}}{2}-\delta_i}f^{0}_{\Delta_{-i}}(\delta_{-i})d\delta_{-i}\\
=&\frac{\mu_i-\mu_{-i}}{2}\Delta f_{\Delta_i}(\frac{\mu_i-\mu_{-i}}{2}+\delta_i)\int_{\frac{\mu_i+\mu_{-i}}{2}-\delta_i}^{\frac{\mu_i+\mu_{-i}}{2}+\delta_i}f^{0}_{\Delta_{-i}}(\delta_{-i})d\delta_{-i}\\
&+\delta_i\Delta f_{\Delta_i}(\frac{\mu_i-\mu_{-i}}{2}+\delta_i)\int_{-\frac{\mu_i+\mu_{-i}}{2}+\delta_i}^{\frac{\mu_i+\mu_{-i}}{2}+\delta_i}f^{0}_{\Delta_{-i}}(\delta_{-i})d\delta_{-i}.
\end{split}
\end{equation}
The above equality holds since $\Delta f(\frac{\mu_i-\mu_{-i}}{2}+\delta_i)=-\Delta f(\frac{\mu_i-\mu_{-i}}{2}-\delta_i)$ and it is easy to verify that the upper formula is no less than zero. Hence we have:
$U(\mu_i)-U(\mu_{-i})\geq0$ when $\mu_i<-\mu_{-i}$. This implies that when $\mu_{-i}>0$, the optimal strategy $\mu_i$ for utility $i$ can not be less than $-\mu_{-i}$.

In order to show that the optimal $\mu_i$ is within ($-\mu_{-i}$, 0), we first prove that the cost function is left-continuous at $\mu_i=0$ and right continuous at $\mu_i=-\mu_{-i}$. Based on the above observation, we show that bidding a little bit smaller than $0$ or a little bit larger than $-\mu_{-i}$ is always better than bidding at $0$ and $-\mu_{-i}$, respectively. We will first prove the continuity part.

i) Left continuous at $\mu_i=0$. Here, what we need to prove is:
$${\lim_{\mu_i \to 0^-}}U(\mu_i)=U(0).$$
Consider the difference between $U(\mu_i)$ and $U(0)$, where $\mu_i<0$, we have
\begin{equation}\label{s-equation24}
\begin{split}
U(\mu_i)-U(0)=&\int_{0}^{+\infty}\frac{\mu_i}{2}\Delta f_{\Delta_i}(\frac{\mu_i}{2}+\delta_i)\int_{\frac{\mu_i}{2}-\delta_i+\mu_{-i}}^{\frac{\mu_i}{2}+\delta_i+\mu_{-i}}f^{0}_{\Delta_{-i}}(\delta_{-i})d\delta_{-i}d\delta_{i}\\
&+\int_{0}^{+\infty}\delta_i\Delta f_{\Delta_i}(\frac{\mu_i}{2}+\delta_i)\int_{-\frac{\mu_i}{2}+\delta_i-\mu_{-i}}^{\frac{\mu_i}{2}+\delta_i+\mu_{-i}}f^{0}_{\Delta_{-i}}(\delta_{-i})d\delta_{-i}d\delta_{i}.
\end{split}
\end{equation}
Let $K_{max}^{\Delta_{-i}}\triangleq f_{\Delta_{-i}}(0)$ and $K_{max}^{\Delta_{i}}\triangleq f_{\Delta_{i}}(0)$. By an abuse the notation, we use $\Delta f_{\Delta_i}(\delta_i)$ to denote $f^{\mu_i}_{\Delta_i}(\delta_i)-f^{0}_{\Delta_i}(\delta_i)$. It is easy to see that
$$\int_{0}^{+\infty}\frac{\mu_i}{2}\Delta f_{\Delta_i}(\frac{\mu_i}{2}+\delta_i)\int_{\frac{\mu_i}{2}-\delta_i+\mu_{-i}}^{\frac{\mu_i}{2}+\delta_i+\mu_{-i}}f^{0}_{\Delta_{-i}}(\delta_{-i})d\delta_{-i}d\delta_{i}\geq0,$$
and
\begin{equation}
\begin{split}
&\int_{0}^{+\infty}\frac{\mu_i}{2}\Delta f_{\Delta_i}(\frac{\mu_i}{2}+\delta_i)\int_{\frac{\mu_i}{2}-\delta_i+\mu_{-i}}^{\frac{\mu_i}{2}+\delta_i+\mu_{-i}}f^{0}_{\Delta_{-i}}(\delta_{-i})d\delta_{-i}d\delta_{i}\\
\leq&\int_{0}^{+\infty}\frac{\mu_i}{2}\Delta f_{\Delta_i}(\frac{\mu_i}{2}+\delta_i)d\delta_{i}=\frac{\mu_i}{2}\int_{-\frac{\mu_i}{2}}^{\frac{\mu_i}{2}}f^0_{\Delta_i}(\delta_i)d\delta_id\delta_{i}\\
\leq& K_{max}^{\Delta_{i}}\cdot\frac{(\mu_i)^2}{2}.
\end{split}
\end{equation}
For the second term of Eq.~(\ref{s-equation24}), when $\mu_i \to 0^-$, we have $\frac{\mu_i}{2}+\mu_{-i}>-\frac{\mu_i}{2}-\mu_{-i}$, then the following relationships hold.
$$\int_{0}^{+\infty}\delta_i\Delta f_{\Delta_i}(\frac{\mu_i}{2}+\delta_i)\int_{-\frac{\mu_i}{2}+\delta_i-\mu_{-i}}^{\frac{\mu_i}{2}+\delta_i+\mu_{-i}}f^{0}_{\Delta_{-i}}(\delta_{-i})d\delta_{-i}d\delta_{i}\leq0,$$
and
\begin{equation}
\begin{split}
&\int_{0}^{+\infty}\delta_i\Delta f_{\Delta_i}(\frac{\mu_i}{2}+\delta_i)\int_{-\frac{\mu_i}{2}+\delta_i-\mu_{-i}}^{\frac{\mu_i}{2}+\delta_i+\mu_{-i}}f^{0}_{\Delta_{-i}}(\delta_{-i})d\delta_{-i}d\delta_{i}\\
\geq& \int_{0}^{+\infty}\delta_i\Delta f_{\Delta_i}(\frac{\mu_i}{2}+\delta_i)d\delta_{i}=\frac{\mu_i}{2}+\int_{-\frac{\mu_i}{2}}^{\frac{\mu_i}{2}}\delta_if^0_{\Delta_i}(\delta_i)d\delta_i\\
=&\frac{\mu_i}{2}.
\end{split}
\end{equation}
Combining the above inequalities, we have:
$$\frac{\mu_i}{2}\leq U(\mu_i)-U(0)\leq K_{max}^{\Delta_{i}}\cdot\frac{(\mu_i)^2}{2}.$$
Therefore, we conclude 
$${\lim_{\mu_i \to 0^-}}U(\mu_i)=U(0).$$ 
This proves that the {cost function} is left continuous at $\mu_i=0$.

ii) Right continuous at $\mu_i=-\mu_{-i}$. Here, what we need to prove is:
$${\lim_{\mu_i \to -\mu_{-i}^+}}U(\mu_i)=U(-\mu_{-i}).$$
Consider the difference between $U(\mu_i)$ and $U(-\mu_{-i})$, where $\mu_i>-\mu_{-i}$, we have
\begin{equation}\label{s-equation27}
\begin{split}
U(\mu_i)-U(-\mu_{-i})=&\int_{0}^{+\infty}\frac{\mu_i-\mu_{-i}}{2}\Delta f_{\Delta_i}\int_{-\delta_i+\frac{\mu_i+\mu_{-i}}{2}}^{\delta_i+\frac{\mu_i+\mu_{-i}}{2}}f^{0}_{\Delta_{-i}}(\delta_{-i})d\delta_{-i}d\delta_{i}\\
&+\int_{0}^{+\infty}\delta_i\Delta f_{\Delta_i} \int_{\delta_i-\frac{\mu_i+\mu_{-i}}{2}}^{\delta_i+\frac{\mu_i+\mu_{-i}}{2}}f^{0}_{\Delta_{-i}}(\delta_{-i})d\delta_{-i}d\delta_{i}.
\end{split}
\end{equation}
Let $K_{max}^{\Delta_{-i}}\triangleq f_{\Delta_{-i}}(0)$, $K_{max}^{\Delta_{i}}\triangleq f_{\Delta_{i}}(0)$. By an abuse the notation, we use $\Delta f_{\Delta_i}(\delta_i)$ to denote $f^{\mu_i}_{\Delta_{i}}(\frac{\mu_i-\mu_{-i}}{2}+\delta_i)-f^{-\mu_{-i}}_{\Delta_{i}}(\frac{\mu_i-\mu_{-i}}{2}+\delta_i)$. It is easy to see that
$$\int_{0}^{+\infty}\frac{\mu_i-\mu_{-i}}{2}\Delta f_{\Delta_i}(\delta_i)\int_{-\delta_i+\frac{\mu_i+\mu_{-i}}{2}}^{\delta_i+\frac{\mu_i+\mu_{-i}}{2}}f^{0}_{\Delta_{-i}}(\delta_{-i})d\delta_{-i}d\delta_{i}\leq0,$$
and
\begin{equation}
\begin{split}
&\int_{0}^{+\infty}\frac{\mu_i-\mu_{-i}}{2}\Delta f_{\Delta_i}(\delta_i)\int_{-\delta_i+\frac{\mu_i+\mu_{-i}}{2}}^{\delta_i+\frac{\mu_i+\mu_{-i}}{2}}f^{0}_{\Delta_{-i}}(\delta_{-i})d\delta_{-i}d\delta_{i}\\
\geq& \frac{\mu_i-\mu_{-i}}{2}\int_{0}^{+\infty}\Delta f_{\Delta_i}(\delta_i) d\delta_{i}=\frac{\mu_i-\mu_{-i}}{2}\int_{-\frac{\mu_i+\mu_{-i}}{2}}^{\frac{\mu_i+\mu_{-i}}{2}}f^{0}_{\Delta_{i}}(\delta_{i})d\delta_{i}\\
\geq& K_{max}^{\Delta_{i}}(\mu_i+\mu_{-i})\cdot\frac{\mu_i-\mu_{-i}}{2}.
\end{split}
\end{equation}
As to the second term of Eq.~(\ref{s-equation27}), when $\mu_i \to -\mu_{-i}^+$, we have $\frac{\mu_i+\mu_{-i}}{2}>-\frac{\mu_i+\mu_{-i}}{2}$, then the following relationships hold.
$$\int_{0}^{+\infty}\delta_i\Delta f_{\Delta_i}(\delta_i) \int_{\delta_i-\frac{\mu_i+\mu_{-i}}{2}}^{\delta_i+\frac{\mu_i+\mu_{-i}}{2}}f^{0}_{\Delta_{-i}}(\delta_{-i})d\delta_{-i}d\delta_{i}\geq0,$$
and
\begin{equation}
\begin{split}
&\int_{0}^{+\infty}\delta_i\Delta f_{\Delta_i}(\delta_i) \int_{\delta_i-\frac{\mu_i+\mu_{-i}}{2}}^{\delta_i+\frac{\mu_i+\mu_{-i}}{2}}f^{0}_{\Delta_{-i}}(\delta_{-i})d\delta_{-i}d\delta_{i}\\
\leq& K_{max}^{\Delta_{-i}}(\mu_i+\mu_{-i})\cdot\int_{0}^{+\infty}\delta_i\Delta f_{\Delta_i} d\delta_i\\
\leq& K_{max}^{\Delta_{-i}}\frac{(\mu_i+\mu_{-i})^2}{2}.
\end{split}
\end{equation}
Combining the above inequalities, we have:
$$K_{max}^{\Delta_{i}}(\mu_i+\mu_{-i})\cdot\frac{\mu_i-\mu_{-i}}{2}\leq U(\mu_i)-U(-\mu_{-i})\leq K_{max}^{\Delta_{-i}}\frac{(\mu_i+\mu_{-i})^2}{2}.$$
Therefore, we conclude 
$${\lim_{\mu_i \to -\mu_{-i}^+}}U(\mu_i)=U(-\mu_{-i}).$$ 
This proves that the {cost function} is right continuous at $\mu_i=-\mu_{-i}$.

Next, we want to show that the minimum value of the above {cost function} for utility $i$ can only be attained within $a_i(\mu_i, \mu_{-i})\cdot \mu_{-i}$, where $a_i(\mu_i, \mu_{-i})\in (-1,0)$. For example, when $\mu_{-i}>0$, the optimal $\mu_i$ must be within ($-\mu_{-i}$, 0). Without causing confusion, we will use $a_i$ to denote $a_i(\mu_i, \mu_{-i})$.

i) When $\mu_i \to 0^-$, we have
$$\mu_{-i}\cdot\mu_i<\mu_{i}\mu_{-i}+\mu_{i}^{2}< (\mu_{-i}+2\mu_i)\cdot\mu_i.$$
Therefore,
$$\left(\frac{a_1+a_2}{2}\mu_{-i}+\frac{b_1-b_2}{2}\right)\cdot \mu_i<\Delta \mathbb{E}[\mathbf{\textsf{ABC}}_{i}]<\left(\frac{a_1+a_2}{2}(\mu_{-i}+2\mu_i)+\frac{b_1-b_2}{2} K_{max}^{\Delta_{i}}\mu_i\right)\cdot \mu_i,$$
where $\Delta \mathbb{E}[\mathbf{\textsf{ABC}}_{i}]\triangleq \mathbb{E}[\mathbf{\textsf{ABC}}_{i}](\mu_i)- \mathbb{E}[\mathbf{\textsf{ABC}}_{i}](0)$ is the cost difference of utility $i$ for choosing $\mu_i$ and $0$, respectively.
Hence we have $\Delta \mathbb{E}[\mathbf{\textsf{ABC}}_{i}]<0$,

It is clear when $\mu_i$ decrease \emph{a little bit} from 0, the cost function value will also decrease. In other words, it is equivalent to state that the left derivative of $\mathbb{E}[\mathbf{\textsf{ABC}}_{i}]$ is positive at $\mu_i=0$.

ii) When $\mu_i \to -\mu_{-i}^+$, we have
$$-\mu_{-i}\cdot(\mu_i+\mu_{-i})<\mu_{i}\mu_{-i}+\mu_{i}^{2}< (\mu_{-i}+2\mu_i)\cdot(\mu_i+\mu_{-i}).$$
Therefore,
\begin{equation}
\begin{split}
&\left(\frac{a_1+a_2}{2}(-\mu_{-i})+\frac{b_1-b_2}{2}K_{max}^{\Delta_{i}}(\mu_i-\mu_{-i})\right)\cdot (\mu_i+\mu_{-i})< \Delta \mathbb{E}[\mathbf{\textsf{ABC}}_{i}],\mu_{-i}+2\mu_i,
\end{split}
\end{equation}
and
\begin{equation}
\begin{split}
&\Delta \mathbb{E}[\mathbf{\textsf{ABC}}_{i}]<\left(\frac{a_1+a_2}{2}\cdot(\mu_{-i}+2\mu_i)+\frac{b_1-b_2}{2}K_{max}^{\Delta_{-i}}\cdot (\mu_i+\mu_{-i})\right)\cdot(\mu_i+\mu_{-i}),
\end{split}
\end{equation}
where $\Delta \mathbb{E}[\mathbf{\textsf{ABC}}_{i}]\triangleq \mathbb{E}[\mathbf{\textsf{ABC}}_{i}](\mu_i)- \mathbb{E}[\mathbf{\textsf{ABC}}_{i}](-\mu_{-i})$ is used to denote the cost difference of utility $i$ for choosing $\mu_i$ and $-\mu_{-i}$, respectively. 

It is clear when $\mu_i$ decrease \emph{a little bit} from 0, the cost function value will also decrease. In other words, it is equivalent to state that the right derivative of $\mathbb{E}[\mathbf{\textsf{ABC}}_{i}]$ is negative at $\mu_i=-\mu_{-i}$.

From the above analysis, we see that given $\mu_{-i}>0$, utility $i's$ optimal strategy $\mu^*_i$ that minimizes $\mathbb{E}[\mathbf{\textsf{ABC}}_{i}]$ must be within $(-\mu_{-i}, 0)$. Similar analysis can be constructed when $\mu_{-i}<0$ and utility $i's$ optimal strategy $\mu^*_i$ that minimizes $\mathbb{E}[\mathbf{\textsf{ABC}}_{i}]$ must be within $(0, -\mu_{-i})$ given $\mu_{-i}<0$.

This completes the proof of Theorem~\ref{theorem2}. 
\endproof
\subsection{Proof of Theorem~\ref{theorem3}}
\proof{}
Previous we have shown that $(u^*_{1}, u^*_{2},..., u^*_{N})$ where $u^*_{i}=0$ for $\forall i\in\{1, 2,...,N\}$ is a pure strategy Nash Equilibrium. In this subsection, we prove the uniqueness of the equilibrium by contradiction. Assume that there exists another strategy profile $(\tilde{u}_{1}, \tilde{u}_{2},...,\tilde{u}_{N})$ that constitutes another Nash Equilibrium different from $\mu^*$, where there exist at least one $\tilde{u}_{i}\not=0$. From the best response perspective of Nash Equilibrium, it is natural to derive the following condition of the Nash Equilibrium strategy profile from Theorem~\ref{theorem2}:
\begin{equation}
\begin{split}
\left[\begin{array}{ccccc} 
     0 & a_1 & \cdots &a_1 & a_1 \\ 
    a_2 & 0 & \cdots & a_2 & a_2\\ 
    \vdots & \vdots & \ddots & \vdots & \vdots \\
a_{N-1} &a_{N-1}&\cdots&0&a_{N-1}\\
a_{N} & a_{N}&\cdots&a_{N}&0
\end{array}\right]
\begin{bmatrix}
  \tilde{u}_{1} \\
 \tilde{u}_{2}  \\
  \vdots   \\
  \tilde{u}_{N-1}\\
\tilde{u}_{N}
 \end{bmatrix}=
\begin{bmatrix}
  \tilde{u}_{1} \\
 \tilde{u}_{2}  \\
  \vdots   \\
  \tilde{u}_{N-1}\\
\tilde{u}_{N}
 \end{bmatrix},
\end{split}
\end{equation}
$\Rightarrow$
\begin{equation}
\begin{split}
\left[\begin{array}{ccccc} 
     1 & -a_1&\cdots&-a_1&-a_1 \\ 
    -a_2 & 1&\cdots&-a_2&-a_2\\ 
    \vdots& \vdots& \ddots& \vdots& \vdots \\
-a_{N-1} &-a_{N-1}&\cdots&1&-a_{N-1}\\
-a_{N} & -a_{N}&\cdots&-a_{N}&1
\end{array}\right]
\begin{bmatrix}
  \tilde{u}_{1} \\
 \tilde{u}_{2}  \\
  \vdots   \\
  \tilde{u}_{N-1}\\
\tilde{u}_{N}
 \end{bmatrix}=
\begin{bmatrix}
  0 \\
 0  \\
  \vdots   \\
  0\\
0
 \end{bmatrix},
\end{split}
\end{equation}
where $a_i=\frac{\tilde{\mu_i}}{\tilde{\mu_{-i}}}\in(-1, 0)$. If we want to show that $(u^*_{1}, u^*_{2},..., u^*_{N})$ where $u^*_{i}=0$ for $\forall i\in\{1,2,...,N\}$ is the unique Nash Equilibrium, it is equivalent to show that there does not exist another Nash Equilibrium strategy profile $(\tilde{u}_{1}, \tilde{u}_{2},...,\tilde{u}_{N})$ in which there exists at least one $\tilde{u}_{i}\not=0$. Assume we have a such strategy profile with total $m$ elements be non-zero. We use $(\tilde{u}_{1},..., \tilde{u}_{j},..., \tilde{u}_{m})$ to denote such an equilibrium strategy profile, where $1\leq m \leq N$. We have
\begin{equation}\label{s-equation34}
\begin{split}
\left[\begin{array}{cccc} 
     1 & -a_1&\cdots&-a_1 \\ 
    -a_2 & 1&\cdots&-a_2\\ 
    \vdots& \vdots& \ddots& \vdots \\
-a_{N} & -a_{N}&\cdots&1
\end{array}\right]
\begin{bmatrix}
  \tilde{u}_{1} \\
  \vdots   \\
\vdots   \\
\tilde{u}_{m}
 \end{bmatrix}=
\begin{bmatrix}
  0 \\
  \vdots   \\
\vdots   \\
0
 \end{bmatrix}.
\end{split}
\end{equation}
In order to show that there does not exit a profile $(\tilde{u}_{1},...,\tilde{u}_{j},...,\tilde{u}_{m})$ such that Eq.~(\ref{s-equation34}) holds. We have the following proposition:
 \begin{proposition}\label{s-proposition1}
The matrix M with the below form is of full rank, where $b_i \in (0, 1)$, $ \forall i\in\{1,2,...,m\}$.
\[ 
M=\left[\begin{array}{cccc} 
     1 & b_1 &\cdots &b_1 \\ 
    b_2 & 1&\cdots&b_2\\ 
    \vdots& \vdots& \ddots& \vdots \\
b_{m} &b_{m}&\cdots&1
\end{array}\right].
\]
\end{proposition}
\proof{}
 Consider the determinant of the matrix.
\[ 
det(M)=\left|\begin{array}{cccc} 
     1 & b_1 &\cdots &b_1 \\ 
    b_2 & 1&\cdots&b_2\\ 
    \vdots& \vdots& \ddots& \vdots \\
b_{m} &b_{m}&\cdots&1
\end{array}\right| 
=
\left|\begin{array}{cccc} 
    1 &   b_1-1    & \cdots&b_1-1 \\ 
    b_2 &   1-b_2   & \cdots& 0\\ 
    \vdots & 0 & \ddots& 0 \\
b_{m} & 0 & \cdots& 1-b_m
\end{array}\right|.
\]
Applying the property in Schur complement, let
\[ 
A=\left[1\right],
\quad B=\left[\begin{array}{cccc} 
    b_1-1 &\cdots & b_1-1
\end{array}\right],
\]
\[ 
C=\left[\begin{array}{cccc} 
    b_2  \\ 
    \vdots \\ 
    b_{m}  
\end{array}\right],
\quad \textup{and} \ D=\left[\begin{array}{cccc} 
      1-b_2   & \cdots& 0\\ 
     0 & \ddots& 0 \\
 0 & \cdots& 1-b_m
\end{array}\right].
\]
We know that $det(M)=det(D)det(A-BD^{-1}C)$, where
\[ 
D^{-1}=
\left[\begin{array}{cccc} 
     \frac{1}{1-b_2}   & \cdots& 0\\ 
     0 & \ddots& 0 \\
 0 & \cdots& \frac{1}{1-b_m}
\end{array}\right].
\]
It is easy to see that $det(M)>0$. Therefore,  matrix $M$ is a full rank matrix. 
\endproof

The above observation implies that there do not exist such a profile with non-zero entries while satisfying the necessary equilibrium conditions proposed in Theorem~\ref{theorem2}. Therefore, the strategy profile $(u^*_{1}, u^*_{2},...,u^*_{N})$ in which $u^*_{i}=0$ for $\forall i\in\{1,2,...,N\}$ is the unique pure strategy  Nash Equilibrium.

This completes the proof of Theorem~\ref{theorem3}. 
\endproof

\subsection{Proof of Corollary~\ref{corollary1}}
\proof{}
Under the piece-wise linear symmetric pricing model and independent prediction errors, given $\mu_{-i}=0$, the second term in (\ref{equation14}) is a quadratic function of $\mu_i$ which take the minimum value at $\mu^*_i=0$. Therefore,  Corollary~\ref{corollary1} is actually a direct conclusion of Theorem~\ref{theorem1} and Lemma~\ref{lemma2}.

This completes the proof of Corollary~\ref{corollary1}.
\endproof

\subsection{Proof of Theorem~\ref{theorem4}}

\proof{} It is easy to compute $\mathbb{E}[\mathsf{ABC}_{total}]$
\begin{align*}
\mathbb{E}[\mathsf{ABC}_{total}] & =p_{d}+\frac{\mathbb{E}[p_{s}\cdot\Delta]}{D_{total}}-\frac{p_{d}}{D_{total}}\mu,
\end{align*}
where $\Delta=\sum_{i=1}^{N}\Delta_{i}$, $D_{total}=\sum_{i=1}^{N}D_{i}$ and $\mu=\sum_{i=1}^{N}\mu_i$.
It remains to compute $\mathbb{E}[p_{s}\cdot\Delta]$: 
$$\mathbb{E}[p_{s}\cdot\Delta]	=	\mathbb{E}[(\xi_{1}\Delta+\xi_{2})\Delta]=	\mathbb{E}[\xi_{1}\Delta^{2}+\xi_{2}\Delta],$$
where $\xi_1$ and $\xi_2$ are the same as the ones in the proof of Theorem~\ref{theorem1}. By applying Lemma~\ref{lemma1}, we know that the total mismatch is symmetric distributed and centralized at $\mu=\sum_{i=1}^{N}\mu_i$. Let $f^{\mu}_{\Delta}$ be the probability density function of $\Delta$ centered at $\mu$. When $a_1=a_2$, we have
\begin{equation}
\begin{split}
\mathbb{E}[\xi_{1}\Delta^{2}]&=	\int_{-\infty}^{0}a_2p_d\delta^{2}f^{\mu}_{\Delta}(\delta)d\delta+\int_{0}^{+\infty}a_1p_d\delta^{2}f^{\mu}_{\Delta}(\delta)d\delta_{i}\\
&=\frac{(a_{1}+a_{2})p_{d}}{2}(\sigma^{2}+\mu^{2}).
\end{split}
\end{equation}
Secondly, let us compute $\mathbb{E}[\xi_{2}\Delta_{i}].$
\begin{equation}
\begin{split}
\mathbb{E}[\xi_{2}\Delta]&=\int_{-\infty}^{0}b_2p_d\delta f^{\mu}_{\Delta}(\delta)d\delta+\int_{0}^{+\infty}b_1p_d\delta f^{\mu}_{\Delta}(\delta)d\delta_{i}\\
&=b_1p_d\mu+({b_2-b_1})p_d\int_{-\infty}^{0}\delta f^{\mu}_{\Delta}(\delta)d\delta\\
&=\frac{b_1+b_2}{2}p_d\mu+({b_2-b_1})p_d\cdot\left[\int_{-\infty}^{-\mu}\delta f^{0}_{\Delta}(\delta)d\delta+\mu\int_{0}^{-\mu}f^{0}_{\Delta}(\delta)d\delta\right].
\end{split}
\end{equation}
By an abuse of the notation, we denote
$$U(\mu)=\int_{-\infty}^{-\mu}\delta f^{0}_{\Delta}(\delta)d\delta+\mu\int_{0}^{-\mu}f^{0}_{\Delta}(\delta)d\delta.$$
It is easy to see that $U(\mu)=U(-\mu)$, hence it is sufficient to consider the case of $\mu>0$. When $\mu>0$, we have
\begin{equation}
\begin{split}
U(\mu)-U(0)&=\int_{0}^{-\mu}\delta f^{0}_{\Delta}(\delta)d\delta+\mu\int_{0}^{-\mu}f^{0}_{\Delta}(\delta)d\delta\\
&=\int_{\mu}^{0}\delta f^{\mu}_{\Delta}(\delta)d\delta<0,
\end{split}
\end{equation}
where the last inequality comes from $f^{\mu}_{\Delta}(\cdot)$ is centralized at $\mu$.

Next we prove the strictly increasing property of $U(\cdot)$. Consider $\mu_1>\mu_2\geq0$, we have
\begin{equation}
\begin{split}
U(\mu_1)-U(\mu_2)&=-\int_{0}^{\mu_1}\delta f^{\mu_1}_{\Delta}(\delta)d\delta+\int_{0}^{\mu_2}\delta f^{\mu_2}_{\Delta}(\delta)d\delta\\
&=-\left(\left(\int_{0}^{\mu_1-\mu_2}+\int_{\mu_1-\mu_2}^{\mu_1}\right)\delta f^{\mu_1}_{\Delta}(\delta)d\delta-\int_{0}^{\mu_2}\delta f^{\mu_2}_{\Delta}(\delta)d\delta \right)\\
&=-\int_{0}^{\mu_1-\mu_2}\delta f^{\mu_1}_{\Delta}(\delta)d\delta+(\mu_2-\mu_1)\int_{-\mu_2}^{0}f^{0}_{\Delta}(\delta)d\delta<0,
\end{split}
\end{equation}
where the last inequality comes from that $f^{0}_{\Delta}(\cdot)$ is centralized at $0$. It is easy to see that $-\int_{0}^{\mu_1-\mu_2}\delta f^{\mu_1}_{\Delta}(\delta)d\delta\leq0$ and $(\mu_2-\mu_1)\int_{-\mu_2}^{0}f^{0}_{\Delta}(\delta)d\delta<0$.

From the above proof we see that the social cost $\mathbb{E}[\mathsf{ABC}_{total}]$ is minimized at $\mu=\sum_{i=1}^{N}\mu_i=0$. Therefore, the unique pure strategy Nash Equilibrium $(u^*_{1}, u^*_{2},..., u^*_{N})$ in which $u^*_{i}=0$ for $\forall i \in \{1,2,..., N\}$ is efficient from the social cost minimization perspective. In addition, the more the total strategy $\mu$ deviates from $0$, the more inefficient the market is. 

This completes the proof of Theorem~\ref{theorem4}. 
\endproof

\subsection{Proof of Theorem~\ref{theorem5}}
More precisely, Theorem~\ref{theorem5} can be expressed as
\proof{} It is easy to verify that when the price model is symmetric and the utility $j$ bids according to prediction, i.e., $\mu_j=0$, then:
$$\mathbb{E}[\xi_{1}\Delta_{j}^{2}]=\frac{(a_{1}+a_{2})p_{d}}{2}\sigma_{j}^{2},$$
and
$$\mathbb{E}[\xi_{1}\Delta_{j}\Delta_{-j}]=\frac{a_{1}+a_{2}}{2}p_{d}\mu_{j}\mu_{-j}=0,$$
where $\xi_1$ and $\xi_2$ are the same as the ones in the proof of Theorem~\ref{theorem1}. It remains to calculate $\mathbb{E}[\xi_{2}\Delta_{j}]$:
\begin{equation}
\begin{split}
\mathbb{E}[\xi_{2}\Delta_{j}] & =\mathbb{E}\left[b_{1}p_{d}\Delta_{j}\left(1-\int_{-\infty}^{-\Delta_{j}}f_{\Delta_{-j}}(\delta_{-j})d\delta_{-j})+b_{2}p_{d}\Delta_{j}\int_{-\infty}^{-\Delta_{j}}f_{\Delta_{-j}}(\delta_{-j})d\delta_{-j}\right)\right]\\
 & =b_{1}p_{d}\mu_{j}+\mathbb{E}\left[\Delta_{i}(b_{2}-b_{1})p_{d}\int_{-\infty}^{-\Delta_{j}}f_{\Delta_{-j}}(\delta_{-j})d\delta_{-j})\right]\\
 & =b_{1}p_{d}\mu_{j}+\mathbb{E}\left[\Delta_{i}(b_{2}-b_{1})p_{d}(\int_{-\infty}^{\mu_{-j}}f_{\Delta_{-j}}(\delta_{-j})d\delta_{-j}+\int_{\mu_{-j}}^{-\Delta_{j}}f_{\Delta_{-j}}(\delta_{-j})d\delta_{-j})\right]\\
 & =p_{d}\mu_{j}\frac{b_{1}+b_{2}}{2}+(b_{1}-b_{2})p_{d}\mathbb{E}\left[\Delta_{j}\hat{F}(\Delta_j)\right],
\end{split}
\end{equation}
where $\hat{F}(\Delta_j)\triangleq -\int_{\mu_{-j}}^{-\Delta_{j}}f_{\Delta_{-j}}(\delta_{-j})d\delta_{-j}=\int_{-\Delta_{j}}^{\mu_{-j}}f_{\Delta_{-j}}(\delta_{-j})d\delta_{-j}$. Let us define
\begin{equation}
\begin{split}
V(\mu_{-j})&=\mathbb{E}\left[\Delta_{j}\hat{F}(\Delta_j)\right]\\&=\int_{-\infty}^{+\infty}\delta_j f^0_{\Delta_j}(\delta_j)\cdot\int_{-\delta_{j}}^{\mu_{-j}}f^{\mu_{-j}}_{\Delta_{-j}}(\delta_{-j})d\delta_{-j}d\delta_{j}\\
&=\int_{-\infty}^{+\infty}\delta_j f^0_{\Delta_j}(\delta_j)\cdot\int_{-\delta_{j}-\mu_{-j}}^{0}f^{0}_{\Delta_{-j}}(\delta_{-j})d\delta_{-j}d\delta_{j}\\
&=\int_{-\infty}^{+\infty}\delta_j f^0_{\Delta_j}(\delta_j)\cdot(\int_{-\delta_{j}}^{0}+\int_{-\delta_{j}-\mu_{-j}}^{-\delta_{j}})f^{0}_{\Delta_{-j}}(\delta_{-j})d\delta_{-j}d\delta_{j}.
\end{split}
\end{equation}
Consider
\begin{equation}
\begin{split}
V(-\mu_{-j})&=\mathbb{E}\left[\Delta_{i}\hat{F}(\Delta_j)\right]\\
&=\int_{-\infty}^{+\infty}\delta_j f^0_{\Delta_j}(\delta_j)\cdot\int_{-\delta_{j}}^{-\mu_{-j}}f^{-\mu_{-j}}_{\Delta_{-j}}(\delta_{-j})d\delta_{-j}d\delta_{j}\\
&=\int_{-\infty}^{+\infty}(-\delta_j) f^0_{\Delta_j}(-\delta_j)\cdot\int_{\delta_{j}+\mu_{-j}}^{0}f^{0}_{\Delta_{-j}}(\delta_{-j})d\delta_{-j}d\delta_{j}\\
&=\int_{-\infty}^{+\infty}\delta_j f^0_{\Delta_j}(\delta_j)\cdot\int_{-\delta_{j}-\mu_{-j}}^{0}f^{0}_{\Delta_{-j}}(\delta_{-j})d\delta_{-j}d\delta_{j}\\
&=V(\mu_{-j}).
\end{split}
\end{equation}
Hence the above $V(\cdot)$ is an even function. It is sufficient to consider the case of $\mu_{-j}>0$.
Let us compute
\begin{equation}
\begin{split}
V(\mu_{-j})-V(0)=\int_{-\infty}^{+\infty}\delta_{j}f_{\Delta_j}(\delta_j)\int_{-\delta_{j}-\mu_{-j}}^{-\delta_{j}}f^{0}_{\Delta_{-j}}(\delta_{-j})d\delta_{-j}d\delta_{j}.
\end{split}
\end{equation}
Consider two points  $\delta^*_j$ and $-\delta^*_j$ with $\delta^*_j\geq0$. We have
\begin{equation}
\begin{split}
&\delta^*_j f_{\Delta_j}(\delta^*_j)\int_{-\delta^*_j-\mu_{-j}}^{-\delta^*_j}f^{0}_{\Delta_{-j}}(\delta_{-j})d\delta_{-j}+(-\delta^*_j)\cdot f_{\Delta_j}(-\delta^*_j)\int_{\delta^*_j-\mu_{-j}}^{\delta^*_j}f^{0}_{\Delta_{-j}}(\delta_{-j})d\delta_{-j}\\
=&\delta^*_j f_{\Delta_j}(\delta^*_j)(\int_{-\delta^*_j-\mu_{-j}}^{-\delta^*_j}f^{0}_{\Delta_{-j}}(\delta_{-j})d\delta_{-j}-\int_{\delta^*_j-\mu_{-j}}^{\delta^*_j}f^{0}_{\Delta_{-j}}(\delta_{-j})d\delta_{-j}).
\end{split}
\end{equation}
Since $\delta^*_j f_{\Delta_j}(\delta^*_j)\geq 0$ and $f^{0}_{\Delta_{-j}}(\delta_{-j})$ satisfy the symmetric unimodal distribution conditions. We have
$$\int_{-\delta^*_j-\mu_{-j}}^{-\delta^*_j}f^{0}_{\Delta_{-j}}(\delta_{-j})-\int_{\delta^*_j-\mu_{-i}}^{\delta^*_j}f^{0}_{\Delta_{-j}}(\delta_{-j}) \leq 0.$$
Then we have
$$V(\mu_{-j})-V(0)\leq 0.$$

To sum up, when the pricing model is symmetric, i.e., $a_{1}=a_{2}$, $b_{1}+b_{2}=2$, then given $\mu_{j}=0$, The expectation of $\mathbf{\textsf{ABC}}_{j}$ is given as:
\begin{equation}
\begin{split}
\mathbb{E}[\mathbf{\textsf{ABC}}_{j}]&=p_{d}+\frac{p_{d}}{D_{j}}\left[\frac{a_{1}+a_{2}}{2}\sigma_{j}^{2}+(b_{1}-b_{2})\mathbb{E}\left[\Delta_{i}\hat{F}(\Delta_j)\right]\right],
\end{split}
\end{equation}
where $\hat{F}$ is defined as above and $\mathbb{E}[\mathbf{\textsf{ABC}}_{j}]$ takes the maximum value when $\mu_{-j}=0$.

Next we prove the strictly decreasing property of $U(\cdot)$. if strictly central dominant condition is satisfied for either $\Delta_{-j}$ or $\Delta_j$. Consider $\mu_1>\mu_2\geq0$, we have
\begin{equation}
\begin{split}
V(\mu_{1})-V(\mu_2)=\int_{-\infty}^{+\infty}\delta_{j}f_{\Delta_j}(\delta_j)\int_{-\delta_{j}-\mu_{1}}^{-\delta_{j}-\mu_{2}}f^{0}_{\Delta_{-j}}(\delta_{-j})d\delta_{-j}d\delta_{j}.
\end{split}
\end{equation}
Consider two points  $\delta^*_j$ and $-\delta^*_j$ with $\delta^*_j\geq0$. We have
\begin{equation}
\begin{split}
&\delta^*_j f_{\Delta_j}(\delta^*_j)\int_{-\delta^*_j-\mu_{1}}^{-\delta^*_j-\mu_{2}}f^{0}_{\Delta_{-j}}(\delta_{-j})d\delta_{-j}+(-\delta^*_j)\cdot f_{\Delta_j}(-\delta^*_j)\int_{\delta^*_j-\mu_{-j}}^{\delta^*_j-\mu_{2}}f^{0}_{\Delta_{-j}}(\delta_{-j})d\delta_{-j}\\
=&\delta^*_j f_{\Delta_j}(\delta^*_j)(\int_{-\delta^*_j-\mu_{1}}^{-\delta^*_j-\mu_{2}}f^{0}_{\Delta_{-j}}(\delta_{-j})d\delta_{-j}-\int_{\delta^*_j-\mu_{1}}^{\delta^*_j-\mu_{2}}f^{0}_{\Delta_{-j}}(\delta_{-j})d\delta_{-j})\leq 0,
\end{split}
\end{equation}
where the last strict inequality holds when either $f_{\Delta_j}(\cdot)$ or $f^{0}_{\Delta_{-j}}(\cdot)$ is strictly central dominant. For example, when $f_{\Delta_j}(\cdot)$ is strictly central dominant, we have $\delta^*_j f_{\Delta_j}(\delta^*_j)>0$ and $(\int_{-\delta^*_j-\mu_{1}}^{-\delta^*_j-\mu_{2}}f^{0}_{\Delta_{-j}}(\delta_{-j})d\delta_{-j}-\int_{\delta^*_j-\mu_{1}}^{\delta^*_j-\mu_{2}}f^{0}_{\Delta_{-j}}(\delta_{-j})d\delta_{-j})\leq0$ and is less than zero for some $\delta^*_j$. When $f^{0}_{\Delta_{-j}}(\cdot)$ is strictly central dominant,  we have $(\int_{-\delta^*_j-\mu_{1}}^{-\delta^*_j-\mu_{2}}f^{0}_{\Delta_{-j}}(\delta_{-j})d\delta_{-j}-\int_{\delta^*_j-\mu_{1}}^{\delta^*_j-\mu_{2}}f^{0}_{\Delta_{-j}}(\delta_{-j})d\delta_{-j})<0$ and $f_{\Delta_j}(\cdot)\geq0$ and is great than zero for some $\delta^*_j$. Hence we have:
$V(\mu_1)-V(\mu_2)<0$ when $\mu_{1}>\mu_{2}>0$.

This completes the proof of Theorem~\ref{theorem5}.
\endproof

\subsection{Proof of Theorem~\ref{theorem6}}
\proof{}

We prove the statements in Theorem~\ref{theorem6} one by one.

\subsubsection{Proof of Theorem~\ref{theorem6} (1)}

\proof{}
According to pricing model in Eq.~(\ref{equation20}), given the day-ahead market price $p_{d}$, the spot price $p_{s}$ is a step function of $\Delta$. Namely, 
\begin{equation}
\begin{split}
p_{s} & =1_{\{\Delta>0\}}\cdot (p(\Delta)+b_1p_d)+1_{\{\Delta<0\}}\cdot(p(\Delta)+b_1p_d)+1_{\{\Delta=0\}}\cdot p_d\\
 & =\xi_{1}+\xi_{2},
\end{split}
\end{equation}
where $1_{\{\cdot\}}$ is indicator function, and $$\xi_{1}\triangleq p(\Delta),$$and $$\xi_{2}\triangleq b_{1}p_{d}1_{\{\Delta>0\}}+b_{2}p_{d}1_{\{\Delta<0\}}+p_{d}1_{\{\Delta=0\}}.$$Then we can compute the expectation of $\mathsf{ABC}_{i}$ in the following way: 
\begin{equation}\label{general-cost}
\begin{split}
\mathbb{E}[\mathsf{ABC}_{i}]  =\mathbb{E}\left[p_{d}+\frac{\Delta_{i}(p_{s}-p_{d})}{D_{i}}\right]=p_{d}+\frac{\mathbb{E}[p_{s}\cdot\Delta_{i}]}{D_{i}}-\frac{p_{d}}{D_{i}}\mu_{i}.
\end{split}
\end{equation}
It remains to compute $\mathbb{E}[p_{s}\cdot\Delta_{i}]$:
\begin{equation}
\begin{split}
\mathbb{E}[p_{s}\cdot\Delta_{i}]	&=	\mathbb{E}[(\xi_{1}+\xi_{2})\Delta_{i}]=\mathbb{E}[\xi_{1}\Delta_{i}+\xi_{2}\Delta_{i}].
\end{split}
\end{equation}
Therefore, $\mathbb{E}[p_{s}\cdot\Delta_{i}]$ can be divided into two terms. In the following, we will compute these two terms one by one.

 It is easy to verify that when the price model is symmetric as defined in Eq.~(\ref{equation20}), $\mathbb{E}[\xi_{2}\Delta_{i}]$ has the same expression as linear price model case.

It remains to compute $\mathbb{E}[\xi_{1}\Delta_{i}]$:
\begin{equation}
\begin{split}
\mathbb{E}[\xi_{1}\Delta_{i}]&=\int_{-\infty}^{+\infty}\delta_{i}f^{\mu_i}_{\Delta_{i}}(\delta_{i})\int_{-\infty}^{+\infty}p(\delta_i+\delta_{-i})f_{\Delta_{-i}}(\delta_{-i})d\delta_{-i}d\delta_{i}.
\end{split}
\end{equation}
By an abuse of notation, we denote $$U(\mu_i)=\int_{-\infty}^{+\infty}\delta_{i}f^{\mu_i}_{\Delta_{i}}(\delta_{i})\int_{-\infty}^{+\infty}p(\delta_i+\delta_{-i})f_{\Delta_{-i}}(\delta_{-i})d\delta_{-i}d\delta_{i}.$$
Given $\mu_{-i}=0$, we have
\begin{equation}
\begin{split}
U(-\mu_{i})=&\int_{-\infty}^{+\infty}\delta_{i}f^{-\mu_i}_{\Delta_{i}}(\delta_{i})\int_{-\infty}^{+\infty}p(\delta_i+\delta_{-i})f^0_{\Delta_{-i}}(\delta_{-i})d\delta_{-i}d\delta_{i}\\
=&\int_{-\infty}^{+\infty}(-\delta_{i})f^{-\mu_i}_{\Delta_{i}}(-\delta_{i})\int_{-\infty}^{+\infty}p(-\delta_i-\delta_{-i})f^0_{\Delta_{-i}}(-\delta_{-i})d\delta_{-i}d\delta_{i}\\
=&\int_{-\infty}^{+\infty}\delta_{i}f^{\mu_i}_{\Delta_{i}}(\delta_{i})\int_{-\infty}^{+\infty}p(\delta_i+\delta_{-i})f^0_{\Delta_{-i}}(\delta_{-i})d\delta_{-i}d\delta_{i}\\
=&U(\mu_i).
\end{split}
\end{equation}
Hence it is sufficient to consider the case when $\mu_i>0$. Considering
\begin{equation}
\begin{split}
U(\mu_i)-U(0)=\int_{-\infty}^{+\infty}\delta_{i}(f^{\mu_i}_{\Delta_{i}}(\delta_{i})-f^{0}_{\Delta_{i}}(\delta_{i}))\int_{-\infty}^{+\infty}p(\delta_i+\delta_{-i})f^0_{\Delta_{-i}}(\delta_{-i})d\delta_{-i}d\delta_{i}.
\end{split}
\end{equation}
Consider two symmetric points $\frac{\mu_i}{2}+\delta_i$ and $\frac{\mu_i}{2}-\delta_i$, where $\delta_i>0$. Let us use $\Delta f_{\Delta_i}(\delta_i)$ to denote $f^{\mu_i}_{\Delta_i}(\frac{\mu_i}{2}+\delta_i)-f^{0}_{\Delta_i}(\frac{\mu_i}{2}+\delta_i)$. We have
\begin{equation}
\begin{split}
&(\frac{\mu_i}{2}+\delta_i)\Delta f_{\Delta_i}(\delta_i)\int_{-\infty}^{+\infty}p(\delta_i+\frac{\mu_i}{2}+\delta_{-i})f_{\Delta_{-i}}(\delta_{-i})d\delta_{-i}\\&-(\frac{\mu_i}{2}-\delta_i)\Delta f_{\Delta_i}\int_{-\infty}^{+\infty}p(-\delta_i+\frac{\mu_i}{2}+\delta_{-i})f_{\Delta_{-i}}(\delta_{-i})d\delta_{-i}\\
=&\frac{\mu_i}{2}\Delta f_{\Delta_i}\int_{-\infty}^{+\infty}[p(\delta_i+\frac{\mu_i}{2}+\delta_{-i})-p(-\delta_i+\frac{\mu_i}{2}+\delta_{-i})]\cdot f_{\Delta_{-i}}(\delta_{-i})d\delta_{-i}
\\&+\delta_i\Delta f_{\Delta_i}\int_{-\infty}^{+\infty}[p(\delta_i+\frac{\mu_i}{2}+\delta_{-i})+p(-\delta_i+\frac{\mu_i}{2}+\delta_{-i})]f_{\Delta_{-i}}(\delta_{-i})d\delta_{-i}\\
\geq& 0.
\end{split}
\end{equation}
Hence we have:
$U(\mu_i)-U(0)\geq0$ when $\mu_{i}>0$. Furthermore, considering $\mu_1>\mu_2\geq0$, we have
\begin{equation}
\begin{split}
U(\mu_1)-U(\mu_2)=\int_{-\infty}^{+\infty}\delta_{i}(f^{\mu_1}_{\Delta_{i}}(\delta_{i})-f^{\mu_2}_{\Delta_{i}}(\delta_{i}))\int_{-\infty}^{+\infty}p(\delta_i+\delta_{-i})f_{\Delta_{-i}}(\delta_{-i})d\delta_{-i}d\delta_{i}.
\end{split}
\end{equation}
Consider $\frac{\mu_1+\mu_2}{2}+\delta_i$ and $\frac{\mu_1+\mu_2}{2}-\delta_i$, where $\delta_i>0$. By an abuse of notation,  denote $f^{\mu_1}_{\Delta_{i}}(\frac{\mu_1+\mu_2}{2}+\delta_i)-f^{\mu_2}_{\Delta_{i}}(\frac{\mu_1+\mu_2}{2}+\delta_i)$ as $\Delta f_{\Delta_i}(\delta_i)$. We have
\begin{equation}
\begin{split}
&(\frac{\mu_1+\mu_2}{2}+\delta_i)\Delta f_{\Delta_i}(\delta_i)\int_{-\infty}^{+\infty}p(\delta_i+\frac{\mu_1+\mu_2}{2}+\delta_{-i})\cdot f_{\Delta_{-i}}(\delta_{-i})d\delta_{-i}\\
&-(\frac{\mu_1+\mu_2}{2}-\delta_i)\Delta f_{\Delta_i}(\delta_i)\cdot \int_{-\infty}^{+\infty}p(-\delta_i+\frac{\mu_1+\mu_2}{2}+\delta_{-i})f_{\Delta_{-i}}(\delta_{-i})d\delta_{-i}\\
=&\frac{\mu_1+\mu_2}{2}\Delta f_{\Delta_i}(\delta_i)\int_{-\infty}^{+\infty}[p(\delta_i+\frac{\mu_1+\mu_2}{2}+\delta_{-i})-p(-\delta_i+\frac{\mu_1+\mu_2}{2}+\delta_{-i})]\cdot f_{\Delta_{-i}}(\delta_{-i})d\delta_{-i}\\&+\delta_i\Delta f_{\Delta_i}(\delta_i)\int_{-\infty}^{+\infty}[p(\delta_i+\frac{\mu_1+\mu_2}{2}+\delta_{-i})+p(-\delta_i+\frac{\mu_1+\mu_2}{2}+\delta_{-i})]f_{\Delta_{-i}}(\delta_{-i})d\delta_{-i}\\\geq &0.
\end{split}
\end{equation}
It is easy to verify that when $p(\cdot)$ is strictly increasing, we will always have $U(\mu_1)>U(\mu_2)$. Together with Lemma~\ref{lemma2}, we get the desired results.

This completes the proof of Theorem~\ref{theorem6} (1). 
\endproof

\subsubsection{Proof of Theorem~\ref{theorem6} (2)}
\proof{}
Without loss of generality, assume $\mu_{-i}>0$.
Let us focus on $\mathbb{E}[\xi_{1}\Delta_{i}]$:
\begin{equation}
\begin{split}
\mathbb{E}[\xi_{1}\Delta_{i}]&=\int_{-\infty}^{+\infty}\delta_{i}f^{\mu_i}_{\Delta_{i}}(\delta_{i})\int_{-\infty}^{+\infty}p(\delta_i+\delta_{-i})f^{\mu_{-i}}_{\Delta_{-i}}(\delta_{-i})d\delta_{-i}d\delta_{i}.
\end{split}
\end{equation}
By an abuse of notation, we denote $$U(\mu_i)=\int_{-\infty}^{+\infty}\delta_{i}f^{\mu_i}_{\Delta_{i}}(\delta_{i})\int_{-\infty}^{+\infty}p(\delta_i+\delta_{-i})f^{\mu_{-i}}_{\Delta_{-i}}(\delta_{-i})d\delta_{-i}d\delta_{i}.$$
When $\mu_i>0$, consider 
\begin{equation}
\begin{split}
&U(\mu_i)-U(0)=\int_{-\infty}^{+\infty}\delta_{i}(f^{\mu_i}_{\Delta_{i}}(\delta_{i})-f^{0}_{\Delta_{i}}(\delta_{i}))\cdot \int_{-\infty}^{+\infty}p(\delta_i+\delta_{-i})f^{\mu_{-i}}_{\Delta_{-i}}(\delta_{-i})d\delta_{-i}d\delta_{i}.
\end{split}
\end{equation}
Consider two symmetric points $\frac{\mu_i}{2}+\delta_i$ and $\frac{\mu_i}{2}-\delta_i$, where $\delta_i>0$. Let us use $\Delta f_{\Delta_i}(\delta_i)$ to denote $f^{\mu_i}_{\Delta_i}(\frac{\mu_i}{2}+\delta_i)-f^{0}_{\Delta_i}(\frac{\mu_i}{2}+\delta_i)$. We have
\begin{equation}
\begin{split}
&(\frac{\mu_i}{2}+\delta_i)\Delta f_{\Delta_i}(\delta_i)\int_{-\infty}^{+\infty}p(\delta_i+\frac{\mu_i}{2}+\delta_{-i})f^{\mu_{-i}}_{\Delta_{-i}}(\delta_{-i})d\delta_{-i}\\&-(\frac{\mu_i}{2}-\delta_i)\Delta f_{\Delta_i}(\delta_i)\int_{-\infty}^{+\infty}p(-\delta_i+\frac{\mu_i}{2}+\delta_{-i})f^{\mu_{-i}}_{\Delta_{-i}}(\delta_{-i})d\delta_{-i}\\
=&\frac{\mu_i}{2}\Delta f_{\Delta_i}(\delta_i)\int_{-\infty}^{+\infty}[p(\delta_i+\frac{\mu_i}{2}+\delta_{-i}+\mu_{-i})-p(-\delta_i+\frac{\mu_i}{2}+\delta_{-i}+\mu_{-i})]f_{\Delta_{-i}}(\delta_{-i})d\delta_{-i}\\&+\delta_i\Delta f_{\Delta_i}(\delta_i)\int_{-\infty}^{+\infty}[p(\delta_i+\frac{\mu_i}{2}+\delta_{-i}+\mu_{-i})+p(-\delta_i+\frac{\mu_i}{2}+\delta_{-i}+\mu_{-i})]f_{\Delta_{-i}}(\delta_{-i})d\delta_{-i}.
\end{split}
\end{equation}
It is easy to verify that when $p(\cdot)$ is strictly increasing, we will always have $U(\mu_i)>U(0)$ if $\mu_i>0$.
Furthermore, considering $\mu_i<-\mu_{-i}$, following similar approach, we can prove that
$$U(\mu_1)-U(-\mu_{-i})>0.$$
The above observation reveals that given $\mu_{-i}>0$,  utility $i$ will not choose to bid at $\mu_i>0$ or $\mu_i<-\mu_{-i}$. Then let us consider the case when bid \emph{a little bit less than} zero and bid \emph{a little bit more than} $-\mu_{-i}.$ In other words, let us consider the left derivative at $\mu_{i}=0^{-}$ and right derivative at $\mu_{i}={-\mu_{-i}}^{+}$.

i) When $\mu_i \to 0^-$, we have
\begin{equation}
\begin{split}
U(\mu_i)-U(0)=&\int_{-\infty}^{+\infty}\delta_{i}f^{0}_{\Delta_{i}}(\delta_{i})\int_{-\infty}^{+\infty}[p(\delta_i+\delta_{-i}+\mu_i+\mu_{-i})-p(\delta_i+\delta_{-i}+\mu_{-i})]f^{0}_{\Delta_{-i}}(\delta_{-i})d\delta_{-i}d\delta_{i}\\&+\mu_{i}\int_{-\infty}^{+\infty}f^{0}_{\Delta_{i}}(\delta_{i})\cdot \int_{-\infty}^{+\infty}p(\delta_i+\delta_{-i}+\mu_i+\mu_{-i})f^{0}_{\Delta_{-i}}(\delta_{-i})d\delta_{-i}d\delta_{i}.
\end{split}
\end{equation}
For the second term, we can prove that
\begin{equation}
\begin{split}
\mu_{i}\int_{-\infty}^{+\infty}f^{0}_{\Delta_{i}}(\delta_{i})\int_{-\infty}^{+\infty}p(\delta_i+\delta_{-i}+\mu_i+\mu_{-i})\cdot f^{0}_{\Delta_{-i}}(\delta_{-i})d\delta_{-i}d\delta_{i}\leq\mu_{i}p(\mu_i+\mu_{-i})<0.
\end{split}
\end{equation}
For the first term, we have
\begin{equation}
\begin{split}
&\int_{-\infty}^{+\infty}\delta_{i}f^{0}_{\Delta_{i}}(\delta_{i})\int_{-\infty}^{+\infty}[p(\delta_i+\delta_{-i}+\mu_i+\mu_{-i})-p(\delta_i+\delta_{-i}+\mu_{-i})]f^{0}_{\Delta_{-i}}(\delta_{-i})d\delta_{-i}d\delta_{i}\\
=&\int_{0}^{+\infty}\delta_{i}f^{0}_{\Delta_{i}}(\delta_{i})\int_{-\infty}^{+\infty}[p(\delta_i+\delta_{-i}+\mu_i+\mu_{-i})-p(\delta_i+\delta_{-i}+\mu_{-i})]f^{0}_{\Delta_{-i}}(\delta_{-i})d\delta_{-i}d\delta_{i}\\
&-\int_{0}^{+\infty}\delta_{i}f^{0}_{\Delta_{i}}(\delta_{i})\int_{-\infty}^{+\infty}[p(-\delta_i+\delta_{-i}+\mu_i+\mu_{-i})-p(-\delta_i+\delta_{-i}+\mu_{-i})]f^{0}_{\Delta_{-i}}(\delta_{-i})d\delta_{-i}d\delta_{i}\\
=&\mu_{i}\cdot\int_{0}^{+\infty}\delta_{i}f^{0}_{\Delta_{i}}(\delta_{i})\int_{-\infty}^{+\infty}[p'(\delta_i+\delta_{-i}+\mu_{-i})-p'(-\delta_i+\delta_{-i}+\mu_{-i})]f^{0}_{\Delta_{-i}}(\delta_{-i})d\delta_{-i}d\delta_{i}.
\end{split}
\end{equation}
Consider two symmetric points $-\mu_{-i}+\epsilon$ and $-\mu_{-i}-\epsilon$, where $\epsilon\geq 0$. The inner integral part  is 
\begin{equation}
\begin{split}&\int_{0}^{+\infty}[p'(\delta_i+\epsilon)-p'(-\delta_i+\epsilon)]f^{0}_{\Delta_{-i}}(-\mu_{-i}+\epsilon)d\epsilon \\&+\int_{0}^{+\infty}[p'(\delta_i-\epsilon)-p'(-\delta_i-\epsilon)]f^{0}_{\Delta_{-i}}(-\mu_{-i}-\epsilon)d\epsilon.
\end{split}
\end{equation}
It is easy to prove that this term is non-negative. Hence when $\mu_{i}\to0^{-}$, $U(\mu_i)-U(0)<0$.

We know that 
\begin{equation}
\begin{split}U(-\mu_{-i})=&\int_{-\infty}^{+\infty}\delta_{i}f^{0}_{\Delta_{i}}(\delta_{i})\int_{-\infty}^{+\infty}p(\delta_i+\delta_{-i})f^{0}_{\Delta_{-i}}(\delta_{-i})d\delta_{-i}d\delta_{i}\\&-\mu_{-i}\int_{-\infty}^{+\infty}f^{0}_{\Delta_{i}}(\delta_{i})\int_{-\infty}^{+\infty}p(\delta_i+\delta_{-i})f^{0}_{\Delta_{-i}}(\delta_{-i})d\delta_{-i}d\delta_{i}.
\end{split}
\end{equation}
It is easy to prove that $$\mu_{-i}\int_{-\infty}^{+\infty}f^{0}_{\Delta_{i}}(\delta_{i})\int_{-\infty}^{+\infty}p(\delta_i+\delta_{-i})f^{0}_{\Delta_{-i}}(\delta_{-i})d\delta_{-i}d\delta_{i}=0.$$

ii) When  $\mu_{i}\to-\mu_{-i}^{+}$, we have
\begin{equation}
\begin{split}
&U(\mu_i)-U(-\mu_{-i})\\=&\int_{-\infty}^{+\infty}\delta_{i}f^{0}_{\Delta_{i}}(\delta_{i})\cdot\int_{-\infty}^{+\infty}[p(\delta_i+\delta_{-i}+\mu_i+\mu_{-i})-p(\delta_i+\delta_{-i})]\cdot f^{0}_{\Delta_{-i}}(\delta_{-i})d\delta_{-i}d\delta_{i}\\&+\mu_{i}\int_{-\infty}^{+\infty}f^{0}_{\Delta_{i}}(\delta_{i})\cdot\int_{-\infty}^{+\infty}p(\delta_i+\delta_{-i}+\mu_i+\mu_{-i})f^{0}_{\Delta_{-i}}(\delta_{-i})d\delta_{-i}d\delta_{i}.
\end{split}
\end{equation}
For the first term, we can prove that 
\begin{equation}
\begin{split}
&\int_{-\infty}^{+\infty}\delta_{i}f^{0}_{\Delta_{i}}(\delta_{i})\int_{-\infty}^{+\infty}[p(\delta_i+\delta_{-i}+\mu_i+\mu_{-i})-p(\delta_i+\delta_{-i})]\cdot f^{0}_{\Delta_{-i}}(\delta_{-i})d\delta_{-i}d\delta_{i}\\
=&(\mu_i+\mu_{-i})\int_{0}^{+\infty}\delta_{i}f^{0}_{\Delta_{i}}(\delta_{i})\int_{0}^{+\infty}[p'(\delta_i+\delta_{-i})-p'(-\delta_i+\delta_{-i})\\&+p'(-\delta_i+\delta_{-i})-p'(-\delta_i-\delta_{-i})]\cdot f^{0}_{\Delta_{-i}}(\delta_{-i})d\delta_{-i}d\delta_{i}\\
=&0.
\end{split}
\end{equation}
For the second term, we have
\begin{equation}
\begin{split}
&\mu_{i}\int_{-\infty}^{+\infty}f^{0}_{\Delta_{i}}(\delta_{i})\int_{-\infty}^{+\infty}p(\delta_i+\delta_{-i}+\mu_i+\mu_{-i})\cdot f^{0}_{\Delta_{-i}}(\delta_{-i})d\delta_{-i}d\delta_{i}\\
=&\mu_{i}\int_{-\infty}^{+\infty}f^{0}_{\Delta_{i}}(\delta_{i})\int_{-\infty}^{+\infty}[p(\delta_i+\delta_{-i}+\mu_i+\mu_{-i}) -p(\delta_i+\delta_{-i})]\cdot f^{0}_{\Delta_{-i}}(\delta_{-i})d\delta_{-i}d\delta_{i}\\
=&\mu_{i}(\mu_i+\mu_{-i})\int_{-\infty}^{+\infty}f^{0}_{\Delta_{i}}(\delta_{i})\cdot \int_{-\infty}^{+\infty}p'(\delta_i+\delta_{-i})f^{0}_{\Delta_{-i}}(\delta_{-i})d\delta_{-i}d\delta_{i}\\
<&0.
\end{split}
\end{equation}
Hence when $\mu_{i}\to-\mu_{-i}^{+}$, $U(\mu_i)-U(-\mu_{-i})<0$. Similar analysis can be constructed when $\mu_{-i}<0$ and utility $i's$ optimal strategy $\mu^*_i$ that minimizes $\mathbb{E}[\mathbf{\textsf{ABC}}_{i}]$ must be within $(0, -\mu_{-i})$ given $\mu_{-i}<0$. Therefore, the necessary condition for a strategy profile to be Nash Equilibrium is the same as the ones in Theorem~\ref{theorem2}. Hence, the uniqueness of the Nash Equilibrium is a direct extension result of Theorem~\ref{theorem3}.

This completes the proof of Theorem~\ref{theorem6} (2). 
\endproof

\subsubsection{Proof of Theorem~\ref{theorem6} (3)}
\proof{}
It is easy to compute $\mathbb{E}[\mathsf{ABC}_{total}]$
\begin{align*}
\mathbb{E}[\mathsf{ABC}_{total}] & =p_{d}+\frac{\mathbb{E}[p_{s}\cdot\Delta]}{D_{total}}-\frac{p_{d}}{D_{total}}\mu.
\end{align*}
where $\Delta=\sum_{i=1}^{N}\Delta_{i}$, $D_{total}=\sum_{i=1}^{N}D_{i}$ and $\mu=\sum_{i=1}^{N}\mu_i$.
It remains to compute $\mathbb{E}[p_{s}\cdot\Delta]$: 
\begin{equation}
\begin{split}
\mathbb{E}[p_{s}\cdot\Delta]	&=	\mathbb{E}[(\xi_{1}+\xi_{2})\Delta] =	\mathbb{E}[\xi_{1}\Delta+\xi_{2}\Delta],
\end{split}
\end{equation}
where $\xi_1$ and $\xi_2$ are the same as the ones in the proof of Theorem~\ref{theorem4}. By applying Lemma~\ref{lemma1}, we know that the total mismatch is symmetric distributed and centralized at $\mu=\sum_{i=1}^{N}\mu_i$. Let $f^{\mu}_{\Delta}$ be the probability density function of $\Delta$ centered at $\mu$. Therefore,   $\mathbb{E}[p_{s}\cdot\Delta]$ can be divided into two terms. In the following, we will compute these two terms one by one.

 It is easy to verify that when the price model is symmetric as defined in Eq.~(\ref{equation20}), $\mathbb{E}[\xi_{2}\Delta]$ has the same expression as linear price model case.

It remains to compute $\mathbb{E}[\xi_{1}\Delta]$:
\begin{equation}
\begin{split}
\mathbb{E}[\xi_{1}\Delta]&=\int_{-\infty}^{+\infty}\delta f^{\mu}_{\Delta}(\delta)p(\delta)d\delta.
\end{split}
\end{equation}
By an abuse of notation, we denote $$U(\mu)=\int_{-\infty}^{+\infty}\delta f^{\mu}_{\Delta}(\delta)p(\delta)d\delta.$$
We have
\begin{equation}
\begin{split}
U(-\mu)&=\int_{-\infty}^{+\infty}\delta f^{-\mu}_{\Delta}(\delta)p(\delta)d\delta=\int_{-\infty}^{+\infty}\delta f_{\Delta}(\delta+\mu)p(\delta)d\delta\\&=\int_{-\infty}^{+\infty}\delta f_{\Delta}(-\delta+\mu)p(\delta)d\delta=U(\mu).
\end{split}
\end{equation}
Therefore, it is sufficient to consider the case of $\mu>0$. When $\mu>0$, we have $$U(\mu)-U(0)=\int_{-\infty}^{+\infty}\delta(f^{\mu}_{\Delta}(\delta)-f^{0}_{\Delta}(\delta))p(\delta)d\delta.$$
Consider two symmetric points $\frac{\mu}{2}+\delta$ and $\frac{\mu}{2}-\delta$, where $\delta_i>0$. Let us use $\Delta f_{\Delta}(\delta)$ to denote $f^{\mu}_{\Delta}(\frac{\mu}{2}+\delta_i)-f^{0}_{\Delta}(\frac{\mu}{2}+\delta)$. We have
\begin{equation}
\begin{split}
&(\frac{\mu}{2}+\delta)\Delta f_{\Delta}(\delta) p(\delta+\frac{\mu}{2})-(\frac{\mu}{2}-\delta)\Delta f_{\Delta}(\delta) p(-\delta+\frac{\mu}{2})\\
=&\frac{\mu}{2}\Delta f_{\Delta}(\delta) [p(\delta+\frac{\mu}{2})-p(-\delta+\frac{\mu}{2})]+\delta\Delta f_{\Delta}(\delta)[p(\delta+\frac{\mu}{2})+p(-\delta+\frac{\mu}{2})]\\
\geq& 0.
\end{split}
\end{equation}
Hence we have:
$U(\mu)-U(0)\geq0$ when $\mu>0$. Furthermore, considering $\mu_1>\mu_2\geq0$, we have
$$U(\mu_1)-U(\mu_2)=\int_{-\infty}^{+\infty}\delta(f^{\mu_1}_{\Delta}(\delta)-f^{\mu_2}_{\Delta}(\delta))p(\delta)d\delta.$$
Consider $\frac{\mu_1+\mu_2}{2}+\delta$ and $\frac{\mu_1+\mu_2}{2}-\delta$, where $\delta_i>0$. By an abuse of notation,  denote $f^{\mu_1}_{\Delta}(\frac{\mu_1+\mu_2}{2}+\delta)-f^{\mu_2}_{\Delta}(\frac{\mu_1+\mu_2}{2}+\delta)$ as $\Delta f_{\Delta}(\delta)$. We have
\begin{equation}
\begin{split}
&(\frac{\mu_1+\mu_2}{2}+\delta)\Delta fp(\delta+\frac{\mu_1+\mu_2}{2})-(\frac{\mu_1+\mu_2}{2}-\delta)\Delta fp(-\delta+\frac{\mu_1+\mu_2}{2})\\
=&\frac{\mu_1+\mu_2}{2}\Delta f[p(\delta+\frac{\mu_1+\mu_2}{2})-p(-\delta+\frac{\mu_1+\mu_2}{2}]\\&+\delta\Delta f[p(\delta+\frac{\mu_1+\mu_2}{2})+p(-\delta+\frac{\mu_1+\mu_2}{2})]\\
\geq &0.
\end{split}
\end{equation}
It is easy to verify that when $p(\cdot)$ is strictly increasing, we will always have $U(\mu_1)>U(\mu_2)$. Together with Theorem~\ref{theorem4}, we get the desired results.

This completes the proof of Theorem~\ref{theorem6} (3). 
\endproof

\subsubsection{Proof of Theorem~\ref{theorem6} (4)}
\proof{}
As to the (0, $N-1$) fault immune robustness part, we only need to focus on the first term of the cost function under the pricing function of Eq.~(\ref{equation20}). Define
 $$U(\mu_{-j})=\int_{-\infty}^{+\infty}\delta_{j}f^{0}_{\Delta_{j}}(\delta_{j})\int_{-\infty}^{+\infty}p(\delta_i+\delta_{-j})f^{\mu_{-j}}_{\Delta_{-j}}(\delta_{-j})d\delta_{-j}d\delta_{j}.$$
Given $\mu_j=0$, we have
\begin{equation}
\begin{split}
U(-\mu_{-j})&=\int_{-\infty}^{+\infty}\delta_{j}f^{0}_{\Delta_{j}}(\delta_{j})\int_{-\infty}^{+\infty}p(\delta_j+\delta_{-j})f^{-\mu_{-j}}_{\Delta_{-j}}(\delta_{-j})d\delta_{-j}d\delta_{j}\\
&=\int_{-\infty}^{+\infty}(-\delta_{j})f^{0}_{\Delta_{j}}(-\delta_{j})\int_{-\infty}^{+\infty}p(-\delta_j-\delta_{-j})f^{-\mu_{-j}}_{\Delta_{-j}}(-\delta_{-j})d\delta_{-j}d\delta_{j}\\
&=\int_{-\infty}^{+\infty}\delta_{j}f^{0}_{\Delta_{j}}(\delta_{j})\int_{-\infty}^{+\infty}p(\delta_i+\delta_{-j})f^{\mu_{-j}}_{\Delta_{-j}}(\delta_{-j})d\delta_{-j}d\delta_{j}\\
&=U(\mu_{-j}).
\end{split}
\end{equation}
Hence it is sufficient to consider the case when $\mu_{-j}>0$. Considering
\begin{equation}
\begin{split}
U(\mu_{-j})-&U(0)=\int_{-\infty}^{+\infty}\delta_{j}f^{0}_{\Delta_{j}}(\delta_{j})\int_{-\infty}^{+\infty}p(\delta_i+\delta_{-j})\cdot (f^{\mu_{-j}}_{\Delta_{-j}}(\delta_{-j})-f^{0}_{\Delta_{-i}}(\delta_{-j}))d\delta_{-j}d\delta_{j}\\
=&\int_{0}^{+\infty}\delta_{j}f^{0}_{\Delta_{j}}(\delta_{i})\int_{-\infty}^{+\infty}(p(\delta_i+\delta_{-j})-p(-\delta_i+\delta_{-j}))\cdot (f^{\mu_{-j}}_{\Delta_{-j}}(\delta_{-j})-f^{0}_{\Delta_{-j}}(\delta_{-j}))d\delta_{-j}d\delta_{j}.
\end{split}
\end{equation}
Considering two points $\frac{\mu_{-j}}{2}+\delta_{-j}$ and $\frac{\mu_{-j}}{2}-\delta_{-j}$, where $\delta_{-j}>0$. Denote $f^{\mu_{-j}}_{\Delta_{-j}}(\frac{\mu_{-j}}{2}+\delta_{-j})-f^{0}_{\Delta_{-j}}(\frac{\mu_{-j}}{2}+\delta_{-j})$ as $\Delta f_{\Delta_{-j}}(\delta_{-j})$. We have
\begin{equation}
\begin{split}
&\int_{0}^{+\infty}\delta_{j}f^{0}_{\Delta_{j}}(\delta_{j})\int_{-\infty}^{+\infty}(p(\delta_j+\delta_{-j})-p(-\delta_j+\delta_{-j}))\cdot (f^{\mu_{-j}}_{\Delta_{-j}}(\delta_{-j})-f^{0}_{\Delta_{-j}}(\delta_{-j}))d\delta_{-j}d\delta_{j}\\
=&\int_{0}^{+\infty}\delta_{j}f^{0}_{\Delta_{j}}(\delta_{i})\int_{0}^{+\infty}[p(\delta_j+\frac{\mu_{-j}}{2}+\delta_{-j})-p(-\delta_j+\frac{\mu_{-j}}{2}+\delta_{-j})\\&-p(\delta_j+\frac{\mu_{-j}}{2}-\delta_{-j})+p(-\delta_j+\frac{\mu_{-j}}{2}-\delta_{-j})]\Delta f_{\Delta_{-j}}(\delta_{-j})d\delta_{-j}d\delta_{j}\\
\leq& 0.
\end{split}
\end{equation}
The last strictly inequality holds if the strictly inequality in derivative of the pricing function exists, i.e., Eq.~(\ref{equation23}) holds. The decreasing part can be proved by the similar approach without requiring the strictly central
dominant condition to be satisfied. Briefly, consider $\mu_1>\mu_2\geq0$. We have
\begin{equation}
\begin{split}
U(\mu_1)-&U(\mu_2)=\int_{-\infty}^{+\infty}\delta_{j}f^{0}_{\Delta_{j}}(\delta_{j})\int_{-\infty}^{+\infty}p(\delta_i+\delta_{-j})\cdot (f^{\mu_1}_{\Delta_{-j}}(\delta_{-j})-f^{\mu_2}_{\Delta_{-i}}(\delta_{-j}))d\delta_{-j}d\delta_{j}\\
=&\int_{0}^{+\infty}\delta_{j}f^{0}_{\Delta_{j}}(\delta_{i})\int_{-\infty}^{+\infty}(p(\delta_i+\delta_{-j})-p(-\delta_i+\delta_{-j}))\cdot (f^{\mu_{1}}_{\Delta_{-j}}(\delta_{-j})-f^{\mu_{2}}_{\Delta_{-j}}(\delta_{-j}))d\delta_{-j}d\delta_{j}.
\end{split}
\end{equation}
Considering two points $\frac{\mu_1+\mu_2}{2}+\delta_{-j}$ and $\frac{\mu_1+\mu_2}{2}-\delta_{-j}$, where $\delta_{-j}>0$. By an abuse of notation, denote $f^{\mu_1}_{\Delta_{-j}}(\frac{\mu_1+\mu_2}{2}+\delta_{-j})-f^{\mu_2}_{\Delta_{-j}}(\frac{\mu_1+\mu_2}{2}+\delta_{-j})$ as $\Delta f_{\Delta_{-j}}(\delta_{-j})$. We have
\begin{equation}
\begin{split}
&\int_{0}^{+\infty}\delta_{j}f^{0}_{\Delta_{j}}(\delta_{j})\int_{-\infty}^{+\infty}(p(\delta_j+\delta_{-j})-p(-\delta_j+\delta_{-j}))\cdot (f^{\mu_1}_{\Delta_{-j}}(\delta_{-j})-f^{\mu_2}_{\Delta_{-j}}(\delta_{-j}))d\delta_{-j}d\delta_{j}\\
=&\int_{0}^{+\infty}\delta_{j}f^{0}_{\Delta_{j}}(\delta_{i})\int_{0}^{+\infty}[p(\delta_j+\frac{\mu_1+\mu_2}{2}+\delta_{-j})-p(-\delta_j+\frac{\mu_1+\mu_2}{2}+\delta_{-j})\\&-p(\delta_j+\frac{\mu_1+\mu_2}{2}-\delta_{-j})+p(-\delta_j+\frac{\mu_1+\mu_2}{2}-\delta_{-j})]\Delta f_{\Delta_{-j}}(\delta_{-j})d\delta_{-j}d\delta_{j}\\
\leq& 0.
\end{split}
\end{equation}
The last strictly inequality holds if the strictly inequality in derivative of the pricing function exists, i.e., Eq.~(\ref{equation23}) holds.

This completes the proof of Theorem~\ref{theorem6} (4). \endproof

Thus we complete the proof of Theorem~\ref{theorem6}.
\endproof

\subsection{Proof of Theorem~\ref{theorem7}} 
\proof{}
Previous analysis focuses on the scenario that the load forecasting
error of utilities are mutually independent. In this section, we relax
this assumption to consider more general case, i.e., the load forecasting
error of utilities are correlated. More specifically, we assume the
load forecasting error of utility $i$, denoted as $\Delta_{i}$,
follows Gaussian distribution with mean $\mu_{i}$ and variance $\sigma_{i}^{2}$
and they are jointly normal. For simplicity, we assume that the correlation
between $\Delta_{i}$ and $\Delta_{j}$ are $\rho_{ij}$ for any $i$
and $j$ and the covariance matrix of random vector $[\Delta_{1},\Delta_{2},...,\Delta_{N}]$
can be expressed as:
\[
\left[\begin{array}{cccc}
\sigma_{1}^{2} & \rho_{12}\sigma_{1}\sigma_{2} & \cdots & \rho_{1N}\sigma_{1}\sigma_{N}\\
\rho_{12}\sigma_{1}\sigma_{2} & \sigma_{2}^{2} & \cdots & \rho_{2N}\sigma_{2}\sigma_{N}\\
\vdots & \vdots & \ddots & \vdots\\
\rho_{1N}\sigma_{1}\sigma_{N} & \rho_{2N}\sigma_{2}\sigma_{N} & \cdots & \sigma_{N}^{2}
\end{array}\right].
\]
For simplicity, let us consider $\rho_{ij}\ge0,\forall i,j$, i.e.,
we assume there are positive correlation among utilities. 

Under these assumptions, we define $\Delta_{-i}=\sum_{j,j\neq i}^{N}\Delta_{j}$
and have
\[
\Delta_{-i}\sim\mathcal{N}(\mu_{-i},\sigma_{-i}^{2}),
\]
where $\mu_{-i}=\sum_{j\neq i}\mu_{j}$ and $\sigma_{-i}^{2}=\sum_{j\neq i}\sigma_{j}^{2}+2\sum_{1\le j_{1}<j_{2}\leq N,j_{1}\neq i,j_{2}\neq i}\rho_{j_{1}j_{2}}\sigma_{j_{1}}\sigma_{j_{2}}$.

Obviously, $[\Delta_{i},\Delta_{-i}]$ follows bivariate normal distribution
and their joint PDF is 
\begin{equation}
\begin{split}
f(\delta_{i},\delta_{-i})=&\frac{1}{2\pi\sigma_{i}\sigma_{-i}\sqrt{1-\rho_{i}^{2}}}exp(-\frac{1}{2(1-\rho_{i}^{2})}[\frac{(\delta_{i}-\mu_{i})^{2}}{\sigma_{i}^{2}}\\&+\frac{(\delta_{-i}-\mu_{-i})^{2}}{\sigma_{-i}^{2}}-\frac{2\rho_{i}(\delta_{i}-\mu_{i})(\delta_{-i}-\mu_{-i})}{\sigma_{i}\sigma_{-i}}]),\label{eq:jointPDF-1}
\end{split}
\end{equation}
where $\rho_{i}$ is the correlation between $\Delta_{i}$ and $\Delta_{-i}$
and it can be expressed as
\begin{align*}
\rho_{i} & =\frac{\mathbb{E}[\Delta_{i}\Delta_{-i}]-\mathbb{E}[\Delta_{i}]\mathbb{E}[\Delta_{-i}]}{\sigma_{i}\sigma_{-i}}\\
 & =\frac{\sum_{j\neq i}(\rho_{ij}\sigma_{i}\sigma_{j}+\mu_{i}\mu_{j})-\mu_{i}(\sum_{j\neq i}\mu_{j})}{\sigma_{i}\sqrt{\sum_{j\neq i}\sigma_{j}^{2}+2\sum_{1\le j_{1}<j_{2}\leq N,j_{1}\neq i,j_{2}\neq i}\rho_{j_{1}j_{2}}\sigma_{j_{1}}\sigma_{j_{2}}}}\\
 & =\frac{\sum_{j\neq i}\rho_{ij}\sigma_{j}}{\sqrt{\sum_{j\neq i}\sigma_{j}^{2}+2\sum_{1\le j_{1}<j_{2}\leq N,j_{1}\neq i,j_{2}\neq i}\rho_{j_{1}j_{2}}\sigma_{j_{1}}\sigma_{j_{2}}}}.
\end{align*}
The following theorem characters the expected $\textsf{ABC}$s of
utilities.
\begin{theorem}\label{s-theorem8}
When the pricing model is symmetric, i.e., $a_{1}=a_{2},$ $b_{1}+b_{2}=2$,
the expected $\textsf{ABC}$ of utility $i$ can be expressed as:
\begin{align*}
\mathbb{E}[\textsf{ABC}_{i}] & =p_{d}+\frac{p_{d}}{D_{i}}[\frac{a_{1}+a_{2}}{2}(\sigma_{i}^{2}+\mu_{i}^{2}+\mu_{i}\mu_{-i}+\rho_{i}\sigma_{i}\sigma_{-i})\\
 & \quad+\frac{b_{1}-b_{2}}{2}\mathbb{E}[\Delta_{i}erf(\frac{\frac{\Delta_{i}+\mu_{-i}}{\sigma_{-i}}+\rho_{i}\frac{\Delta_{i}-\mu_{i}}{\sigma_{i}}}{\sqrt{2(1-\rho_{i}^{2})}})]].
\end{align*}
\end{theorem}
\proof{}
We know that
\begin{equation}
\mathbb{E}[\mathsf{ABC}_{i}]=p_{d}+\frac{\mathbb{E}[p_{s}\cdot\Delta_{i}]}{D_{i}}-\frac{p_{d}}{D_{i}}\mu_{i}.\label{eq:theorem5_1-1}
\end{equation}
where $\mathbb{E}[p_{s}\cdot\Delta_{i}]$ can be expressed as:
\begin{equation}
\mathbb{E}[p_{s}\cdot\Delta_{i}]=\mathbb{E}[\xi_{1}\Delta_{i}^{2}+\xi_{1}\Delta_{i}\Delta_{-i}+\xi_{2}\Delta_{i}].\label{eq:theorem5_2-1}
\end{equation}
It then remains to compute $\mathbb{E}[\xi_{1}\Delta_{i}^{2}]$, $\mathbb{E}[\xi_{1}\Delta_{i}\Delta_{-i}]$,
and $\mathbb{E}[\xi_{2}\Delta_{i}]$.

Let $f(\delta_{i},\delta_{-i})$ in Eq. (\ref{eq:jointPDF-1}) be
the joint PDF of $\Delta_{i}$ and $\Delta_{-i}$. Let us first compute
$\mathbb{E}[\xi_{1}\Delta_{i}^{2}]$:
\begin{align*}
\mathbb{E}[\xi_{1}\Delta_{i}^{2}]=&  \int_{-\infty}^{+\infty}\int_{-\infty}^{+\infty}\mathbb{E}[\xi_{1}\Delta_{i}^{2}|\Delta_{i}=\delta_{i},\Delta_{-i}=\delta_{-i}]\cdot f(\delta_{i},\delta_{-i})d\delta_{i}d\delta_{-i}\\
= & \int_{-\infty}^{+\infty}\delta_{i}^{2}\left(\int_{-\delta_{i}}^{+\infty}a_{1}p_{d}f(\delta_{i},\delta_{-i})d\delta_{-i}+\int_{-\infty}^{-\delta_{i}}a_{2}p_{d}f(\delta_{i},\delta_{-i})d\delta_{-i}\right)d\delta_{i}\\
= & \int_{-\infty}^{+\infty}\delta_{i}^{2}\left(a_{1}p_{d}f_{\Delta_{i}}(\delta_{i})(\frac{1}{2}+\frac{1}{2}erf(\frac{\frac{\delta_{i}+\mu_{-i}}{\sigma_{-i}}+\rho_{i}\frac{\delta_{i}-\mu_{i}}{\sigma_{i}}}{\sqrt{2(1-\rho_{i}^{2})}}))\right.\\
 & \left.+a_{2}p_{d}f_{\Delta_{i}}(\delta_{i})(\frac{1}{2}-\frac{1}{2}erf(\frac{\frac{\delta_{i}+\mu_{-i}}{\sigma_{-i}}+\rho_{i}\frac{\delta_{i}-\mu_{i}}{\sigma_{i}}}{\sqrt{2(1-\rho_{i}^{2})}}))\right)d\delta_{i}\\
= & \frac{a_{1}+a_{2}}{2}p_{d}(\sigma_{i}^{2}+\mu_{i}^{2}).
\end{align*}

Secondly, let us compute $\mathbb{E}[\xi_{2}\Delta_{i}]:$
\begin{align*}
\mathbb{E}&[\xi_{2}\Delta_{i}]=  \int_{-\infty}^{+\infty}\int_{-\infty}^{+\infty}\mathbb{E}[\xi_{2}\Delta_{i}|\Delta_{i}=\delta_{i},\Delta_{-i}=\delta_{-i}]\cdot f(\delta_{i},\delta_{-i})d\delta_{i}d\delta_{-i}\\
= & \int_{-\infty}^{+\infty}\delta_{i}\left(\int_{-\delta_{i}}^{+\infty}b_{1}p_{d}f(\delta_{i},\delta_{-i})d\delta_{-i}+\int_{-\infty}^{-\delta_{i}}b_{2}p_{d}f(\delta_{i},\delta_{-i})d\delta_{-i}\right)d\delta_{i}\\
= & \int_{-\infty}^{+\infty}\delta_{i}\left(b_{1}p_{d}f_{\Delta_{i}}(\delta_{i})(\frac{1}{2}+\frac{1}{2}erf(\frac{\frac{\delta_{i}+\mu_{-i}}{\sigma_{-i}}+\rho_{i}\frac{\delta_{i}-\mu_{i}}{\sigma_{i}}}{\sqrt{2(1-\rho_{i}^{2})}}))\right.\\
 & \left.+b_{2}p_{d}f_{\Delta_{i}}(\delta_{i})(\frac{1}{2}-\frac{1}{2}erf(\frac{\frac{\delta_{i}+\mu_{-i}}{\sigma_{-i}}+\rho_{i}\frac{\delta_{i}-\mu_{i}}{\sigma_{i}}}{\sqrt{2(1-\rho_{i}^{2})}}))\right)d\delta_{i}\\
= & \frac{b_{1}+b_{2}}{2}p_{d}\mu_{i}+\frac{b_{1}-b_{2}}{2}p_{d}\mathbb{E}[\Delta_{i}erf(\frac{\frac{\Delta_{i}+\mu_{-i}}{\sigma_{-i}}+\rho_{i}\frac{\Delta_{i}-\mu_{i}}{\sigma_{i}}}{\sqrt{2(1-\rho_{i}^{2})}})].
\end{align*}

Lastly, let us compute $\mathbb{E}[\xi_{1}\Delta_{i}\Delta_{-i}]$.
For ease of presentation, let us denote $A=\frac{\frac{\delta_{i}+\mu_{-i}}{\sigma_{-i}}+\rho_{i}\frac{\delta_{i}-\mu_{i}}{\sigma_{i}}}{\sqrt{2(1-\rho_{i}^{2})}}$. We have
\begin{align*}
\mathbb{E}[\xi_{1}\Delta_{i}\Delta_{-i}]=&
  \int_{-\infty}^{+\infty}\int_{-\infty}^{+\infty}\mathbb{E}[\xi_{1}\Delta_{i}\Delta_{-i}|\Delta_{i}=\delta_{i},\Delta_{-i}=\delta_{-i}]\cdot f(\delta_{i},\delta_{-i})d\delta_{i}d\delta_{-i}\\
= & \int_{-\infty}^{+\infty}\delta_{i}\left(a_{1}p_{d}f_{\Delta_{i}}(\delta_{i})(\frac{\mu_{-i}+\sigma_{-i}\rho_{i}\frac{\delta_{i}-\mu_{i}}{\sigma_{i}}}{2}(1+erf(A))+\sigma_{-i}\sqrt{\frac{1-\rho_{i}^{2}}{2\pi}}e^{-\frac{1}{2(1-\rho_{i}^{2})}A^{2}})\right.\\
 & \left.+a_{2}p_{d}f_{\Delta_{i}}(\delta_{i})(\frac{\mu_{-i}+\sigma_{-i}\rho_{i}\frac{\delta_{i}-\mu_{i}}{\sigma_{i}}}{2}(1-erf(A))-\sigma_{-i}\sqrt{\frac{1-\rho_{i}^{2}}{2\pi}}e^{-\frac{1}{2(1-\rho_{i}^{2})}A^{2}}\right)d\delta_{i}\\
= & \frac{a_{1}+a_{2}}{2}p_{d}\int_{-\infty}^{+\infty}\delta_{i}f_{\Delta_{i}}(\delta_{i})(\mu_{-i}+\sigma_{-i}\rho_{i}\frac{\delta_{i}-\mu_{i}}{\sigma_{i}})d\delta_{i}\\
= & \frac{a_{1}+a_{2}}{2}p_{d}(\mu_{i}\mu_{-i}+\rho_{i}\sigma_{i}\sigma_{-i}).
\end{align*}

Then plug in Eqs.~(\ref{eq:theorem5_1-1}) and (\ref{eq:theorem5_2-1}),
we can get the final result. 
\endproof
What we need to prove is that, given $\mu_{-i}=0$, the optimal $\mu_{i}^{*}$ that minimizes
$\mathbb{E}[\mathbf{\textsf{ABC}}_{i}]$ is $0$, and $\mathbb{E}[\mathbf{\textsf{ABC}}_{i}]$ is strictly increasing w.r.t. $|\mu_i|$.
\proof{}
Given $\mu_{-i}=0$, the expected ABC of utility $i$ can be simplified
as:
\begin{align*}
\mathbb{E}[\textsf{ABC}_{i}] & =p_{d}+\frac{p_{d}}{D_{i}}[\frac{a_{1}+a_{2}}{2}(\sigma_{i}^{2}+\mu_{i}^{2}+\rho_{i}\sigma_{i}\sigma_{-i})\\
 & \quad+\frac{b_{1}-b_{2}}{2}\mathbb{E}[\Delta_{i}erf(\frac{\frac{\Delta_{i}}{\sigma_{-i}}+\rho_{i}\frac{\Delta_{i}-\mu_{i}}{\sigma_{i}}}{\sqrt{2(1-\rho_{i}^{2})}})]].
\end{align*}

To prove that $\mathbb{E}[\textsf{ABC}_{i}]$ obtains its minimum
when $\mu_{i}=0,$ we rewrite it as
\[
\mathbb{E}[\mathbf{\textsf{ABC}}_{i}]=g_{1}(\mu_{i})+g_{2}(\mu_{i}),
\]
where 
\[
g_{1}(\mu_{i})=p_{d}+\frac{p_{d}}{D_{i}}[\frac{a_{1}+a_{2}}{2}(\sigma_{i}^{2}+\mu_{i}^{2}+\rho_{i}\sigma_{i}\sigma_{-i})],
\]
and 
\[
g_{2}(\mu_{i})=\frac{p_{d}}{D_{i}}\frac{b_{1}-b_{2}}{2}\mathbb{E}[\Delta_{i}erf(\frac{\frac{\Delta_{i}}{\sigma_{-i}}+\rho_{i}\frac{\Delta_{i}-\mu_{i}}{\sigma_{i}}}{\sqrt{2(1-\rho_{i}^{2})}})].
\]
It is straightforward to verify that $argmin_{\mu_{i}}g_{1}(\mu_{i})=0$ and it is increasing w.r.t.$|\mu_i|$.
So it remains to show that $argmin_{\mu_{i}}g_{2}(\mu_{i})=0$ and it is increasing w.r.t $|\mu_i|$. Since
$b_{1}>b_{2}$, it is equivalent to show
\[
argmin_{\mu_{i}}\mathbb{E}[\Delta_{i}erf(\frac{\frac{\Delta_{i}}{\sigma_{-i}}+\rho_{i}\frac{\Delta_{i}-\mu_{i}}{\sigma_{i}}}{\sqrt{2(1-\rho_{i}^{2})}})]=0.
\]

Denote $U(\mu_{i})=\mathbb{E}[\Delta_{i}erf(\frac{\frac{\Delta_{i}}{\sigma_{-i}}+\rho_{i}\frac{\Delta_{i}-\mu_{i}}{\sigma_{i}}}{\sqrt{2(1-\rho_{i}^{2})}})]$.
It remains to prove that $argmin_{\mu_{i}}U(\mu_{i})=0$.

Firstly, we note that $U(\mu_{i})$ is an even function. To see this,
consider the following equations:
\begin{align*}
U(-\mu_{i}) & =\frac{1}{\sqrt{2\pi}\sigma_{i}}\int_{-\infty}^{+\infty}x\cdot erf(\frac{\frac{x}{\sigma_{-i}}+\rho_{i}\frac{x+\mu_{i}}{\sigma_{i}}}{\sqrt{2(1-\rho_{i}^{2})}})e^{-\frac{(x+\mu_{i})^{2}}{2\sigma_{i}^{2}}}dx\\
 & =\frac{1}{\sqrt{2\pi}\sigma_{i}}\int_{-\infty}^{+\infty}(-x)\cdot erf(-\frac{\frac{x}{\sigma_{-i}}+\rho_{i}\frac{x-\mu_{i}}{\sigma_{i}}}{\sqrt{2(1-\rho_{i}^{2})}})\cdot e^{-\frac{(-x+\mu_{i})^{2}}{2\sigma_{i}^{2}}}dx\\
 & =U(\mu_{i}).
\end{align*}
Then it remains to prove that $\forall\mu>0$, $U(\mu)-U(0)>0$.
When $\mu>0$, we have
\begin{align*}
&U(\mu)-U(0)   =\mathbb{E}[X_{\mu}erf(\frac{\frac{X_{\mu}}{\sigma_{-i}}+\rho_{i}\frac{X_{\mu}-\mu}{\sigma_{i}}}{\sqrt{2(1-\rho_{i}^{2})}})]-\mathbb{E}[X_{0}erf(\frac{\frac{X_{0}}{\sigma_{-i}}+\rho_{i}\frac{X_{0}}{\sigma_{i}}}{\sqrt{2(1-\rho_{i}^{2})}})],
\end{align*}
where $X_{\mu}\sim\mathcal{N}(\mu,\sigma_{i}^{2})$ and $X_{0}\sim\mathcal{N}(0,\sigma_{i}^{2})$.

Denote $\mu_{0}=\frac{\rho_{i}\sigma_{-i}}{\sigma_{i}+\rho\sigma_{-i}}\mu$,
obviously $\mu_{0}>0$ and 
\begin{equation}
\begin{split}
\mathbb{E}[X_{\mu}erf(\frac{\frac{X_{\mu}}{\sigma_{-i}}+\rho_{i}\frac{X_{\mu}-\mu}{\sigma_{i}}}{\sqrt{2(1-\rho_{i}^{2})}})] & =\mathbb{E}[(X_{\mu}-\mu_{0})erf(\frac{\frac{X_{\mu}}{\sigma_{-i}}+\rho_{i}\frac{X_{\mu}-\mu}{\sigma_{i}}}{\sqrt{2(1-\rho_{i}^{2})}})]\\&\quad+\mu_{0}\mathbb{E}[erf(\frac{\frac{X_{\mu}}{\sigma_{-i}}+\rho_{i}\frac{X_{\mu}-\mu}{\sigma_{i}}}{\sqrt{2(1-\rho_{i}^{2})}})]\\
 & =\mathbb{E}[(X_{\mu}-\mu_{0})erf(\frac{(\frac{1}{\sigma_{-i}}+\rho_{i}\frac{1}{\sigma_{i}})(X_{\mu}-\mu_{0})}{\sqrt{2(1-\rho_{i}^{2})}})]\\&\quad+\mu_{0}\mathbb{E}[erf(\frac{(\frac{1}{\sigma_{-i}}+\rho_{i}\frac{1}{\sigma_{i}})(X_{\mu}-\mu_{0})}{\sqrt{2(1-\rho_{i}^{2})}})].
\end{split}
\end{equation}

Denote random variable $Y=X_{\mu}-\mu_{0}$, then $Y\sim\mathcal{N}(\mu_{y},\sigma_{i}^{2})$
and $\mu_{y}=\mu-\mu_{0}>0$. Further, let $k=\frac{\frac{1}{\sigma_{-i}}+\rho_{i}\frac{1}{\sigma_{i}}}{\sqrt{2(1-\rho_{i}^{2})}}>0$,
we have 
\[
U(\mu)-U(0)=\mathbb{E}[Yerf(kY)]+\mu_{0}\mathbb{E}[erf(kY)]-\mathbb{E}[X_{0}erf(kX_{0})].
\]

Since $\mu_{y}>0$, we know that
$\mathbb{E}[Yerf(kY)]-\mathbb{E}[X_{0}erf(kX_{0})]>0$ and larger $|\mu|$ is, larger the difference. Further,
since $\mathbb{E}[erf(k(Y-\mu_{y}))]=0$, we have that 
\begin{align*}
\mathbb{E}[erf(kY)]  &=\mathbb{E}[erf(kY)]-\mathbb{E}[erf(k(Y-\mu_{y}))]\\
 & =\mathbb{E}[erf(kY)-erf(k(Y-\mu_{y}))]\\
 & =\frac{1}{\sqrt{2\pi}\sigma_{i}}\int_{-\infty}^{+\infty}(erf(ky)-erf(k(y-\mu_{y})))e^{-\frac{(y-\mu_{y})^{2}}{2\sigma_{i}^{2}}}dy\\
 & >0,
\end{align*}
where the last inequality follows from the fact that $erf(x)$ is
increasing in $x$ and $erf(ky)-erf(k(y-\mu_{y}))\geq0$. In addition, we have when $|\mu|$ increases, $\mu_{0}\mathbb{E}[erf(kY)]$ increases. Then we
prove that for any $\mu>0$, 
\[
U(\mu)-U(0)>0.
\]
Thus $argmin_{\mu_{i}}U(\mu_{i})=0$. We also have larger $|\mu|$ is, larger the difference. 
\endproof

The efficiency part is the same as the independent case in Theorem~\ref{theorem4}.

As to the (0, $N-1$) fault immune robustness part, we have: given $\mu_{i}=0$, the optimal $\mu_{-i}^{*}$ that maximize
$\mathbb{E}[\mathbf{\textsf{ABC}}_{i}]$ is $0$.
\proof{}
When $a_{1}=a_{2},$ $b_{1}+b_{2}=2$, given $\mu_{i}=0$, the expectation
of $\mathbf{\textsf{ABC}}_{i}$ can be expressed as 
\begin{align*}
\mathbb{E}[\textsf{ABC}_{i}] & =p_{d}+\frac{p_{d}}{D_{i}}[\frac{a_{1}+a_{2}}{2}(\sigma_{i}^{2}+\rho_{i}\sigma_{i}\sigma_{-i})\\
 & \quad+\frac{b_{1}-b_{2}}{2}\mathbb{E}[\Delta_{i}erf(\frac{\frac{\Delta_{i}+\mu_{-i}}{\sigma_{-i}}+\rho_{i}\frac{\Delta_{i}}{\sigma_{i}}}{\sqrt{2(1-\rho_{i}^{2})}})]]\\
 & =p_{d}+\frac{p_{d}}{D_{i}}[\frac{a_{1}+a_{2}}{2}(\sigma_{i}^{2}+\rho_{i}\sigma_{i}\sigma_{-i})\\
 & \quad+\frac{b_{1}-b_{2}}{2}\mathbb{E}[\Delta_{i}erf(k_{1}(\Delta_{i}+k_{2}\mu_{-i}))]],
\end{align*}
where
\[
k_{1}=\frac{\frac{1}{\sigma_{-i}}+\rho_{i}\frac{1}{\sigma_{i}}}{\sqrt{2(1-\rho_{i}^{2})}},k_{2}=\frac{\sigma_{i}}{\sigma_{i}+\rho_{i}\sigma_{-i}}.
\]

We prove that $\mathbb{E}[\mathbf{\textsf{ABC}}_{i}]$ attains its maximum at $\mu_{-i}=0$. Note that $\frac{d}{dz}erf(z)=\frac{2}{\sqrt{\pi}}e^{-z^{2}}$,
the first order derivative of $\mathbb{E}[\mathbf{\textsf{ABC}}_{i}]$
w.r.t. $\mu_{-i}$ can be expressed as
\[
\frac{\partial\mathbb{E}[\mathbf{\textsf{ABC}}_{i}]}{\partial\mu_{-i}}=\frac{p_{d}}{D_{i}}\frac{b_{1}-b_{2}}{2}k_{1}k_{2}\frac{2}{\sqrt{\pi}}\mathbb{E}[\Delta_{i}e^{-k_{1}^{2}(\Delta_{i}+k_{2}\mu_{-i})^{2}}].
\]
When $\mu_{-i}=0,$ we can get that 
\[
\frac{\partial\mathbb{E}[\mathbf{\textsf{ABC}}_{i}]}{\partial\mu_{-i}}=\frac{p_{d}}{D_{i}}\frac{b_{1}-b_{2}}{2\sigma_{-i}}k_{1}k_{2}\frac{2}{\sqrt{\pi}}\mathbb{E}[\Delta_{i}e^{-k_{1}^{2}\Delta_{i}^{2}}]=0,
\]
where the equality follows since $g(\delta_{i})=\delta_{i}e^{-k_{1}^{2}\delta_{i}^{2}}$
is an odd function, and the PDF of $\Delta_{i}$ is even function,
thus $\mathbb{E}[\Delta_{i}e^{-\frac{\Delta_{i}^{2}}{2\sigma_{-i}^{2}}}]=0$.
In order to prove that $\mu_{-i}=0$ is the maximum, we need to show
that $\frac{\partial\mathbb{E}[\mathbf{\textsf{ABC}}_{i}]}{\partial\mu_{-i}}>0$
for $\mu_{i}<0$ and $\frac{\partial\mathbb{E}[\mathbf{\textsf{ABC}}_{i}]}{\partial\mu_{-i}}<0$
for $\mu_{i}>0$. To prove this, we have
\begin{align*}
\mathbb{E}[\Delta_{i}e^{-k_{1}^{2}(\Delta_{i}+k_{2}\mu_{-i})^{2}}]  &=\int_{-\infty}^{+\infty}\delta_{i}e^{-k_{1}^{2}(\delta_{i}+k_{2}\mu_{-i})^{2}-\frac{\delta_{i}^{2}}{2\sigma_{i}^{2}}}d\delta_{i}\\
 & =\int_{-\infty}^{+\infty}\delta_{i}e^{-\frac{\delta_{i}^{2}+2\sigma_{i}^{2}k_{1}^{2}(\delta_{i}+k_{2}\mu_{-i})^{2}}{2\sigma_{i}^{2}}}d\delta_{i}\\
 & =\int_{-\infty}^{+\infty}\delta_{i}e^{-\frac{\sigma_{i}^{2}+2\rho_{i}\sigma_{i}\sigma_{-i}+\sigma_{-i}^{2}}{2(1-\rho_{i}^{2})\sigma_{-i}^{2}\sigma_{i}^{2}}(\delta_{i}+\frac{\mu_{-i}\sigma_{i}(\sigma_{i}+\rho_{i}\sigma_{-i})}{\sigma_{i}^{2}+2\rho_{i}\sigma_{i}\sigma_{-i}+\sigma_{-i}^{2}})^{2}}d\delta_{i}\cdot e^{-\frac{\mu_{-i}^{2}}{2(\sigma_{i}^{2}+2\rho_{i}\sigma_{i}\sigma_{-i}+\sigma_{-i}^{2})}}\\
 & =e^{-\frac{\mu_{-i}^{2}}{2(\sigma_{i}^{2}+2\rho_{i}\sigma_{i}\sigma_{-i}+\sigma_{-i}^{2})}}(-\frac{\mu_{-i}\sigma_{i}(\sigma_{i}+\rho_{i}\sigma_{-i})}{\sigma_{i}^{2}+2\rho_{i}\sigma_{i}\sigma_{-i}+\sigma_{-i}^{2}})\cdot \sqrt{\frac{2\pi\sigma_{-i}^{2}\sigma_{i}^{2}(1-\rho_{i}^{2})}{\sigma_{i}^{2}+2\rho_{i}\sigma_{i}\sigma_{-i}+\sigma_{-i}^{2}}},
\end{align*}
 where the last equality follows by the fact that 
\[
\int_{-\infty}^{+\infty}xe^{-a(x-b)^{2}}dx=b\sqrt{\frac{\pi}{a}}.
\]

Further, we note that 
\[
\sigma_{i}+\rho_{i}\sigma_{-i}\ge0,
\]
thus it is straightforward to verify that $\frac{\partial\mathbb{E}[\mathbf{\textsf{ABC}}_{i}]}{\partial\mu_{-i}}>0,\forall\mu_{-i}<0$
and $\frac{\partial\mathbb{E}[\mathbf{\textsf{ABC}}_{i}]}{\partial\mu_{-i}}<0,\forall\mu_{-i}>0$.
Thus $\mu_{-i}=0$ is the maximum point of $\mathbb{E}[\mathbf{ABC}_{i}]$ and $\mathbb{E}[\mathbf{ABC}_{i}]$ is decreasing w.r.t. $|\mu_{-i}|$.
\endproof

We then prove the uniqueness of the Nash Equilibrium.
\proof{}
As to the unique part, assume $\mu_{-i}>0$, we have
\begin{align*}
U(\mu_{i}) & =\int_{-\infty}^{+\infty}x\cdot erf(\frac{\frac{x+\mu_{-i}}{\sigma_{-i}}+\rho_{i}\frac{x-\mu_{i}}{\sigma_{i}}}{\sqrt{2(1-\rho_{i}^{2})}})e^{-\frac{(x+\mu_{i})^{2}}{2\sigma_{i}^{2}}}dx\\
 &=\int_{-\infty}^{+\infty}x\cdot erf(\frac{(\frac{1}{\sigma_{-i}}+\frac{\rho_{i}}{\sigma_{i}})(x-\frac{\frac{\rho_{i}}{\sigma_{i}}\mu_{i}}{\frac{1}{\sigma_{-i}}+\frac{\rho_{i}}{\sigma_{i}}})+\frac{\mu_{-i}}{\sigma_{-i}}}{\sqrt{2(1-\rho_{i}^{2})}})e^{-\frac{(x+\mu_{i})^{2}}{2\sigma_{i}^{2}}}dx.
\end{align*}
Let $t=x-\frac{\frac{\rho_{i}}{\sigma_{i}}\mu_{i}}{\frac{1}{\sigma_{-i}}+\frac{\rho_{i}}{\sigma_{i}}}$, $k_1=\frac{\frac{1}{\sigma_{-i}}}{\sqrt{2(1-\rho_{i}^{2})}}$, and $k_2=\frac{\frac{\rho_{i}}{\sigma_{i}}}{\sqrt{2(1-\rho_{i}^{2})}}$.

Then
\begin{align*}
U(\mu_{i})  =&\int_{-\infty}^{+\infty}(t+\frac{k_2}{k_1+k_2}\mu_i)\cdot erf((k_1+k_2)t +k_1\mu_{-i})e^{-\frac{(t-\frac{k_1}{k_1+k_2}\mu_i)^{2}}{2\sigma_{i}^{2}}}dx\\
 =&\int_{-\infty}^{+\infty}t\cdot erf((k_1+k_2)t+k_1\mu_{-i})e^{-\frac{(t-\frac{k_1}{k_1+k_2}\mu_i)^{2}}{2\sigma_{i}^{2}}}dx\\
&+\int_{-\infty}^{+\infty}\frac{k_2}{k_1+k_2}\mu_i\cdot erf((k_1+k_2)t+k_1\mu_{-i})e^{-\frac{(t-\frac{k_1}{k_1+k_2}\mu_i)^{2}}{2\sigma_{i}^{2}}}dx.
\end{align*}
For the first term, we have
\begin{align*}
\frac{dU_1(\mu_i)}{d\mu_i}=&\frac{k_1}{k_1+k_2}\int_{-\infty}^{+\infty}t(t-\frac{k_1}{k_1+k_2}\mu_i)e^{-\frac{(t-\frac{k_1}{k_1+k_2}\mu_i)^{2}}{2\sigma_{i}^{2}}}\cdot erf((k_1+k_2)t+k_1\mu_{-i})dt.
\end{align*}
For the second term, we have
\begin{align*}
&\frac{dU_2(\mu_i)}{d\mu_i}=\frac{k_2}{k_1+k_2}\int_{-\infty}^{+\infty}e^{-\frac{(t-\frac{k_1}{k_1+k_2}\mu_i)^{2}}{2\sigma_{i}^{2}}}erf((k_1+k_2)t+k_1\mu_{-i})dt\\
 &+\frac{k_1k_2}{(k_1+k_2)^2}\mu_i\int_{-\infty}^{+\infty}(t-\frac{k_1}{k_1+k_2}\mu_i)\cdot e^{-\frac{(t-\frac{k_1}{k_1+k_2}\mu_i)^{2}}{2\sigma_{i}^{2}}}erf((k_1+k_2)t+k_1\mu_{-i})dt.
\end{align*}
It is easy to prove that 

i) when $\mu_i=0$, we have
\begin{align*}
&\left.\frac{dU_1(\mu_i)}{d\mu_i}\right|_{\mu_i=0} =\frac{k_1}{k_1+k_2}\int_{-\infty}^{+\infty}t^2e^{-\frac{t^{2}}{2\sigma_{i}^{2}}}erf((k_1+k_2)t+k_1\mu_{-i})dt>0,
\end{align*}
and
\begin{align*}
&\left.\frac{dU_2(\mu_i)}{d\mu_i}\right|_{\mu_i=0}=\frac{k_2}{k_1+k_2}\int_{-\infty}^{+\infty}e^{-\frac{t^{2}}{2\sigma_{i}^{2}}}erf((k_1+k_2)t+k_1\mu_{-i})dt>0;
\end{align*}

ii) when $\mu_i=-\mu_{-i}$, we have
\begin{align*}
\left.\frac{dU_1(\mu_i)}{d\mu_i}\right|_{\mu_i=-\mu_{-i}}&\hspace{-0.5em}=\frac{k_1}{k_1+k_2}\int_{-\infty}^{+\infty}t(t+\frac{k_1}{k_1+k_2}\mu_{-i})\cdot e^{-\frac{(t+\frac{k_1}{k_1+k_2}\mu_{-i})^{2}}{2\sigma_{i}^{2}}}erf((k_1+k_2)t+k_1\mu_{-i})dt<0,
\end{align*}
and
\begin{align*}
\begin{split}
\left.\frac{dU_2(\mu_i)}{d\mu_i}\right|_{\mu_i=-\mu_{-i}}&\hspace{-0.5em}=\frac{k_1k_2}{(k_1+k_2)^2}(-\mu_{-i})\hspace{-0.3em}\int_{-\infty}^{+\infty}(t+\frac{k_1}{k_1+k_2}\mu_{-i})\cdot e^{-\frac{(t+\frac{k_1}{k_1+k_2}\mu_{-i})^{2}}{2\sigma_{i}^{2}}} erf((k_1+k_2)t+k_1\mu_{-i})dt\\&\hspace{-0.5em}<0.
\end{split}
\end{align*}

It remains to show that when $\mu_i>0$, $\frac{dU(\mu_i)}{d\mu_i}>0$ and  when $\mu_i<-\mu_{-i}$, $\frac{dU(\mu_i)}{d\mu_i}<0$;

iii) when $\mu_{i}>0$, let $z=t-\frac{k_1}{k_1+k_2}\mu_i$, we have
\begin{align*}
&\frac{dU_1(\mu_i)}{d\mu_i}=\frac{k_1}{k_1+k_2}\int_{-\infty}^{+\infty}(z+\frac{k_1}{k_1+k_2}\mu_i)\cdot ze^{-\frac{z^{2}}{2\sigma_{i}^{2}}}erf((k_1+k_2)((k_1+k_2)z+k_1(\mu_i+\mu_{-i}))dz>0,
\end{align*}
and
\begin{align*}
\frac{dU_2(\mu_i)}{d\mu_i} =&\frac{k_2}{k_1+k_2}\int_{-\infty}^{+\infty}e^{-\frac{z^{2}}{2\sigma_{i}^{2}}}erf((k_1+k_2)z+k_1(\mu_i+\mu_{-i}))dt\\
 &+\frac{k_1k_2}{(k_1+k_2)^2}\mu_i\int_{-\infty}^{+\infty}ze^{-\frac{z^{2}}{2\sigma_{i}^{2}}}erf((k_1+k_2)z +k_1(\mu_i+\mu_{-i}))dt\\>&0.
\end{align*}
This holds since both term 1 and term 2 are greater than zero. Similarly, 

iv) when $\mu_i<-\mu_{-i}$, we have $\frac{dU_1(\mu_i)}{d\mu_i}<0$, and $\frac{dU_2(\mu_i)}{d\mu_i}<0$. 

Therefore, the necessary condition for a strategy profile to be Nash Equilibrium is the same as the ones in Theorem~\ref{theorem2}. Hence, the uniqueness of the Nash Equilibrium is a direct extension result of Theorem~\ref{theorem3}.
 \endproof

This completes the proof when $\rho_i\in[0,1)$. Next, we calculate the cost under $\rho_i=1$. 
\proof{} When $\rho_i=1$, there exist a positive linear relationship between $\Delta_i$ and $\Delta_{-i}$. Previously people have proved that when $\rho_i=1$, under the Gaussian distribution, 
$\delta_i-\mu_i=c\cdot(\delta_{-i}-\mu_{-i})$, {for all possible numerical values} $(\delta_i, \delta_{-i}),$
where $c=\frac{\sigma_i}{\sigma_{-i}}$~\cite{sachs2012applied}.

It is easy to verify that when the pricing model is linear symmetric, what we need to calculate is the term 
$\mathbb{E}[\xi_{2}\Delta_{i}]$:

\begin{equation}
\begin{split}
&\mathbb{E}[\xi_{2}\Delta_{i}]=\int_{-\infty}^{\frac{\sigma_{-i}}{\sigma_{i}+\sigma_{-i}}\mu_{i}-\frac{\sigma_{i}}{\sigma_{i}+\sigma_{-i}}\mu_{-i}}\delta_{i}b_2p_{d}f^{\mu_i}_{\Delta_{i}}(\delta_{i})d\delta_{i}+\int_{\frac{\sigma_{-i}}{\sigma_{i}+\sigma_{-i}}\mu_{i}-\frac{\sigma_{i}}{\sigma_{i}+\sigma_{-i}}\mu_{-i}}^{+\infty}\delta_ib_{1}p_{d}f^{\mu_i}_{\Delta_{i}}(\delta_{i})d\delta_{i}\\
&=b_1p_d\mu_i+(b_2-b_1)p_d\int_{-\infty}^{\frac{\sigma_{-i}}{\sigma_{i}+\sigma_{-i}}\mu_{i}-\frac{\sigma_{i}}{\sigma_{i}+\sigma_{-i}}\mu_{-i}}\delta_{i}f^{0}_{\Delta_{i}}(\delta_{i}-\mu_i)d\delta_{i}\\
&=b_1p_d\mu_i+(b_2-b_1)p_d\left[\int_{-\infty}^{\frac{-\sigma_{i}}{\sigma_{i}+\sigma_{-i}}\mu_{i}-\frac{\sigma_{i}}{\sigma_{i}+\sigma_{-i}}\mu_{-i}}\delta_{i}f^{0}_{\Delta_{i}}(\delta_{i})d\delta_{i}+\int_{-\infty}^{\frac{-\sigma_{i}}{\sigma_{i}+\sigma_{-i}}\mu_{i}-\frac{\sigma_{i}}{\sigma_{i}+\sigma_{-i}}\mu_{-i}}\mu_{i}f^{0}_{\Delta_{i}}(\delta_{i})d\delta_{i}\right]\\
&=\frac{b_1+b_2}{2}p_d\mu_i+(b_2-b_1)p_d\left[\int_{-\infty}^{\frac{-\sigma_{i}}{\sigma_{i}+\sigma_{-i}}\mu_{i}-\frac{\sigma_{i}}{\sigma_{i}+\sigma_{-i}}\mu_{-i}}\delta_{i}\cdot f^{0}_{\Delta_{i}}(\delta_{i})d\delta_{i}+\int_{0}^{\frac{-\sigma_{i}}{\sigma_{i}+\sigma_{-i}}\mu_{i}-\frac{\sigma_{i}}{\sigma_{i}+\sigma_{-i}}\mu_{-i}}\mu_{i}f^{0}_{\Delta_{i}}(\delta_{i})d\delta_{i}\right].
\end{split}
\end{equation}
By an abuse of the notation, we define
\begin{equation}
\begin{split}
&U(\mu_i)=\int_{-\infty}^{-\frac{\sigma_{i}}{\sigma_{i}+\sigma_{-i}}\mu_{i}-\frac{\sigma_{i}}{\sigma_{i}+\sigma_{-i}}\mu_{-i}}\delta_{i}f^{0}_{\Delta_{i}}(\delta_{i})d\delta_{i} +\int_{0}^{-\frac{\sigma_{i}}{\sigma_{i}+\sigma_{-i}}\mu_{i}-\frac{\sigma_{i}}{\sigma_{i}+\sigma_{-i}}\mu_{-i}}\mu_{i}f^{0}_{\Delta_{i}}(\delta_{i})d\delta_{i}.
\end{split}
\end{equation}
Let $k=\frac{\sigma_{i}}{\sigma_{i}+\sigma_{-i}}\in(0, 1)$. Then we have
$$U(\mu_i)=\int_{-\infty}^{-k(\mu_{i}+\mu_{-i})}\delta_{i}f^{0}_{\Delta_{i}}(\delta_{i})d\delta_{i}+\int_{0}^{-k(\mu_{i}+\mu_{-i})}\mu_{i}f^{0}_{\Delta_{i}}(\delta_{i})d\delta_{i}.$$

We can calculate the derivative with respect to $\mu_i$ as
\begin{equation}
\begin{split}
&\frac{dU(\mu_i)}{d\mu_i}=k^2(\mu_{i}+\mu_{-i})f^{0}_{\Delta_{i}}(k(\mu_{i}+\mu_{-i}))-k\mu_if^{0}_{\Delta_{i}}(k(\mu_{i}+\mu_{-i})) -\int_{0}^{k(\mu_{i}+\mu_{-i})}f^{0}_{\Delta_{i}}(\delta_{i})d\delta_{i}.
\end{split}
\end{equation}
Given $\mu_{-i}=0$, we have 
$$\frac{dU(\mu_i)}{d\mu_i}=k^2\mu_{i}f^{0}_{\Delta_{i}}(k\mu_{i})-k\mu_if^{0}_{\Delta_{i}}(k\mu_{i})-\int_{0}^{k\mu_{i}}f^{0}_{\Delta_{i}}(\delta_{i})d\delta_{i}.$$
Then, it can be shown that

i) when $\mu_i>0$, we have
$$k^2\mu_{i}f^{0}_{\Delta_{i}}(k\mu_{i})-k\mu_if^{0}_{\Delta_{i}}(k\mu_{i})-\int_{0}^{k\mu_{i}}f^{0}_{\Delta_{i}}(\delta_{i})d\delta_{i}<k^2\mu_{i}f^{0}_{\Delta_{i}}(k\mu_{i})-{2k\mu_{i}}f^{0}_{\Delta_{i}}(k\mu_{i})<0;$$

ii) when $\mu_{i}<0$, we have
$$k^2\mu_{i}f^{0}_{\Delta_{i}}(k\mu_{i})-k\mu_if^{0}_{\Delta_{i}}(k\mu_{i})-\int_{0}^{k\mu_{i}}f^{0}_{\Delta_{i}}(\delta_{i})d\delta_{i}>k^2\mu_{i}f^{0}_{\Delta_{i}}(k\mu_{i})-{2k\mu_{i}}f^{0}_{\Delta_{i}}(k\mu_{i})>0.$$
This proves the part that bidding according to prediction is the pure strategy Nash Equilibrium.

The efficiency part is the same as the independent case in Theorem~\ref{theorem4}.

As to the (0, $N-1$) fault immune robustness part, given $\mu_i=0$, we have
$$\frac{dU(\mu_{-i})}{d\mu_{-i}}=k^2\mu_{-i}f^{0}_{\Delta_{i}}(k\mu_{-i}).$$
It is easy to justify that when $\mu_{-i}>0$ ($\mu_{-i}<0$ respectively), the above derivative is positive (negative respectively).

Finally let us prove the uniqueness part. Let us first focus on the case that $\mu_{-i}>0$, we have

i) when $\mu_i\geq0$, it can be shown that
\begin{equation}
\begin{split}
\frac{dU(\mu_i)}{d\mu_i}&=k^2(\mu_{i}+\mu_{-i})f^{0}_{\Delta_{i}}(k(\mu_{i}+\mu_{-i}))-k\mu_if^{0}_{\Delta_{i}}(k(\mu_{i}+\mu_{-i}))-\int_{0}^{k(\mu_{i}+\mu_{-i})}f^{0}_{\Delta_{i}}(\delta_{i})d\delta_{i}\\
&<k^2(\mu_{i}+\mu_{-i})f^{0}_{\Delta_{i}}(k(\mu_{i}+\mu_{-i}))-k\mu_if^{0}_{\Delta_{i}}(k(\mu_{i}+\mu_{-i}))-{k(\mu_{i}+\mu_{-i})}f^{0}_{\Delta_{i}}(k(\mu_{i}+\mu_{-i}))\\
&<0;
\end{split}
\end{equation}

ii) when $\mu_i\leq-\mu_{-i}$, we have
\begin{equation}
\begin{split}
\frac{dU(\mu_i)}{d\mu_i}&=k^2(\mu_{i}+\mu_{-i})f^{0}_{\Delta_{i}}(k(\mu_{i}+\mu_{-i}))-k\mu_if^{0}_{\Delta_{i}}(k(\mu_{i}+\mu_{-i}))-\int_{0}^{k(\mu_{i}+\mu_{-i})}f^{0}_{\Delta_{i}}(\delta_{i})d\delta_{i}\\
&\geq k^2(\mu_{i}+\mu_{-i})f^{0}_{\Delta_{i}}(k(\mu_{i}+\mu_{-i}))-k\mu_if^{0}_{\Delta_{i}}(k(\mu_{i}+\mu_{-i}))-{k(\mu_{i}+\mu_{-i})}f^{0}_{\Delta_{i}}(k(\mu_{i}+\mu_{-i}))\\
&>k^2(\mu_{i}+\mu_{-i})f^{0}_{\Delta_{i}}(k(\mu_{i}+\mu_{-i}))-{k(\mu_{i}+\mu_{-i})}f^{0}_{\Delta_{i}}(k(\mu_{i}+\mu_{-i}))\\
&\geq0.
\end{split}
\end{equation}
Therefore, the necessary condition for a strategy profile to be Nash Equilibrium is the same as the ones in Theorem~\ref{theorem2}. Hence, the uniqueness of the Nash Equilibrium is a direct extension result of Theorem~\ref{theorem3}.
\endproof
This completes the proof of Theorem~\ref{theorem7}. 

\endproof
\subsection{Proof of Theorem~\ref{theorem8}}
\proof{} We prove the statements in Theorem~\ref{theorem8} one by one.
\subsubsection{Proof of Theorem~\ref{theorem8} (1)}
\proof{}
The existence of the pure strategy Nash Equilibrium is an direct result of Theorem~\ref{theorem3} and Theorem~\ref{theorem6} (1) since $f_i(p_d)>0, \forall p_d\in R^+$. It is easy to prove that all the third terms of (\ref{general-cost}) are piece-wise continuous on $R^+$, which means that $\mathbb{E}[\mathbf{\textsf{ABC}}_{i}|\, p_d]\cdot f_i(p_d)$ is also piece-wise continuous on $R^+$.
The second claim comes from that if $\mathbb{E}[\mathbf{\textsf{ABC}}_{i}|\, \mu_i(p_d)=\bar{\mu_i}{(p_d)}]>\mathbb{E}[\mathbf{\textsf{ABC}}_{i}|\, \mu_i(p_d)=\hat{\mu_i}{(p_d)}], \forall p_d\in(s, t)$, and $\mathbb{E}[\mathbf{\textsf{ABC}}_{i}|\, \mu_i=\bar{p_d}]\geq\mathbb{E}[\mathbf{\textsf{ABC}}_{i}|\, \mu_i=\hat{p_d}], \forall p_d\in R/(s, t)$, then the integral difference~\cite{rao2018measure}
$$C_i(\bar{\mu}_i(p_d), 0)-C_i(\hat{\mu}_i(p_d), 0)>0.$$
The above relationship comes from that $\mathbb{E}[\mathbf{\textsf{ABC}}_{i}|\, \bar{\mu}_i(p_d)]\cdot f_i(p_d)$ and $\mathbb{E}[\mathbf{\textsf{ABC}}_{i}|\, \hat{\mu}_i(p_d)]\cdot f_i(p_d)$ are both piece-wise continuous on $(s, t)$. Therefore, there exists an interval $[s', t']$ such that $\mathbb{E}[\mathbf{\textsf{ABC}}_{i}|\, \bar{\mu}_i(p_d)]\cdot f_i(p_d)$ and $\mathbb{E}[\mathbf{\textsf{ABC}}_{i}|\, \hat{\mu}_i(p_d)]\cdot f_i(p_d)$ are both continuous and $\mathbb{E}[\mathbf{\textsf{ABC}}_{i}|\, \bar{\mu}_i(p_d)]\cdot f_i(p_d)> \mathbb{E}[\mathbf{\textsf{ABC}}_{i}|\, \hat{\mu}_i(p_d)]\cdot f_i(p_d)$. As such, the above inequality holds~\cite{rao2018measure}.

This completes the proof of Theorem~\ref{theorem8} (1). 
\endproof

\subsubsection{Proof of Theorem~\ref{theorem8} (2)}
\proof{}
The uniqueness of the pure strategy Nash Equilibrium is also similar to the case of quantity bid, as shown in Theorem~\ref{theorem2}, Theorem~\ref{theorem3}, and Theorem~\ref{theorem6} (1) since $f_i(p_d)>0, \forall p_d\in R^+$. Specifically, assume $\boldsymbol{\hat{\mu}^{*}(p_d)}=(\hat{\mu}_{1}(p_d),\hat{\mu}_{2}^{*}(p_d),...,\hat{\mu}_{N}^{*}(p_d))$ is a strategy profile different from $\boldsymbol{\mu^{*}(p_d)}=(\mu_{1}^{*}(p_d),\mu_{2}^{*}(p_d),...,\mu_{N}^{*}(p_d))=\vec{0}$. Since $\mu_i(p_d)$ are all piece-wise continuous functions without any isolated point such that the continuity condition is not satisfied, then there must exist an interval $(s, t)$ such that $\mu_i(p_d)$ are all continuous for $i=1, 2,...,N$ and at least one $\mu_j(p_d)\neq 0$ at the entire domain of $(s, t)$.  Therefore, $\mu_{-i}(p_d)$ are also continuous and can not always be zero in $(s, t)$ for at least one $i=1,2,...,N$. Based on the above observation, we see that there also exists an interval $(s_1, t_1) \subset (s, t)$ such that $\mu_{-i}(p_d)$ are either always zero, always positive, or always negative within $(s_1, t_1)$ for $i=1,2...,N$ and exist at least one $\mu_{-j}(p_d)$ such that $\mu_{-j}(p_d)$ is either always positive or negative within $(s_1, t_1)$. Let $(\mu_1(p_d), \mu_2(p_d),...,\mu_m(p_d))$ be a set of strategies with $\mu_{-i}(p_d)\neq 0 \ \forall p_d\in(s_1, t_1)$ (such set exists). It is easy to prove that any other $\mu_j(p_d)$ that does not in the above set must always be zero since $\mu_{-j}(p_d)=0, \forall p_d\in(s_1, t_1)$ and $\mu_j(p_d)$ is continuous. Since the above strategy profile constitutes a Nash Equilibrium, then all utilities play the best response to others' behaviors. Let us focus on the interval $(s_1, t_1)$,  given the continuous non-zero $\mu_{-1}(p_d)$, there must exist another smaller interval $(s_2, t_2) \subset (s_1, t_1)$ such that $\mu_1(p_d)$ is in the range of $(-\mu_{-1}(p_d), 0)$ or $(0, -\mu_{-1}(p_d))$ depending on whether $\mu_{-1}(p_d)$ is positive or not in $(s_1, t_1) \subset (s, t)$. This argument can be proved by contradiction. Without loss of generality, let us assume $\mu_{-1}(p_d)>0 \ \forall p_d \in(s_1, t_1)$, if there does not exist an interval $(s_2, t_2) \subset (s_1, t_1)$ such that $\mu_1({p_d}) \in(-\mu_{-1}(p_d), 0)$, then we must have $\mu_1({p_d})\geq 0$ or $\mu_1({p_d})\leq -\mu_{-1}(p_d)$ for all $p_d\in(s_1, t_1)$ since it is continuous, which means utility $i$ can always get higher benefit by playing the best response in the interval $(s_1, t_1)$ compared with the assumed case, which cause a contradiction that the strategy profile is a Nash Equilibrium. (Such observation comes from that compared with utility 1 play $\mu^*_1(p_d)$ a little bit more than $-\mu_{-i}(p_d)$ if $\mu_i(p_d)\leq-\mu_{-i}(p_d)$ and play $\mu^*_1(p_d)$ a little bit less than $0$ if $\mu_i(p_d)\geq 0$ while satisfying continuity, we must have $\mathbb{E}[\mathbf{\textsf{ABC}}_{1}|\, {\mu}_i(p_d)]\cdot f_i(p_d)>\mathbb{E}[\mathbf{\textsf{ABC}}_{1}|\, {\mu}^*_i(p_d)]\cdot f_i(p_d)$, and they are both piece-wise continuous on $(s_1, t_1)$ with $\mathbb{E}[\mathbf{\textsf{ABC}}_{i}|\, {\mu}_i(p_d)]$ and $\mathbb{E}[\mathbf{\textsf{ABC}}_{i}|\, {\mu}^*_i(p_d)]$ continuous. Therefore, there exists a smaller interval $(s', t')$ such that $\mathbb{E}[\mathbf{\textsf{ABC}}_{i}|\, \bar{\mu}_i(p_d)]\cdot f_i(p_d)$ and $\mathbb{E}[\mathbf{\textsf{ABC}}_{i}|\, \hat{\mu}_i(p_d)]\cdot f_i(p_d)$ are continuous since $f_i(p_d)$ is piece-wise continuous and $\mathbb{E}[\mathbf{\textsf{ABC}}_{i}|\, \bar{\mu}_i(p_d)]\cdot f_i(p_d)>\mathbb{E}[\mathbf{\textsf{ABC}}_{i}|\, \hat{\mu}_i(p_d)]\cdot f_i(p_d)$). Therefore, we know that there exists an  interval $(s_2, t_2) \subset (s_1, t_1)$  such that utility $1's$ strategy $\mu_1(\hat{p_d}) \in(-\mu_{-1}(p_d), 0)$ if $\mu_{-1}(p_d)>0$ or $\mu_1({p_d}) \in(0, -\mu_{-1}(p_d))$ if $\mu_{-1}(p_d)<0$. By the same analysis, we conclude that there exists a smaller interval $(s_3, t_3) \subset (s_2, t_2)$ such that utility $2's$ strategy $\mu_2({p_d}) \in(-\mu_{-2}(p_d), 0)$ if $\mu_{-2}(p_d)>0$ or $\mu_i({p_d}) \in(0, -\mu_{-2}(p_d))$ if $\mu_{-2}(p_d)<0$. Following this way, we conclude that there exists an interval $(s_{N+1}, t_{N+1})$ such that all utilities' bidding curve satisfy the following condition
$$\mu_i({p_d}) \in(-\mu_{-i}(p_d), 0) \ {\rm{if}}\  \mu_{-i}(p_d)>0 \ {\rm{or}}\  \mu_i({p_d}) \in(0, -\mu_{-i}(p_d)) \ {\rm{if}}\  \mu_{-i}(p_d)<0, \ \forall i=1,2,...,m.$$

Let us consider any point $\hat{p_d} \in(s_{N+1}, t_{N+1}).$ From the result of Theorem~\ref{theorem3}, such $(\mu_1(\hat{p_d}),\\\mu_2(\hat{p_d}),...,\mu_m(\hat{p_d}))$ does not exist, which cause a contradiction. 
Therefore, we conclude that the pure strategy Nash  Equilibrium is unique.

This completes the proof of Theorem~\ref{theorem8} (2). 
\endproof

\subsubsection{ Proof of Theorem~\ref{theorem8} (3)}
\proof{}
The efficiency of the pure strategy Nash Equilibrium is an direct result of Theorem~\ref{theorem4} and Theorem~\ref{theorem6} (3) since $f(p_d)>0, \forall p_d\in R^+$. It is easy to prove that $\mathbb{E}[\mathbf{\textsf{ABC}}_{total}|\, \mu(p_d)]$ is piece-wise continuous, which means $\mathbb{E}[\mathbf{\textsf{ABC}}_{total}|\, \mu(p_d)]\cdot f(p_d)$ is also piece-wise continuous on $R^+$. The second claim comes from that if $\mathbb{E}[\mathbf{\textsf{ABC}}_{total}|\, \mu(p_d)=\bar{\mu}{(p_d)}]>\mathbb{E}[\mathbf{\textsf{ABC}}_{i}|\, \mu_i(p_d)=\hat{\mu_i}{(p_d)}], \forall p_d\in(s, t)$, and $\mathbb{E}[\mathbf{\textsf{ABC}}_{total}|\, \mu=\bar{p_d}]\geq\mathbb{E}[\mathbf{\textsf{ABC}}_{total}|\, \mu=\hat{p_d}], \forall p_d\in R/(s, t)$, then the integral difference~\cite{rao2018measure}
$$C_{total}(\bar{\mu}(p_d))-C_{total}(\hat{\mu}(p_d))>0.$$
The above relationship comes from that $\mathbb{E}[\mathbf{\textsf{ABC}}_{total}|\, \bar{\mu}(p_d)]\cdot f(p_d)$ and $\mathbb{E}[\mathbf{\textsf{ABC}}_{total}|\, \hat{\mu}(p_d)]\cdot f(p_d)$ are both piece-wise continuous on $(s, t)$. Therefore there exists an interval $[s', t']$ such that $\mathbb{E}[\mathbf{\textsf{ABC}}_{total}|\, \bar{\mu}(p_d)]\cdot f(p_d)$ and $\mathbb{E}[\mathbf{\textsf{ABC}}_{total}|\, \hat{\mu}(p_d)]\cdot f(p_d)$ are both continuous and $\mathbb{E}[\mathbf{\textsf{ABC}}_{total}|\, \bar{\mu}(p_d)]\cdot f(p_d)> \mathbb{E}[\mathbf{\textsf{ABC}}_{total}|\, \hat{\mu}(p_d)]\cdot f(p_d)$. As such, the above inequality holds~\cite{rao2018measure}.

This completes the proof of Theorem~\ref{theorem8} (3). 
\endproof

\subsubsection{ Proof of Theorem~\ref{theorem8} (4)}
\proof{}
The (0, $N-1$) fault immune robustness  of the pure strategy Nash Equilibrium is an direct result of Theorem~\ref{theorem5} and Theorem~\ref{theorem6} (4) since $f_j(p_d)>0, \forall p_d\in R^+$. It is easy to prove that $\mathbb{E}[\mathbf{\textsf{ABC}}_{j}|\, p_d]\cdot f_j(p_d)$ is piece-wise continuous on $R^+$. The second claim comes from that if $\mathbb{E}[\mathbf{\textsf{ABC}}_{j}|\, \mu_S(p_d)=\bar{\mu_S}{(p_d)}]<\mathbb{E}[\mathbf{\textsf{ABC}}_{j}|\, \mu_S(p_d)=\hat{\mu_i}{(p_d)}], \forall p_d\in(s, t)$, and $\mathbb{E}[\mathbf{\textsf{ABC}}_{j}|\, \mu_S(p_d)=\bar{\mu}{(p_d)}]\leq\mathbb{E}[\mathbf{\textsf{ABC}}_{j}|\, \mu_S(p_d)=\hat{\mu}{(p_d)}], \forall p_d\in R/(s, t)$, then the integral difference~\cite{rao2018measure}
$$C_{j}(\bar{\mu}_S(p_d),0)-C_{j}(\hat{\mu}_S(p_d), 0)<0.$$
The above relationship comes from that $\mathbb{E}[\mathbf{\textsf{ABC}}_{j}|\, \mu_S(p_d)=\bar{\mu}{(p_d)}]\cdot f_j(p_d)$ and $\mathbb{E}[\mathbf{\textsf{ABC}}_{i}|\, \mu_S(p_d)=\hat{\mu}{(p_d)}]\cdot f_j(p_d)$ are both piece-wise continuous on $(s, t)$. Therefore there exists an interval $[s', t']$ such that $\mathbb{E}[\mathbf{\textsf{ABC}}_{j}|\, \mu_S(p_d)=\bar{\mu}{(p_d)}]\cdot f_j(p_d)$ and $\mathbb{E}[\mathbf{\textsf{ABC}}_{i}|\, \mu_S(p_d)=\hat{\mu}{(p_d)}]\cdot f_j(p_d)$ are both continuous and $\mathbb{E}[\mathbf{\textsf{ABC}}_{j}|\, \mu_S(p_d)=\bar{\mu}{(p_d)}]\cdot f_j(p_d)<\mathbb{E}[\mathbf{\textsf{ABC}}_{i}|\, \mu_S(p_d)=\hat{\mu}{(p_d)}]\cdot f_j(p_d)$. As such, the above inequality holds~\cite{rao2018measure}.

This completes the proof of Theorem~\ref{theorem8} (4). 
\endproof

Thus we complete the proof of Theorem~\ref{theorem8}.
\endproof

\subsection{Proof of Theorem~\ref{theorem9}}
\proof{}
The results of Theorem~\ref{theorem9} are direct extensions of Theorem~\ref{theorem7}
and Theorem~\ref{theorem8} since $f_i(p_d)>0 $ and $f(p_d)>0, \forall p_d\in R^+$. The existence, uniqueness, efficiency, and robustness of the equilibrium can be proved by the same approach as shown in Theorem~\ref{theorem8}. We omit the repeated steps here.

This completes the proof of Theorem~\ref{theorem9}. 
\endproof

\end{appendices}

\end{document}